\documentclass[twocolumn]{aastex62}

\usepackage{xcolor} 
\definecolor{bg-green}{rgb}{0.8588,0.9333,0.8666}
\definecolor{Q-color}{rgb}{0.5,0.1,0.5}

\usepackage{CJK}
\usepackage{epsf, graphicx, subfigure}
\usepackage{amsmath} 
\usepackage{ulem}
\usepackage{natbib}
\usepackage{color}
\usepackage{colortbl}
\usepackage{hyperref}
\hypersetup{
    colorlinks,%
    citecolor=blue,%
    filecolor=black,%
    linkcolor=blue,%
    urlcolor=black
}

\renewcommand{\arraystretch}{1.2}

\shorttitle{AGN Feedback in Galaxy Clusters}
\shortauthors{Qiu et al.}

\begin{document} 

\begin{CJK}{UTF8}{bsmi} 

\title{The Interplay of Kinetic and Radiative Feedback in Galaxy Clusters}

\author{Yu Qiu (邱宇)$^\dagger$}
\email{$^\dagger$geoyuqiu@gatech.edu}
\affiliation{Center for Relativistic Astrophysics, Georgia Institute of Technology, Atlanta, GA 30332}

\author{Tamara Bogdanovi\'c$^\ddagger$}
\email{$^\ddagger$tamarab@gatech.edu}
\affiliation{Center for Relativistic Astrophysics, Georgia Institute of Technology, Atlanta, GA 30332}

\author{Yuan Li}
\affiliation{Center for Computational Astrophysics, Flatiron Institute, New York, NY 10010}
\affiliation{Department of Astronomy, University of California, Berkeley, CA 94720}

\author{KwangHo Park}
\affiliation{Center for Relativistic Astrophysics, Georgia Institute of Technology, Atlanta, GA 30332}

\author{John H. Wise}
\affiliation{Center for Relativistic Astrophysics, Georgia Institute of Technology, Atlanta, GA 30332}


\begin{abstract}
Recent observations provide evidence that some cool-core clusters (CCCs) host quasars in their brightest cluster galaxies (BCGs). Motivated by these findings we use 3D radiation-hydrodynamic simulations with the code \texttt{Enzo} to explore the joint role of the kinetic and radiative feedback from supermassive black holes (SMBHs) in BCGs. We implement kinetic feedback as sub-relativistic plasma outflows and model radiative feedback using the ray-tracing radiative transfer or thermal energy injection. In our simulations the central SMBH transitions between the radiatively efficient and radiatively inefficient states on timescales of a few Gyr, as a function of its accretion rate. The timescale for this transition depends primarily on the fraction of power allocated to each feedback mode, and to a lesser degree on the overall feedback luminosity of the active galactic nucleus (AGN). Specifically, we find that (a) kinetic feedback must be present at both low and high accretion rates in order to prevent the cooling catastrophe, and (b) its contribution likely accounts for $>10\%$ of the total AGN feedback power, since below this threshold simulated BCGs tend to host radio-loud quasars most of the time, in apparent contrast with observations. We also find a positive correlation between the AGN feedback power and the mass of the cold gas filaments in the cluster core, indicating that observations of H$\alpha$ filaments can be used as a measure of AGN feedback. 
\end{abstract}
 


\section{Introduction} \label{sec:intro}

Galaxy clusters are the largest gravitationally bound systems in the universe with mass as high as $\sim10^{14-15}\,\text{M}_\sun$. Contributing over 80\% of the total mass, dark matter is the most dominant constituent of galaxy clusters. In the absence of direct observations of the dark matter, however, observational studies commonly resort to measurements of the luminous baryonic content. Most of the baryonic matter in clusters lies in the hot plasma ($T\gtrsim 10^7\,\text{K}$), also known as the intracluster medium (ICM). The ICM cools mainly by emission of bremsstrahlung radiation, with the luminosity $\propto n_i n_e T^{1/2}$, where $n_i$, $n_e$, $T$ are the ion number density, electron number density, and temperature of the plasma, respectively. Unchecked, ICM cooling would produce a cooling flow of $\gtrsim100\,M_\sun\,\text{yr}^{-1}$ and spur continuous star formation at cluster centers that would result in bluer and brighter BCGs than those seen in observations \citep{Fabian1994}. This discrepancy implies that a heating mechanism must be present to reduce, or possibly shut off, star formation in cluster cores. 

In about a half of all resolved galaxy clusters, known as the cool-core clusters (CCCs), the central ICM temperature is lower than the virial temperature of the gas \citep{Hudson2010}. Because of the higher density of the ICM in these clusters, their central ($r \lesssim100\,\text{kpc}$) cooling times are much shorter than the Hubble time \citep{VoigtFabian2004}. The ICM in cores of CCCs however seems to maintain the temperature corresponding to $\sim30-50\,\%$ of the virial temperature on timescales of several gigayears \citep[][]{Allen2001}. The lack of ICM plasma cooler than $k_\text{B}T\sim2\,\text{keV}$ also points to the existence of an active heating mechanism that counters the radiative cooling. 

Currently, a prevailing paradigm is that the main heating source inside cluster cores are active galactic nuclei (AGNs) within BCGs, powered by accretion onto their central SMBHs \citep{Fabian2012}. Broadly known as the AGN feedback, it can be categorized in two main mechanisms: the {\it radiative (or quasar-mode) feedback} that releases energy through photon emission from the nucleus, and the {\it kinetic (or radio-mode) feedback} that does so through ejection of relativistic particles\footnote{The term ``radio-mode feedback'' refers to the synchrotron emission of relativistic jet plasma observed at radio wavelengths. In this work, we use it interchangeably with ``kinetic feedback" and ``jetted feedback''.}. Of the two, the radio-mode feedback has been extensively studied in simulations in the past \citep[e.g.,][]{Vernaleo2006, Cattaneo2007, Dubois2010, Gaspari2012, Gaspari2013, Li2014b, Li2015, Prasad2015, Yang2016}. These earlier studies find that jetted feedback can deliver a sufficient amount of energy to the cooling flow to prevent or slow down the cooling catastrophe. The details of precisely how kinetic feedback couples to the ICM are still being investigated. 

The impact of radiative feedback has previously been explored in local and cosmological simulations of radiation-regulated black hole accretion and stellar feedback \citep[e.g.,][]{Sijacki2007, Ciotti2010, Choi2012, Vogelsberger2013, Vogelsberger2014, Gan2014, Park2017, Smidt2017, Weinberger2017, Yuan2018, Emerick2018}, but has not been considered in the context of CCCs. This choice was largely motivated by the lack of luminous quasars observed in nearby galaxy clusters. \citet{Green2017}, however, show that this may be a consequence of a selection effect, where an X-ray selected AGN or a quasar are identified as the dominant source, and whether they reside in the BCG of a galaxy cluster has not been investigated. This selection effect is likely to more strongly affect higher redshift objects ($z\geq1$), where association of an AGN with a galaxy cluster becomes more observationally challenging. Evidence that radiative and kinetic feedback may coincide in galaxy clusters is provided by \citet{Russell2013}, who find that about 50\% of the sample of 57 BCGs with prominent X-ray cavities (indicative of radio-mode feedback) also have detectable compact X-ray nuclei.

Understanding the impact of radiative feedback, in addition to jets, is important in light of the large amounts of cold gas that have been observed in central BCGs of galaxy clusters \citep[$>10^{10}\,M_\sun$;][]{ODea2008}. Radiative feedback can in principle affect the thermodynamics of the cold gas through photo-heating and radiation pressure. Hence, even though the fraction of time AGNs in BCGs spend in the radiatively efficient state may be small, the impact of radiative feedback on the evolution of CCCs merits investigation. 

Motivated by these and other observations of CCCs, we perform a suite of 3D radiation-hydrodynamic simulations of a galaxy cluster, with an aim to explore the joint role of kinetic and radiative feedback powered by accretion onto the SMBH in the central cluster galaxy. The layout of this paper is as follows: we introduce numerical methods in Section \ref{sec:method}, present the results in Section \ref{sec:result}, compare our simulations with observations in Section \ref{sec:observation}, discuss the implications in Section \ref{sec:discus}, and conclude in Section \ref{sec:conclusion}.


\section{Methodology} \label{sec:method}

\subsection{Numerical Setup} \label{sec:m_code}

The simulations are performed using a modified version of the adaptive mesh refinement (AMR) hydrodynamic code \texttt{Enzo}\footnote{http://enzo-project.org}, version 2.5 \citep{Enzo2014}, with the ray-tracing radiative transfer package \textsc{moray} \citep{Wise2011}. The hydrodynamic solver we use in our simulations is a 3D adaptation of the \textsc{zeus-2d} code \citep{Stone1992} implemented in \texttt{Enzo}. The isolated cluster is placed at the center of the computational domain with size $(500\,\text{kpc})^3$, in non-comoving coordinates. The domain has outflowing boundaries and is initially divided into a Cartesian grid with $128^3$ cells, resulting in a base grid with the resolution of 3.9\,kpc. On top of the base grid we employ up to four refinement levels, resulting in the finest resolution of 0.24\,kpc. The refinement divides a cell in two equal parts along each axis and is triggered when either of the following criteria is satisfied:
\begin{enumerate}
\item {\it Gas density:} Refinement level $l$ is created when $\rho\ge\rho_l=\rho_{l-1}\times 2^{(\alpha+3)}$. We choose the initial refinement density, $\rho_1 = 5.4\times10^{-26}\,{\rm g\,cm^{-3}}$, that corresponds to the radius of about 40\,kpc at the beginning of the simulation, and $\alpha=-1.2$, resulting in a higher degree (super-Lagrangian) of refinement at higher densities.

\item {\it Cooling time:} Both the cooling time, $t_\text{cool}={e_\text{th}}/({n^2\Lambda})$, and the sound-crossing time, $t_\text{s}={\Delta x}/{c_\text{s}}$, are calculated for each cell, where $e_\text{th}$ is the thermal energy density, $n = n_i + n_e$ is the plasma number density, $\Lambda$ is the cooling function, $\Delta x$ is the size of a cell, and $c_\text{s}$ is the local sound speed. A refinement level is added when the ratio $t_\text{cool}/t_\text{s}<\beta$. Following \citet{Li2012}, we choose $\beta=6$ to better resolve the gas that is rapidly cooling.
\end{enumerate}

Furthermore, the time step for radiative transfer ($dt_\text{P}$) is set by limiting the change of $\textsc{H\,i}$ density in each cell, caused by photoionization, to $<10\%$ or by the light crossing time of the smallest cell, whichever is greater \citep[see][for more detail]{Wise2011}. 

\subsection{Cluster Initialization} \label{sec:m_c}

The initial setup of the simulated cluster is based on the Perseus cluster and is similar to \citet{Li2012}. To model the ICM, the gas density and temperature are initialized with spherically symmetric profiles. We adopt the electron number density profile for the Perseus cluster \citep{Churazov2004,Mathews2006, Li2012}
\begin{equation}
\begin{split}
	n_e(r)=\frac{0.0192}{1+\left(\frac{r_{\rm kpc}}{18}\right)^3}+\frac{0.046}{\left[1+\left(\frac{r_{\rm kpc}}{57}\right)^2\right]^{1.8}}\\
	+\frac{0.0048}{\left[1+\left(\frac{r_{\rm kpc}}{200}\right)^2\right]^{1.1}}\,\text{cm}^{-3},
\end{split}
\end{equation}
where $r_{\rm kpc}$ is the radius from the center of the cluster in kpc. The temperature profile adopted in this study is obtained from the X-ray observations by \citet{Churazov2004}:
\begin{equation}
	k_\text{B}T(r) =7\,\frac{1+\left(\frac{r_{\rm kpc}}{71}\right)^3}{2.3+\left(\frac{r_{\rm kpc}}{71}\right)^3}\,\text{keV}.
\end{equation}

The ICM is composed of the following species: $e^-$, $\textsc{H\,i}$, $\textsc{H\,ii}$, $\text{He}\textsc{\,i}$, $\text{He}\textsc{\,ii}$, and $\text{He}\textsc{\,iii}$. The hydrogen mass fraction is fixed to the solar value, $X = 0.7381$. To account for metal cooling of the ICM that dominates at temperatures $\sim10^5-10^6$\,K, the metallicity of the gas is fixed to $Z = 0.0110$, corresponding to about 80\% of the solar metallicity \citep[based on the solar values in][]{Asplund2009}, and similar to the measurements in the inner regions of the Perseus cluster \citep{Schmidt2002}. The initial fractions of ionized and atomic states are calculated as equilibrium values at the initial temperature specified by the profile above. The chemistry of the gas is subsequently updated during every simulation time step.

The calculation of radiative cooling of the gas utilizes the cooling function implemented in $\texttt{Enzo}$. This cooling function explicitly accounts for the cooling of the H and He species, and is supplemented by a cooling table for metals \citep{Smith2008}. The table provides a cooling function for metal species based on \textsc{CLOUDY} photoionization calculations \citep{Ferland1998}, assuming optically thin gas, and is valid in the temperature range from 10\,K to $10^8\,\text{K}$. In this study, we do not explicitly model molecular gas, although we allow the gas to cool all the way to 10\,K, either radiatively or adiabatically.

The background gravitational potential is assumed to be static (i.e., it does not evolve over time) and includes three components: the dark matter halo, the stellar bulge of the BCG, and the SMBH with mass $M_{\rm BH}\approx 3.8\times 10^8\,M_\odot$ \citep[a factor of 1.13 higher than][]{Wilman2005}. A detailed description of these components is provided in Appendix~\ref{sec:app_acc}. We have verified that this setup for the ICM and the underlying gravitational potential results in a cluster that would be in hydrostatic equilibrium over $\sim10\, \text{Gyr}$ in the absence of any external perturbation, cooling and heating mechanisms.


\subsection{Modeling of Accretion} \label{sec:m_mdot}

The radiative and kinetic feedback in our simulations are powered by accretion onto the central SMBH. Since we do not resolve the nuclear accretion disk in our simulations, the SMBH accretion rate, $\dot{M}_\text{BH}$, is estimated from the properties of the gas surrounding the central SMBH. The feeding mechanism of SMBHs in BCGs is still an open question \citep[see for example a review by][]{McNamara2012}. The leading models are (a) the cold-mode accretion, where streams of cold gas feed the SMBH \citep{Pizzolato2005, Gaspari2017b}, and (b) the hot-mode accretion, where the SMBH accretes hot gas from a steady-state, spherically symmetric flow \citep{Bondi1952}. Observations seem to favor the former, because the hot-mode accretion does not provide sufficiently high accretion rates to sustain systems with powerful outflows \citep[with kinetic energy exceeding $10^{45}\,\text{erg\,s}^{-1}$; e.g.,][]{Rafferty2006}.

In our simulations, we consider both the cold- and hot-mode accretion. In most cases, we take the higher of the two accretion rates as the accretion rate onto the SMBH 
\begin{equation}
	\dot{M}_\text{BH}=\epsilon_{\rm acc} \,\text{max}\left[\frac{M_\text{cg}}{\tau},\,\frac{4\pi G^2\rho_\infty M_\text{BH}^2}{\left(c_\text{s}^2+v_\text{g}^2\right)^{3/2}}\right],
\label{eqn:mdot}
\end{equation}
where $\epsilon_\text{acc} = 10^{-3}$, $10^{-2}$ or $10^{-1}$ is the efficiency of gas accretion in our simulations, which implies that only a fraction of the gas residing within the nominal accretion radius, chosen to be $r_\text{a} =1\,$kpc, will accrete onto the SMBH\footnote{Accretion radius $r_\text{a} =1\,$kpc approximately corresponds to the Bondi radius of $T\approx10^5\,\text{K}$ gas.}. $M_\text{cg}$ is the amount of cold gas (defined as gas with temperature $T<3\times 10^4\,\text{K}$) enclosed within $r_\text{a}$. $\tau = 5\,\text{Myr}$ is the characteristic free-fall timescale of the gas at $\sim1\,\text{kpc}$. $\rho_\infty$ is taken to be the average density of the gas, $c_\text{s}$ is the average sound speed calculated using the mass-weighted gas temperature, and $v_\text{g}$ is the mass-weighted average velocity of the gas, all calculated within $r_\text{a}$. 

The first term within the brackets of equation~\ref{eqn:mdot} represents the {\it cold-mode accretion} fueled by the reservoir of cold gas that accumulates around the SMBH. The second term in the equation accounts for the Bondi accretion of the {\it multi-phase} gas. The two expressions are complementary in the following sense: when the cold gas is present in the central region, the accretion rate is nearly always determined by the first term. During the episodes when the cold gas reservoir is depleted by the AGN feedback, the cold-mode accretion rate can drop to zero, and the second (Bondi) term provides the accretion rate of the warmer and more dilute multi-phase gas. In each time step $dt$, the mass accreted onto the SMBH, $\dot{M}_\text{BH}\,dt$, is removed from the accretion region. We remove the mass from each cell at $r<r_\text{a}$, in proportion to the cell mass.

In order to test the impact of different subgrid prescriptions on the accretion rate, in addition to the model described above (which adopts the higher of the two accretion rates), we also pursue simulations in which the accretion rate is set either by cold-mode only, or by the sum of the cold-mode ($T<3\times 10^4\,\text{K}$) and {\it hot-mode}, Bondi accretion ($T\geq 3\times 10^4\,\text{K}$), using the expressions shown in equation~\ref{eqn:mdot} (see Appendix \ref{sec:app_am} for discussion of these accretion models). Hereafter, we refer to these three approaches to calculation of $\dot{M}_{\rm BH}$ as the cold-mode, multiphase, and hot-mode accretion.

\subsection{Implementation of AGN Feedback} \label{sec:feedback}

The central AGN is the only source of ionizing radiation in our simulations and its total feedback power is defined as
\begin{equation}
	L =\eta\,\dot{M}_\text{BH}\,c^2,
\end{equation}
where $\eta = 0.1$ is the feedback efficiency. Combining the feedback efficiency with the accretion efficiency defined above, we express the overall efficiency as $\epsilon \equiv \eta\,\epsilon_\text{acc} = 10^{-4}$, $10^{-3}$ or $10^{-2}$. We introduce this parameter to facilitate comparisons of the overall efficiency with other works in the literature.

\begin{figure}[t!]
\includegraphics[width=\linewidth]{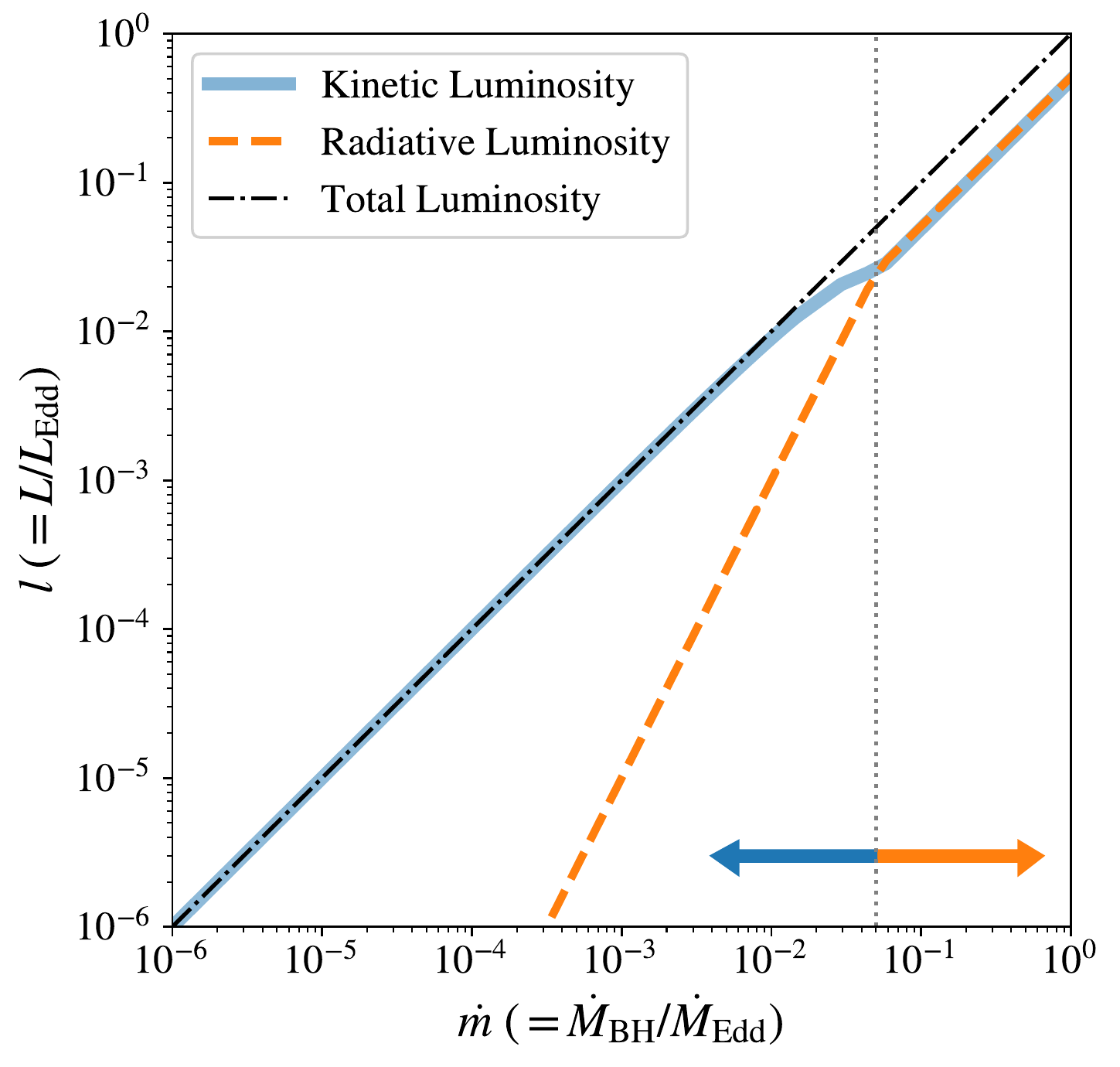}
\caption{Allocation of power to radiative (orange dashed) and kinetic (blue solid) feedback, as a function of SMBH accretion rate, $\dot{m}$. The feedback efficiency is assumed to be 10\%, so that $L_\text{Edd}=0.1\,\dot{M}_\text{Edd}\,c^2$ and $l=\dot{m}$. The vertical grey dotted line shows the transition accretion rate, $\dot{m}_\text{t}=0.05$ (see Equation \ref{eqn:trans}). The arrows mark radiatively efficient (orange) and radiatively inefficient (blue) regimes.} 
\label{fig:model}
\end{figure}

Following the model laid out in \citet{Churazov2005}, the feedback power is allocated between the two modes (radiative and kinetic) as a function of the dimensionless accretion rate, $\dot{m}=\dot{M}_\text{BH}/\dot{M}_\text{Edd}$\footnote{$L_\text{Edd}=\eta\,\dot{M}_\text{Edd}\,c^2 = 1.3\times 10^{46}\,{\rm erg\,s^{-1}}\,(M/10^8\,M_\odot)$ is the Eddington luminosity and $\dot{M}_{\rm Edd}$ is the Eddington accretion rate. For the SMBH in the Perseus cluster, $\dot{M}_\text{Edd} \approx 10\,M_\sun\,\text{yr}^{-1}$.}. This is illustrated in Figure \ref{fig:model}. According to this model, SMBHs accreting at low rates operate in the {\it radiatively inefficient} regime, and channel the bulk of their feedback power into the jet-driven outflows \citep[e.g.,][]{Narayan2008}. SMBHs characterized by higher accretion rates, $\dot{m}\gtrsim0.01$, operate in the {\it radiatively efficient} regime, in which most of their feedback power is released as radiation \citep{Shakura1973}. We too adopt these assumptions and note that the SMBH accretion rate measured in our simulations does not exceed the Eddington rate. 


In this work we modify the \citet{Churazov2005} model and parametrize the division of total feedback power in the radiatively efficient regime, between the kinetic and radiative mode of feedback. This is implemented by assigning a fraction of the total feedback power to jets, $f_\text{J}$, when $\dot{m}$ is larger than some transition rate, and apportioning the rest of the power to emitted radiation. The dimensionless jet power $l_\text{J}$, the radiative luminosity $l_\text{R}$ (both measured in units of $L_\text{Edd}$), and the transition rate $\dot{m}_\text{t}$, are determined as follows
\begin{equation}
\label{eqn:trans}
\begin{aligned}
	l_\text{R}(\dot{m}) &= 10 \dot{m}^2,\ \text{if}\ \dot{m}\le\dot{m}_\text{t}\\
	l_\text{R}(\dot{m}) &= (1-f_\text{J})\dot{m},\ \text{if}\ \dot{m}>\dot{m}_\text{t}\\
	l_\text{J}(\dot{m}) &= \dot{m} - l_\text{R}(\dot{m}) \\
	\dot{m}_\text{t} &= 0.1(1-f_\text{J})\\
\end{aligned}
\end{equation}
The last equation follows from the requirement for continuity of $l_\text{R}$ and $l_\text{J}$ between the radiatively inefficient and efficient regimes. For example, $\dot{m}_\text{t}=0.09$ when $f_\text{J}=0.1$, $\dot{m}_\text{t}=0.05$ when $f_\text{J}=0.5$, and $\dot{m}_\text{t}=0.01$ when $f_\text{J}=0.9$. Note that the allocation of power we adopt suggests that at high accretion rates, the AGNs in our simulations correspond to radio-loud quasars, whereas at low accretion rates they resemble jet-dominated, radio-loud AGNs.

\subsubsection{Radiative Feedback} \label{sec:m_q}

The radiative feedback in our simulations is implemented using two, mutually exclusive approaches: in one, we explicitly calculate radiative transfer (RT) with the ray-tracing module \textsc{moray}, and in the other, we inject thermal energy (TI) commensurate to the energy of the radiation emitted by the central AGN. 
 
{\it Simulations with radiative transfer.} In the RT approach we define the spectral energy distribution (SED) of the emitted radiation as a power-law from 13.6\,eV to 100\,keV
%
\begin{equation}
	L_\nu=\frac{L_\text{R}}{N}\nu^{-1},
\end{equation}
where $\nu$ is the photon frequency, $L_\text{R}\equiv l_\text{R}\, L_\text{Edd}$ is the frequency integrated luminosity of the ionizing radiation, and $N=\int_{13.6\,{\rm eV}}^{100\,\text{keV}}\nu^{-1}d\nu$ is the normalization factor.


\begin{deluxetable}{ccccc}[t!]
\tablecaption{Description of AGN SED}
\tablewidth{0pt}
\tablehead{
\colhead{Bin} & \colhead{$h\nu_i$} & \colhead{$h\nu_j$} & \colhead{$h\nu_{ij}$} & \colhead{$f_i$} \\
\colhead{} & \colhead{(eV)	} & \colhead{(eV)} & \colhead{(eV)} & \colhead{}
}
\startdata
1    & 13.60                         & 24.59                        & 18.02 & 0.4470                 \\ 
2    & 24.59                        & 54.42                        & 35.64 & 0.3032                 \\ 
3    & 54.42                        & 100                          & 72.64 & 0.1139                 \\ 
4    & 100                          & 1000                         & 255.8 & 0.1224                 \\ 
5    & 1000                         & 10000                        & 2558 & 0.01224                 \\ 
6    & 10000                        & 100000                       & 25584 & 0.001224                 \\
\enddata
\label{table:bin}
\tablecomments{$h\nu_i$ and $h\nu_j$ are the starting and ending energy of the bin $i$. $f_i\equiv \Delta N_i/\sum_k \Delta N_k$ is the fraction of photons in a given bin. $h\nu_{ij}$ and $\Delta N_i$ are determined by Equation \ref{eqn:bin}.}
\end{deluxetable}


\begin{deluxetable*}{cccccccccc}[t!]
\tablecaption{Simulation Parameters}
\tablewidth{0pt}
\tablehead{
\colhead{Run} & \colhead{Radiative} & \colhead{$\epsilon$} & \colhead{$f_\text{J}$} & \colhead{Accretion} & \colhead{Resolution} & \colhead{$\langle \dot{M}_{\rm BH}\rangle$} & \colhead{$\langle{L}_\text{J}\rangle$} & \colhead{$\langle{L}_\text{R}\rangle$} & \colhead{$f_\text{QSO}$} \\
\colhead{ID} & \colhead{feedback} & \colhead{ } & \colhead{	 } & \colhead{model} & \colhead{(kpc)} & \colhead{($M_\odot\,{\rm yr^{-1}})$} & \colhead{($10^{45}\,\text{erg\,s}^{-1}$)} & \colhead{($10^{45}\,\text{erg\,s}^{-1}$)} & \colhead{}
}
\startdata
CF01 	& \ldots	& 0.0           		& \ldots		& Cold     & 0.49 	 	&	$1116\pm416$	& \ldots			& \ldots			& 0.00	\\
RT01 	& RT       	& $10^{-3}$       	& 0.5        		& Max     	& 0.49		&	$1.95\pm8.30$	& $5.66\pm23.6$	& $5.44\pm23.7$	& 0.47	\\
RT02	& RT   	& $10^{-3}$    		& 0.5         	& Max    	& 0.24     		&	$0.661\pm2.26$	& $2.04\pm6.43$	& $1.73\pm6.49$	& 0.30	\\
AM01	& TI       	& $10^{-3}$        	& 0.5            	& Max   	& 0.49     		&	$0.320\pm0.496$		& $1.06\pm1.38$	& $0.761\pm1.46$	& 0.24	\\
AM02	& TI      	& $10^{-3}$       	& 0.5          	& Cold    	& 0.49     		&	$0.270\pm0.328$		& $0.944\pm0.905$	& $0.595\pm0.984$	& 0.20	\\
AM03	& TI       	& $10^{-3}$      		& 0.5         	& C+H    	& 0.49    		&	$0.220\pm0.365$		& $0.772\pm1.01$	& $0.483\pm1.08$	& 0.17	\\  
TI01$^*$ 	& TI     	& $10^{-3}$       	& 0.5    		& Max   	& 0.49     		&	$0.320\pm0.496$		& $1.06\pm1.38$	& $0.761\pm1.46$	& 0.24	\\
TI02		& TI      	& $10^{-3}$        	& 0.1              	& Max    	& 0.49   		&	$0.853\pm1.29$		& $0.810\pm0.698$	& $4.06\pm6.76$	& 0.46	\\
TI03		& TI     	& $10^{-3}$       	& 0.9              	& Max  	& 0.49    		&	$0.462\pm0.512$		& $2.37\pm2.62$	& $0.262\pm0.293$	& 0.04	\\
TI04 		& TI      	& $10^{-2}$       	& 0.5              	& C+H   	& 0.49     		&	$0.417\pm1.68$		& $1.20\pm4.80$	& $1.18\pm4.80$	& 0.12	\\
TI07      	& TI     	& $10^{-3}$        	& 0.5         	& C+H  	& 0.24   		&	$0.266\pm0.404$		& $0.962\pm1.12$	& $0.555\pm1.20$	& 0.14	\\
TI08     	& TI     	& $10^{-4}$   		& 0.5          	& Max    	& 0.49   		&	$0.300\pm0.301$		& $1.05\pm0.807$	& $0.659\pm0.926$	& 0.23	\\ 
\enddata
\label{tab:para}
\tablecomments{Radiative feedback: RT -- radiative transfer; TI -- thermal energy injection. $\epsilon$ -- Overall efficiency. $f_\text{J}$ -- Fraction of power allocated to radio-mode feedback when $\dot{m}>\dot{m}_\text{t}$. Accretion model: Max -- the larger of the cold-mode and multiphase accretion rate; Cold -- cold-mode accretion rate only; C+H -- cold-mode plus hot-mode accretion rates. $\langle \dot{M}_{\rm BH}\rangle$ -- Average accretion rate. $\langle{L}_\text{J}\rangle$, $\langle{L}_\text{R}\rangle$ -- Average kinetic and radiative luminosity, respectively, with standard deviations. $f_\text{QSO}$ -- Fraction of time with AGN radiative luminosity $\geq10^{45}$ erg\,s$^{-1}$. $^*$Simulation TI01 is the same as AM01, repeated here for easier comparison.}
\end{deluxetable*}

The RT module \textsc{moray} implemented in \texttt{Enzo} transports photon packages radially out from the source located at the center of the simulation domain. Along the ray it calculates:  (1) photo-ionization rate and (2) X-ray secondary ionization rate of hydrogen and helium atoms and ions, as well as (3) the Compton scattering leading to heating of the electrons. After passing though each cell and depositing momentum and energy, the photon count within a package is attenuated accordingly.  We do not simulate other, metal species of the ICM explicitly, and do not model radiation pressure on dust, which is in principle capable of driving strong outflows \citep[e.g.,][]{Ishibashi2018, Barnes2018}. 

Because it is computationally prohibitive to model the continuous spectrum of an AGN in our simulations, we represent the SED as a discrete function evaluated at six different photon energies. To capture the radiative processes that can take place in the intracluster gas (including photo-ionization, secondary X-ray ionization, and Compton scattering), we divide the photon energies from 13.6 eV to 100 keV into 6 bins. The sizes of the energy bins are determined by the characteristic photo-ionization energy thresholds for $\text{H}\,\textsc{i}$, $\text{He}\,\textsc{i}$, and $\text{He}\,\textsc{ii}$ for photon energies below 100\,eV, and increased by a factor of ten in each bin above 100\,eV. In each time step $\Delta t$, the representative photon energies, $h\nu_{ij}$ (between thresholds $i$ and $j$), and the photon count within a given energy bin, $\Delta N_i$, are calculated from the requirements for energy conservation and photon number conservation:
\begin{equation}
\label{eqn:bin}
\begin{aligned}
	\int_{\nu_i}^{\nu_j}L_\nu d\nu &= h\nu_{ij}\, \frac{\Delta N_i}{\Delta t}\\
	\int_{\nu_i}^{\nu_j}\frac{L_\nu d\nu}{h\nu} &=\frac{\Delta N_i}{\Delta t}
\end{aligned}
\end{equation}
The relevant binning brackets ($h\nu_i,h\nu_j$), the representative photon energies, $h\nu_{ij}$, and associated photon number fractions, $f_i\equiv {\Delta N_i}/{\sum_k \Delta N_k}$ are shown in Table~\ref{table:bin}. 

{\it Simulations with thermal injection.} In the TI approach, the effect of radiative feedback is implemented as injection of thermal energy \citep[similar to, e.g.,][]{Sijacki2007,Choi2012,Yuan2018,Weinberger2017}. In our simulations with thermal injection, the energy of the ionizing radiation emitted by the AGN within one simulation time step, $\Delta E = L_\text{R}\,\Delta t$, is added as the thermal energy to the gas enclosed within the accretion radius, $r_\text{a}$. The thermal energy is distributed among the gas cells in proportion to their mass, $m_i$, so that each cell receives ${\Delta E\,m_i}/ \sum_k m_k$. This results in the change in the specific thermal energy that is uniform across the cells:
\begin{equation}
\label{eqn:prptm}
	\Delta e_i= 
	\frac{\Delta E\,m_i}{\sum\limits_{r_k\le r_\text{a}} m_k}\,
	 \frac{1}{m_i}= \frac{L_\text{R}\, \Delta t}{\sum\limits_{r_k\le r_\text{a}} m_k}\,.
\end{equation}
Since it is less computationally expensive than the calculation of radiative transfer, the TI approach allows us to explore a wider range of model parameters in simulations.


\subsubsection{Kinetic Feedback} \label{sec:m_r}

In this work, the kinetic feedback exerted by AGN jets is approximated by sub-relativistic outflows of plasma \citep[similar to][]{Gaspari2012, Li2015, Prasad2015}. The outflows are modeled by adding kinetic energy to the gas within the region of size $(2r_{\rm a})^3$ centered on the SMBH. The gas in this region is accelerated along the jet axis, which is in all simulations fixed along the $\pm z$ axis (i.e., there is no jet precession). The change in the kinetic energy of a given cell is proportional to its mass, so that:
\begin{equation}
	\Delta k_i = \frac{L_\text{J}\, \Delta t}{\sum\limits_{r_k\le r_\text{a}} m_k},
\end{equation}
where $k_i$ is the specific kinetic energy, $L_\text{J}\equiv l_\text{J}\,L_\text{Edd}$ is the kinetic luminosity. The kinetic energy gain is then expressed as the acceleration along the jet axis
\begin{equation}
	a_z = \frac{\sqrt{v_{z}^2+2\Delta k}-v_{z}}{\Delta t},
\end{equation}
where $v_z$ is the $z$ component of the gas velocity in a cell. Note this distribution of kinetic energy is different from the simulations cited above, which do not apply the mass weighting. This results in somewhat lower outflow velocity of a $\sim{\rm few}\times 10^3\,\text{km\,s}^{-1}$ in our work \citep[relative to $\sim10^4\,\text{km\,s}^{-1}$ in][]{Gaspari2012,Li2015,Prasad2015}. See Section \ref{sec:d_jet} for a discussion of the implications of these choices.


\section{Results} \label{sec:result}

\begin{figure*}[t!]
\includegraphics[height=.21\linewidth]{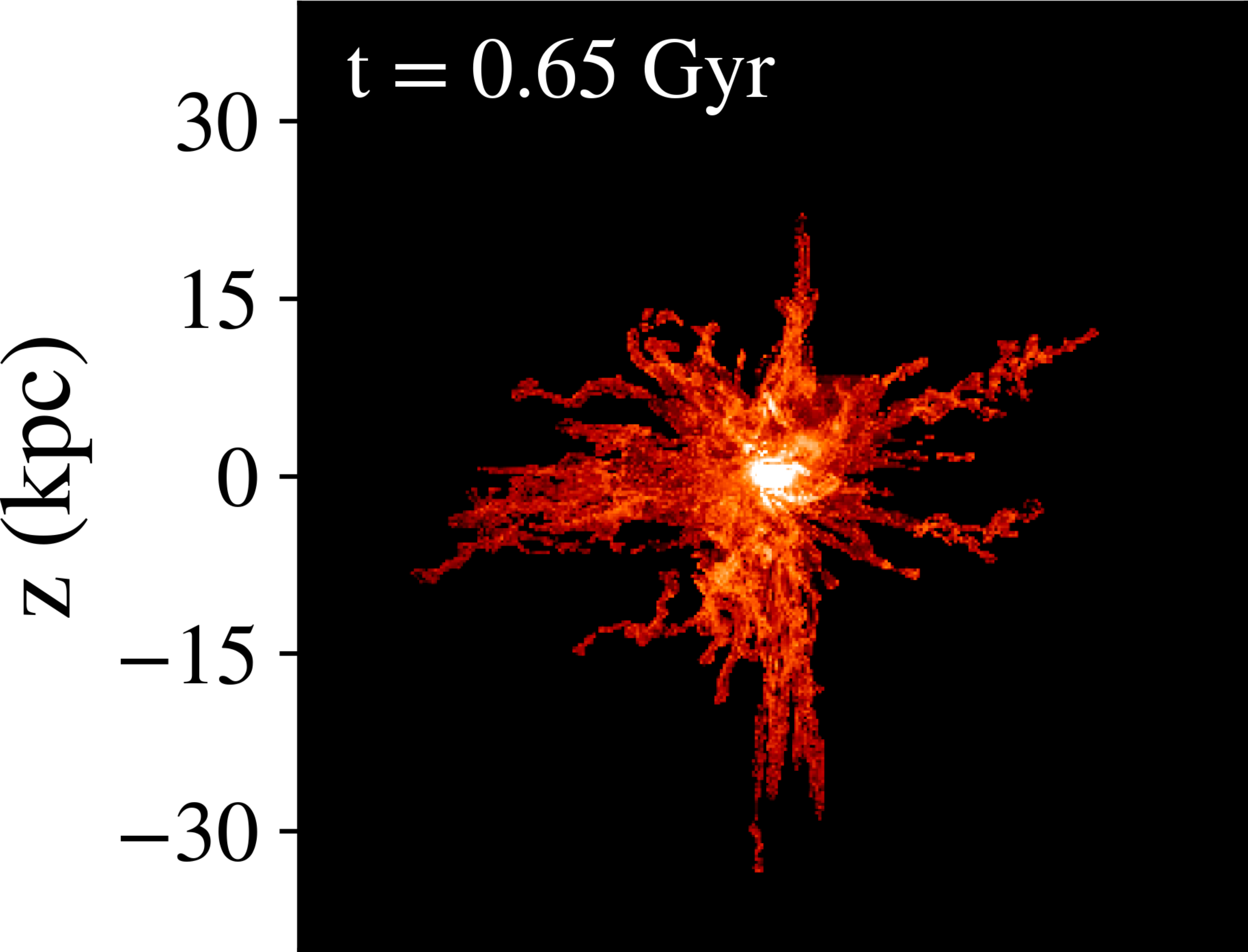}\hspace*{0.01cm}
\includegraphics[height=.21\linewidth]{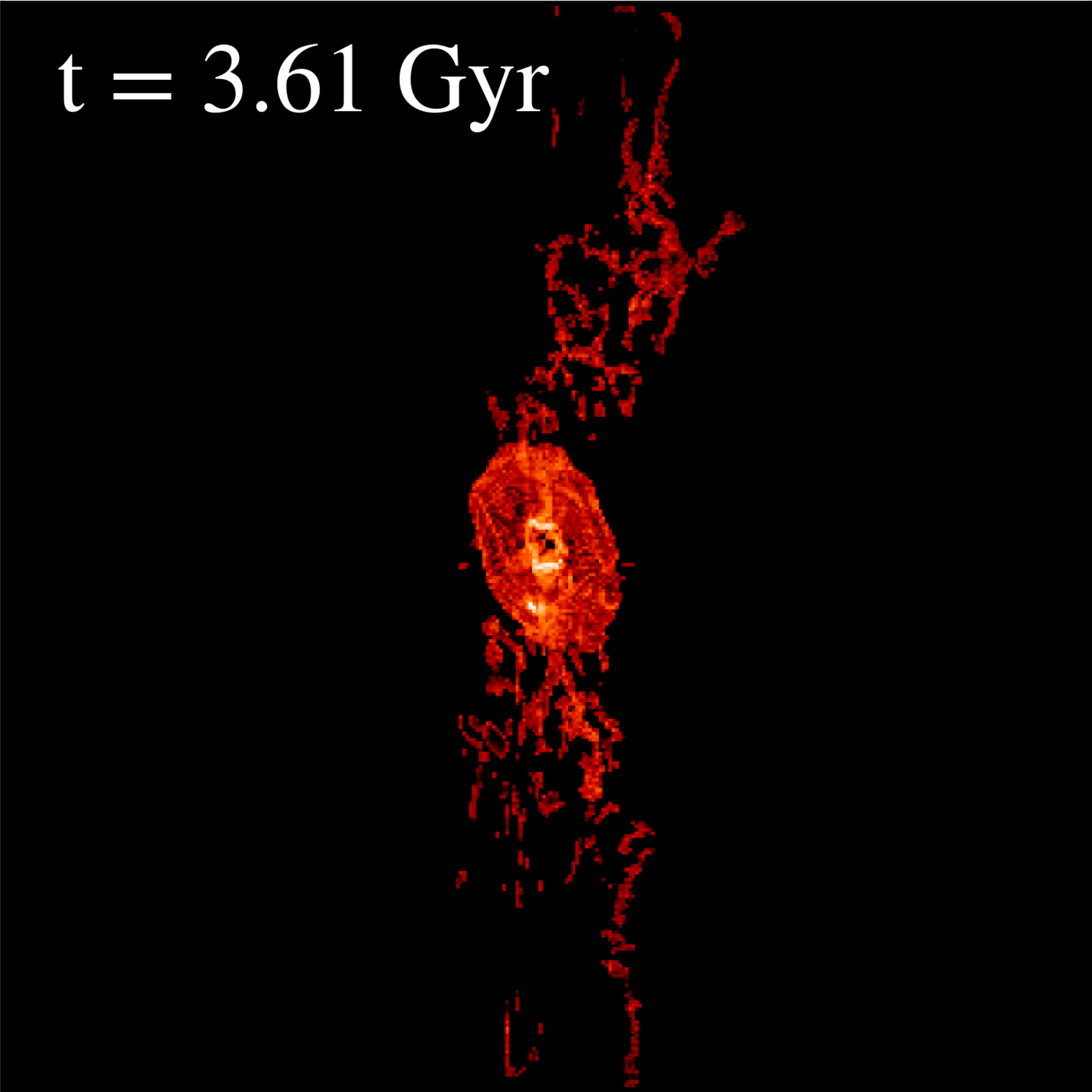}\hspace*{0.01cm}
\includegraphics[height=.21\linewidth]{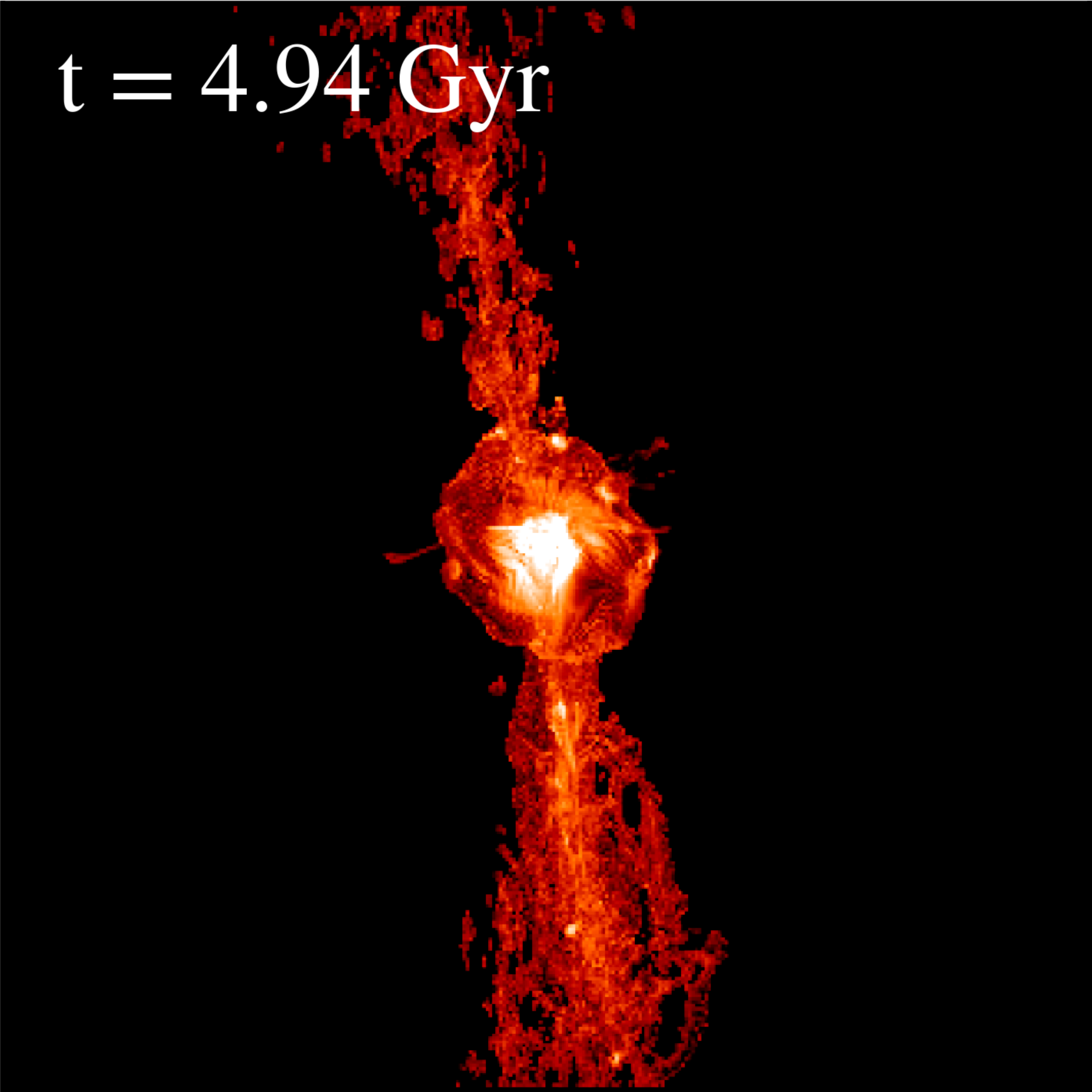}\hspace*{0.01cm}
\includegraphics[height=.21\linewidth]{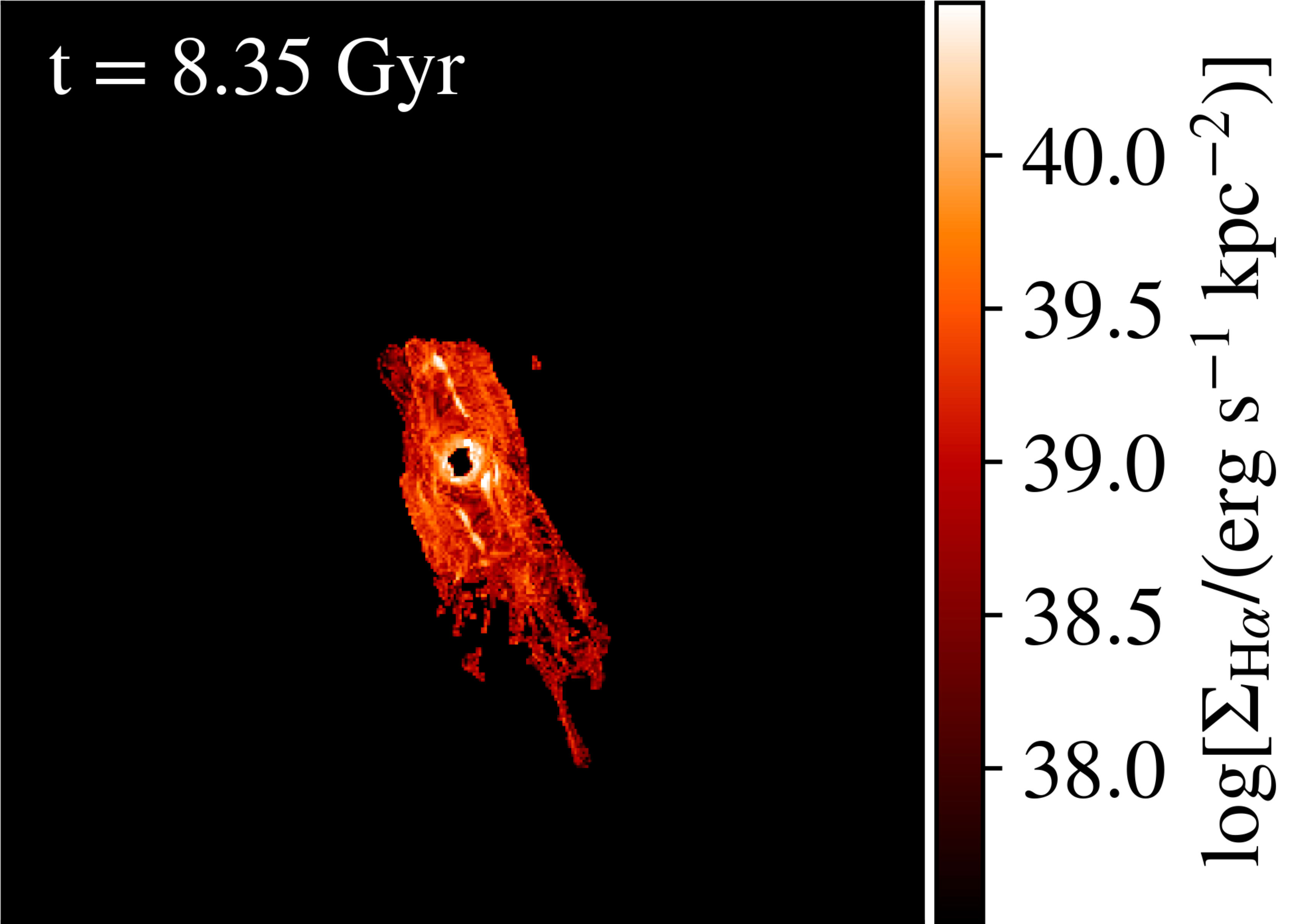}
\includegraphics[height=.21\linewidth]{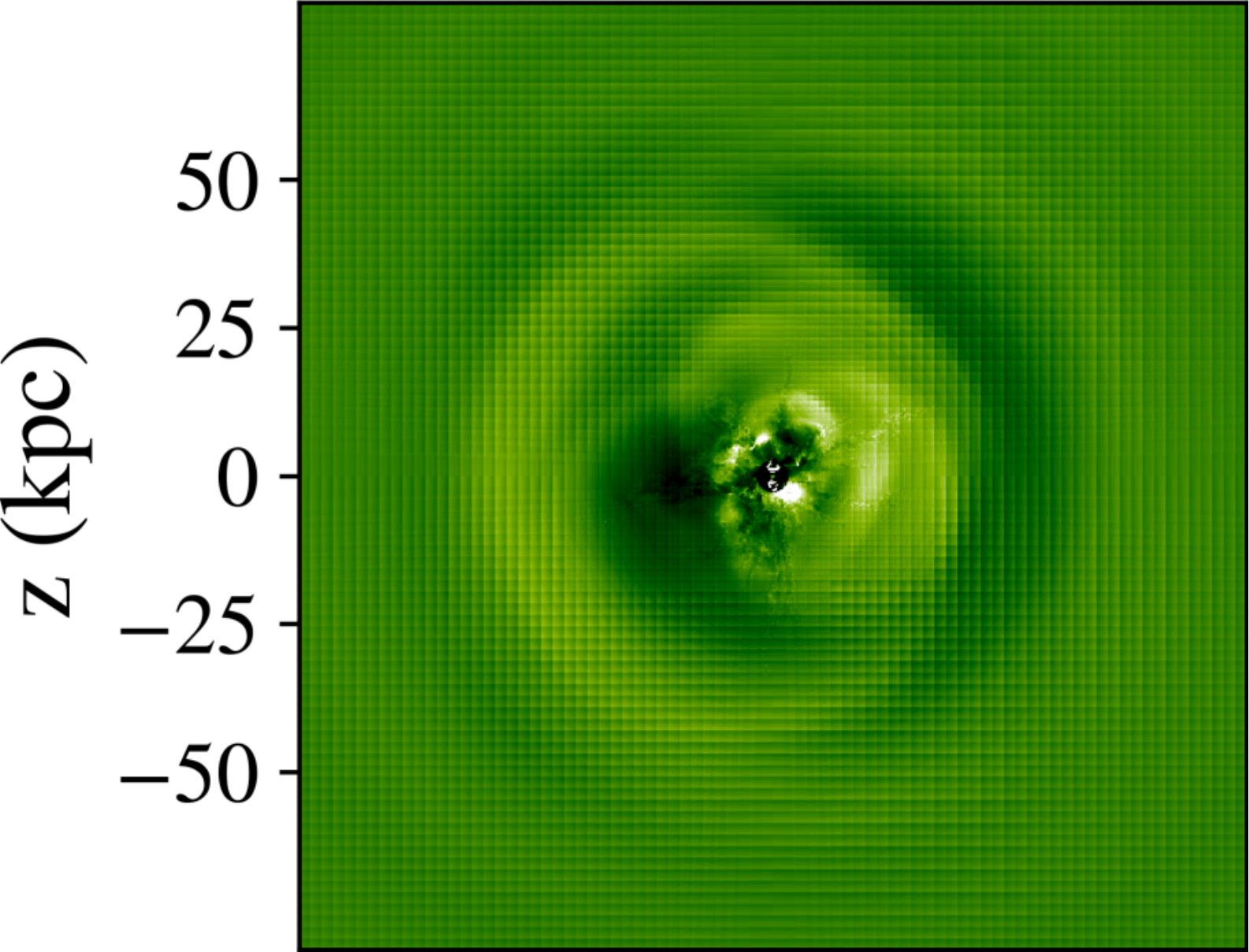}\hspace*{0.01cm}
\includegraphics[height=.21\linewidth]{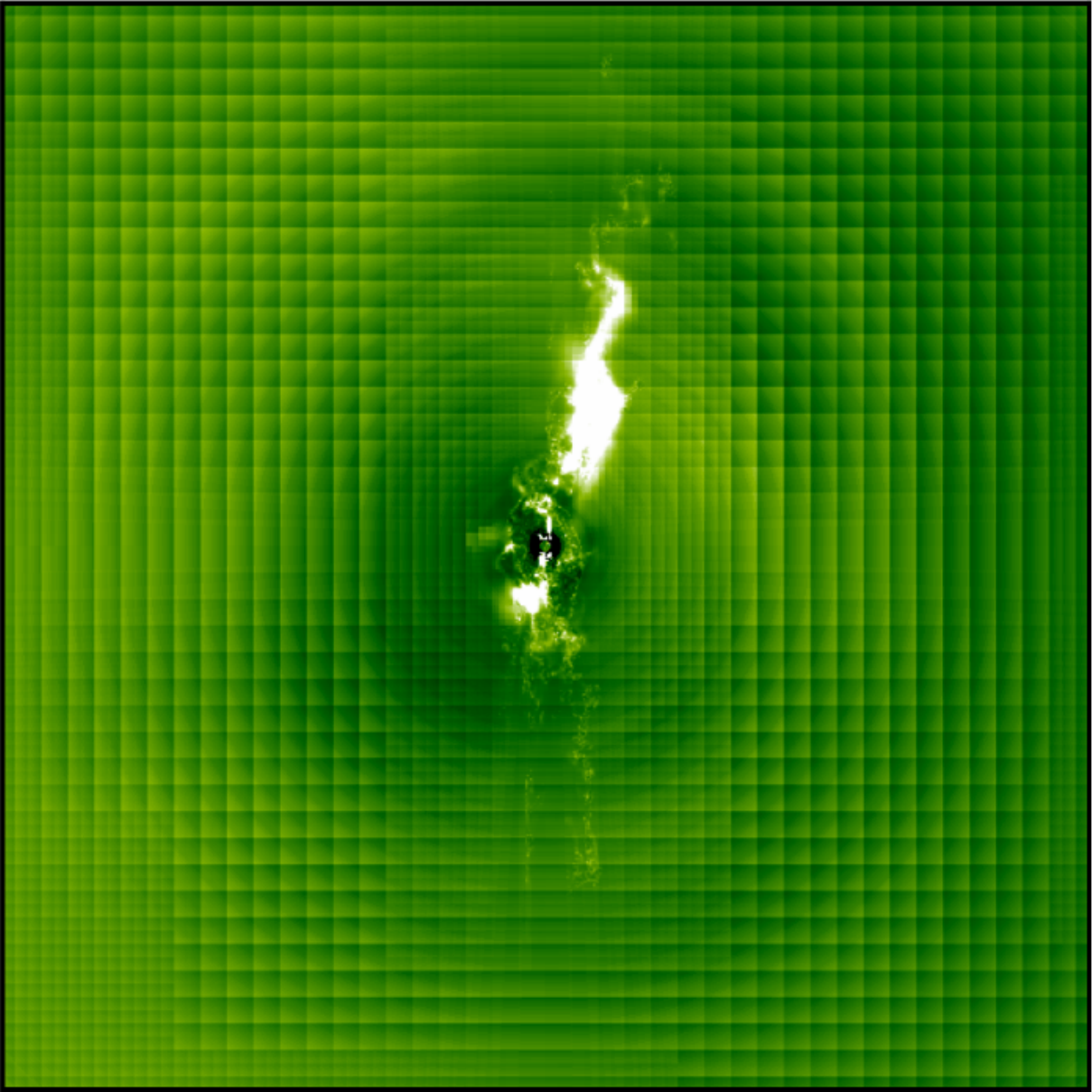}\hspace*{0.01cm}
\includegraphics[height=.21\linewidth]{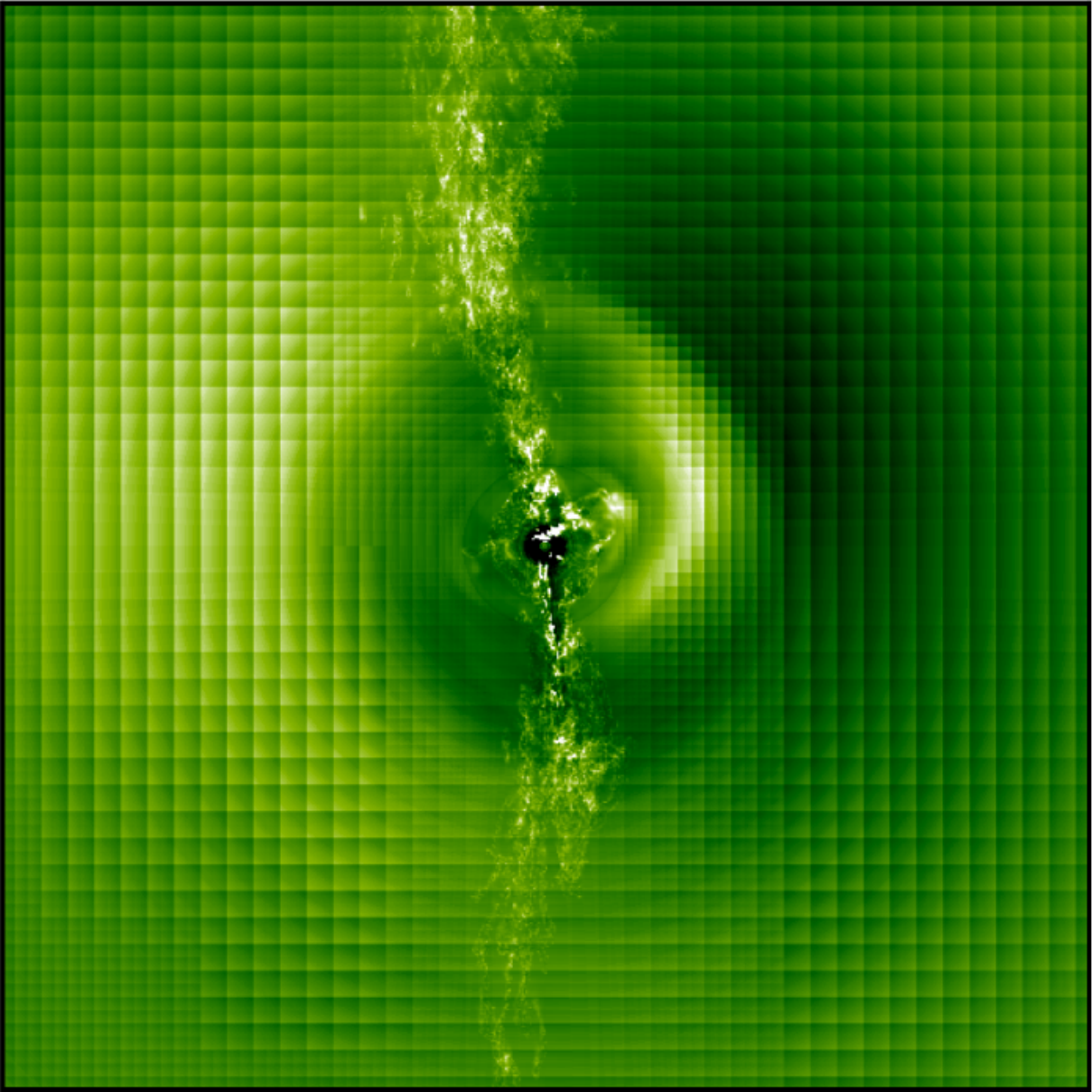}\hspace*{0.01cm}
\includegraphics[height=.21\linewidth]{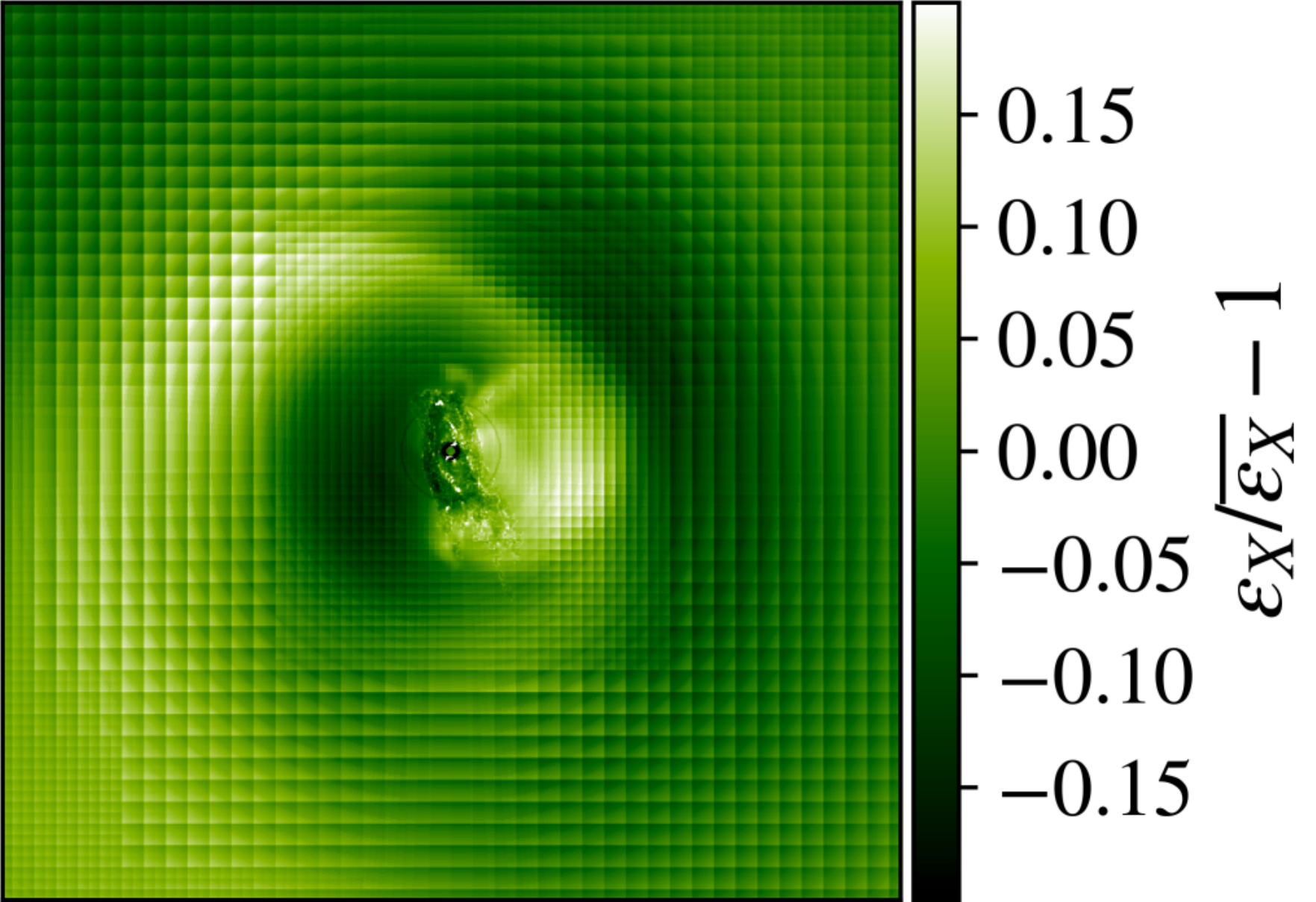}
\caption{Evolution of the ICM in run RT02. {\it Top:} Surface emissivity of H$\alpha$ associated with recombination of hydrogen in the central 80\,kpc of the cluster. The total (surface integrated) H$\alpha$ luminosity ranges between $10^{41}-10^{43}\,{\rm erg\,s}^{-1}$. {\it Bottom:} X-ray surface brightness in the central 160\,kpc of the cluster evaluated from the bremsstrahlung emissivity of the hot gas with $k_\text{B}T>0.5\,\text{keV}$. In order to emphasize the morphology of ripples and X-ray cavities, we show the fractional variance of the surface brightness, $\epsilon_X/\overline{\epsilon_X}-1$. The grid artifacts present in the X-ray emissivity maps arise as a consequence of image processing and visualization and can be ignored.} 
\label{fig:evo}
\end{figure*}

Our suite of simulations is divided in three groups, depending on the implementation of radiative feedback and sub-grid model used to evaluate the accretion rate, namely: the radiative transfer (RT), thermal injection (TI), and accretion model (AM) runs. Table \ref{tab:para} summarizes the parameters used in these runs. In all simulations, the radiative and kinetic feedback are implemented according to the prescriptions described in previous sections. In RT runs, the coupling of the radiative feedback to the ICM is evaluated by calculating the radiative transfer with \textsc{moray}. In AM and TI runs, radiative feedback is implemented as thermal energy injection. In the AM runs, we test three different accretion prescriptions by assuming that the SMBH accretion rate equals (a) the larger of the cold-mode and multiphase accretion rates (Max), (b) cold-mode accretion rate only (Cold), or (c) the sum of cold-mode and hot-mode accretion rates (C+H). Hereafter, we consider the high resolution, radiative transfer run, RT02, as the baseline model, and provide illustrations from it in a number of figures throughout the paper. For comparison with the RT, AM, and TI runs, we also carry out a simulation of a passive cooling flow (CF01), without any form of feedback.

Most simulations are carried out with the overall efficiency $\epsilon=10^{-3}$, jet power fraction $f_\text{J}=0.5$ when $\dot{m}>\dot{m}_\text{t}$, and 0.49\,kpc resolution. We consider this a standard setup for our runs. Additional TI runs are performed to explore parameters $\epsilon=10^{-4}, 10^{-2}$ and $f_\text{J}=0.1,\ 0.9$. For the purposes of a resolution study, we also perform two high resolution runs (0.24\,kpc; RT02, TI07), and describe the impact of numerical resolution on our results in Appendix~\ref{sec:app_res}.

\subsection{Distribution of the hot and cold ICM} \label{sec:r_gas}

Figure~\ref{fig:evo} illustrates the appearance and distribution of the cold and hot components of the ICM in the central region in run RT02. The features seen in this figure are qualitatively representative of those in other simulations in this suite. Figure panels correspond to two feedback dominated episodes ($t=0.65\,\text{Gyr}$ and $4.94\,\text{Gyr}$) and two quiescent episodes, characterized by the lower SMBH accretion rate and equivalently, lower AGN feedback power ($t=3.61\,\text{Gyr}$ and $8.35\,\text{Gyr}$). The top sequence of panels illustrates the distribution of the cold, atomic gas by visualizing the H$\alpha$ surface brightness of radiation associated with the recombination of hydrogen, which is characterized by the emissivity:
\begin{equation}
\begin{split}
\epsilon_{{\rm H}\alpha}&= n_e n_\textsc{H\,ii} T_4^{-0.942-0.031\,\ln T_4}\\
	&\ \ \ \times2.82 \times 10^{-26}\,  {\rm erg\,cm}^3\,{\rm s}^{-1}\,{\rm sr}^{-1}\,,
\end{split}
\end{equation}
where $T_4\equiv T/10^4\,{\rm K}$ \citep{Dong2011,Draine2011}. The volume emissivity $\epsilon_{\rm H\alpha}$ is integrated along the line of sight perpendicular to the jet axis to produce the surface emissivity shown in Figure~\ref{fig:evo}. We find that the total (surface integrated) H$\alpha$ luminosity in this simulation is $\lesssim10^{43}\, {\rm erg\,s}^{-1}$, similar to the luminosities measured in observations of the H$\alpha$ nebulae in CCCs \citep{Voit2015}.

After the first AGN outburst (at $0.65\,\text{Gyr}$), the cold gas that condenses out of the outflowing ICM takes the form of spatially extended filaments. The characteristic free-fall timescale for filaments in our simulations is $\sim 100$\,Myr and they spend most of this time at large radii, since their speed is lowest at the turnaround point of their trajectory. At $\sim1.6\, \text{Gyr}$, the filaments that fall back to the center of the gravitational potential settle into a massive disk ($\sim 10^{12}\,M_\odot$), visible in the second and subsequent panels. As they fall into the cluster center, the component of momentum along the jet axis carried by the filaments cancels out to some degree but not entirely. This causes the filaments to settle into a rotational structure nearly coplanar with the jet axis.


\begin{figure*}[t!]
\begin{center}
\includegraphics[width=0.7\linewidth]{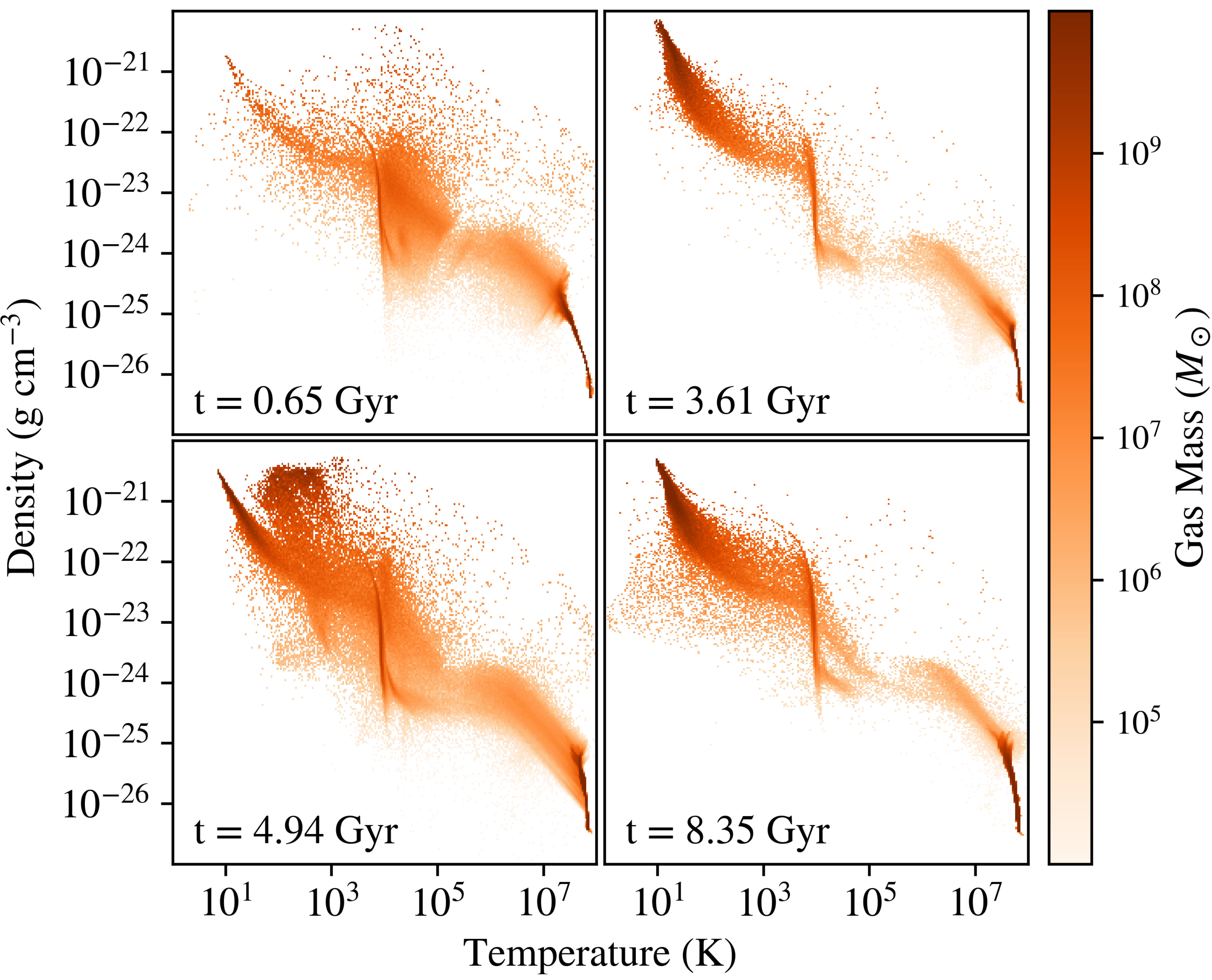}
\end{center}
\caption{Temperature vs. density of the ICM in run RT02 at four different times, matching those shown in Figure~\ref{fig:evo}. Color corresponds to the gas mass within a certain range of $T$ and $\rho$.} 
\label{fig:phase}
\end{figure*}


\begin{figure*}[t!]
\centering
\includegraphics[height=.37\linewidth]{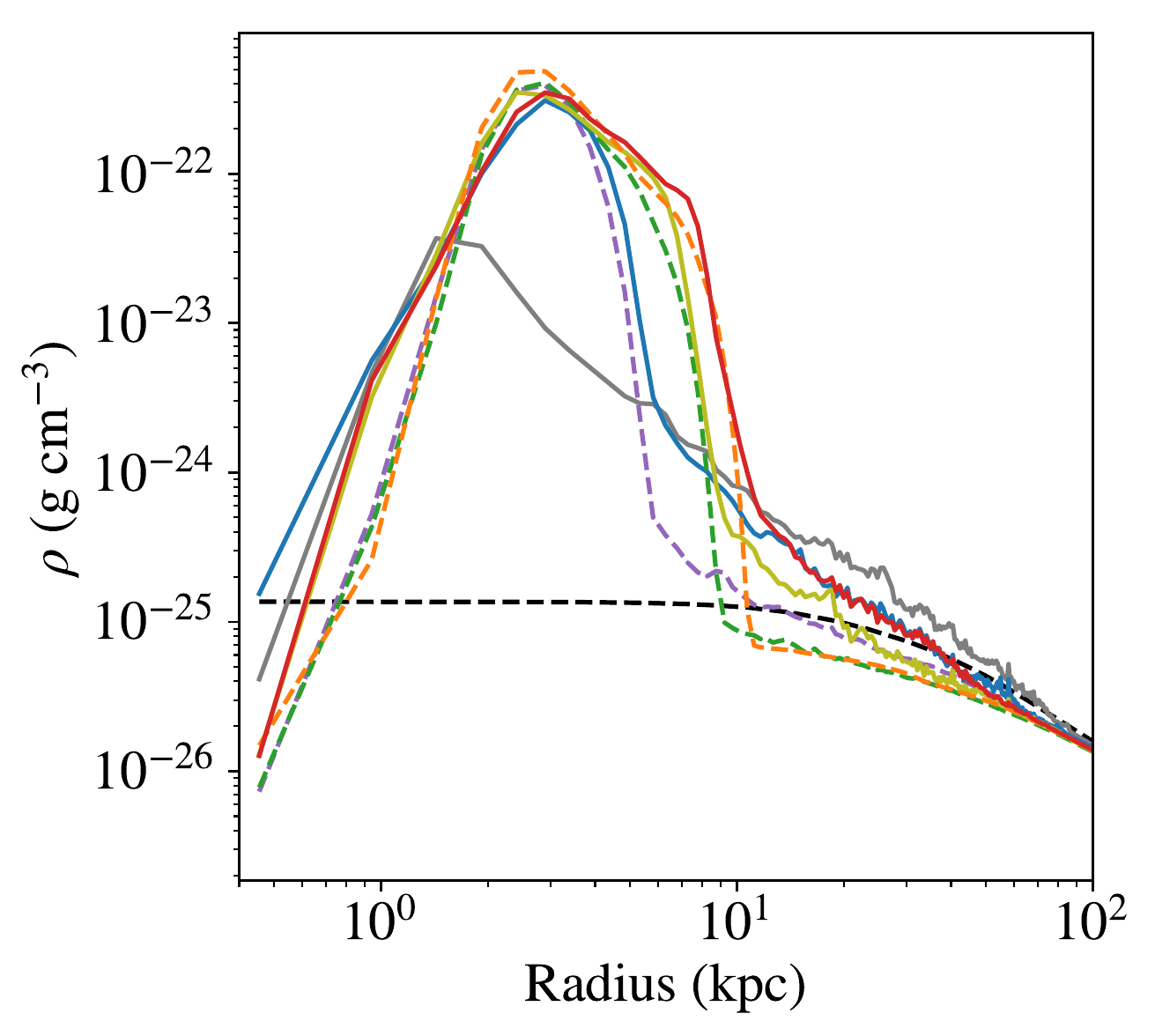}
\includegraphics[height=.37\linewidth]{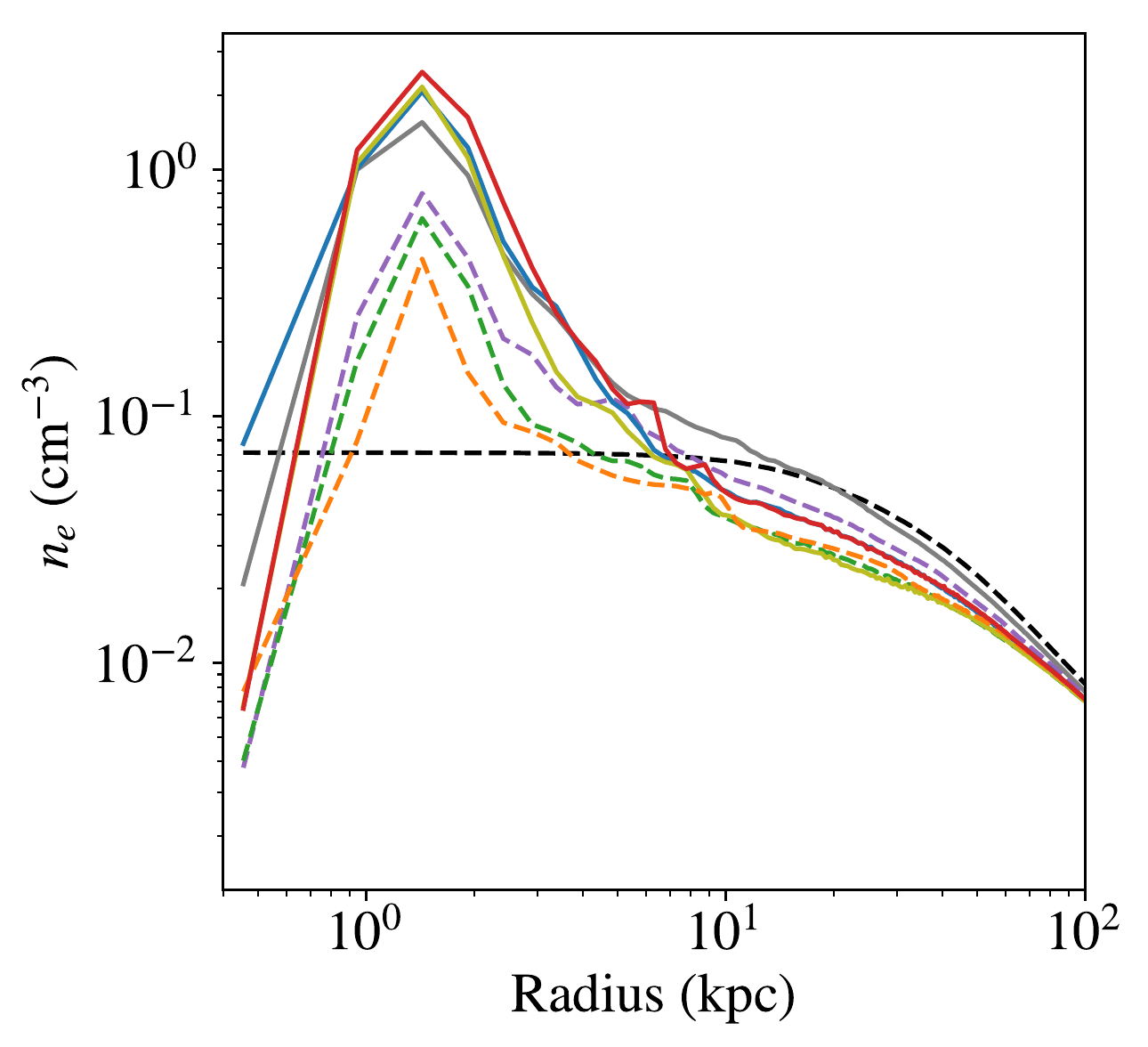}
\hspace*{0.13cm}\includegraphics[height=.37\linewidth]{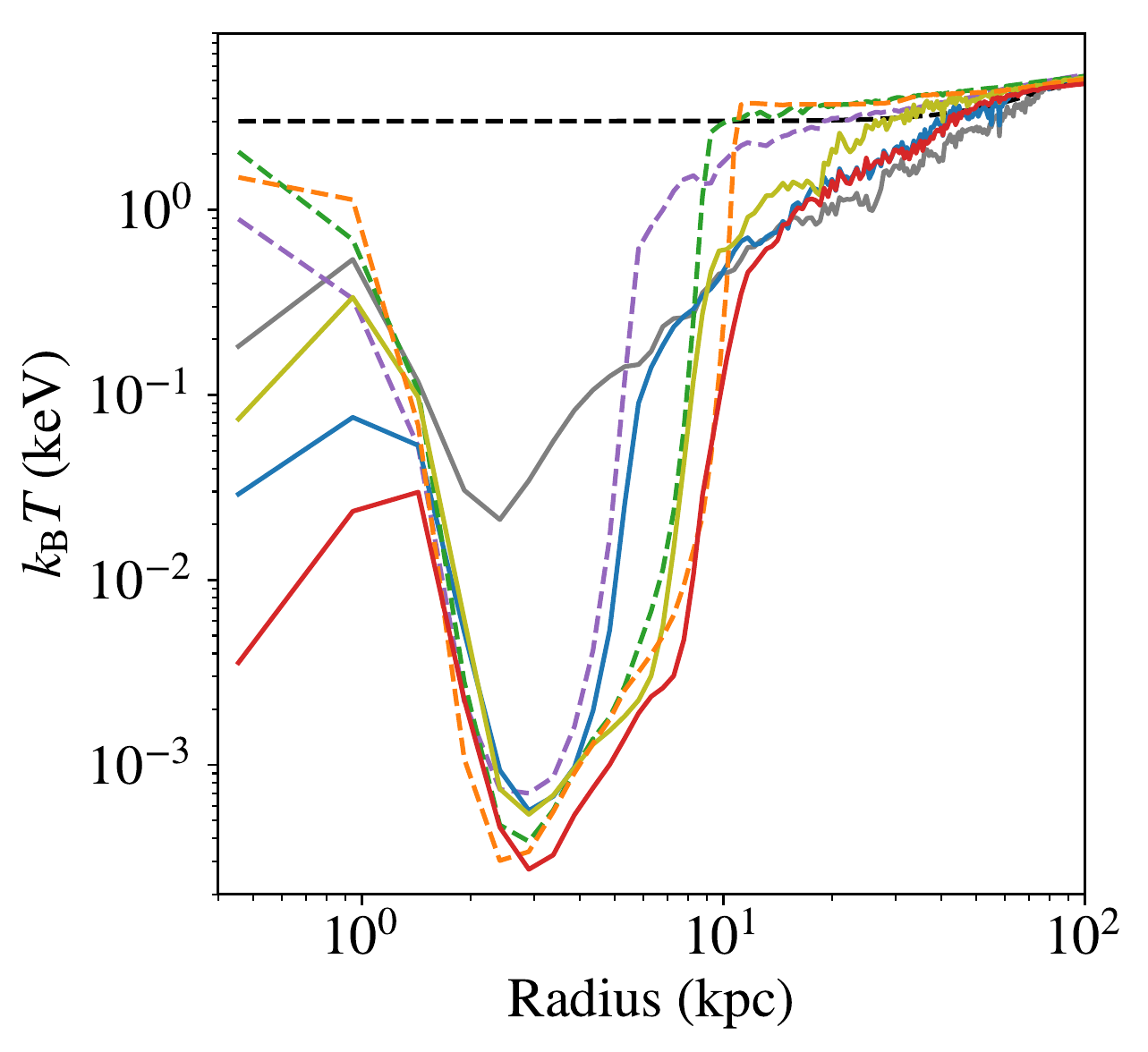}
\includegraphics[height=.37\linewidth]{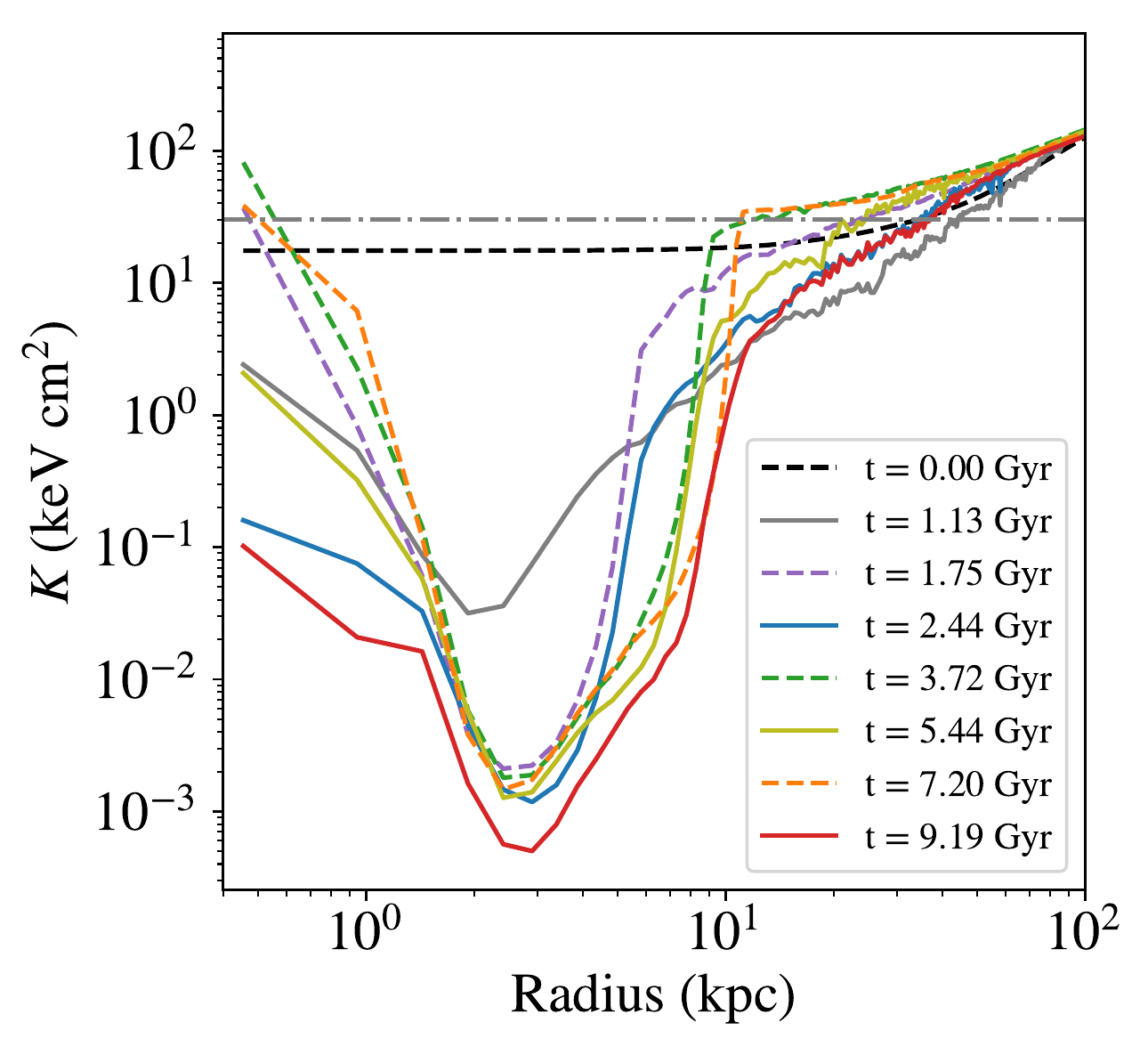}
\caption{Evolution of spherically averaged radial profiles of density ($\rho$), electron number density ($n_e$), mass-weighted temperature ($T$), and entropy ($K$) in simulation RT02. Colors represent different times associated with the maxima (solid) and minima (dashed) of the AGN feedback power shown in Figure \ref{fig:rt02}. The horizontal, grey dash-dot line in the entropy plot at $30\,\text{keV}\,\text{cm}^2$ represents the transition value from \citet{Cavagnolo2008}, below which enhanced H$\alpha$ filament emission is found in observations.}
\label{fig:radial}
\end{figure*}

 During the third AGN outburst (at $4.94\,\text{Gyr}$) the filaments again assume a spatially extended distribution, but this time appear collimated along the jet axis, which is in these images parallel to the $z$ axis. The appearance of the collimated filaments is associated with the presence of the cold gas disk, which directs outflows above and below the disk plane. The subsequent generations of cold filaments (each associated with an AGN outburst) will also fall into the massive disk, contributing to it their angular momentum. The stochastic nature of this process results in a gradual evolution in the orientation of the cold gas disk, and consequently in an evolving distribution of filaments on the timescale of gigayears. All AGN feedback outbursts are followed by a quiescent episode, characterized by the relative absence of the massive filament network.

The bottom row of panels in Figure~\ref{fig:evo} illustrates the X-ray surface brightness of the hot ICM, with $k_\text{B}T>0.5\,\text{keV}$, evaluated from the bremsstrahlung emissivity as $\epsilon_X\propto n_{e} n_i T^{0.5}$. In order to emphasize the characteristic features, such as ripples and X-ray cavities, we use the simulation snapshot to calculate the projected, 2D emissivity map and subtract from it the azimuthal average of the emissivity, $\overline{\epsilon_X}$. The bottom row of panels shows the fractional variance of the surface brightness, $\epsilon_X/\overline{\epsilon_X}-1$. The ripples visible in the images trace the sound waves propagating through the cluster core from the central AGN, which acts as a piston on the surrounding ICM via the pressure imparted by the kinetic and radiative feedback. 

The cavities, depicted as low surface brightness regions in the cluster center, represent the bubbles of low-density plasma inflated in the ICM by the AGN jets. They are scattered about the cluster core (as opposed to being aligned along the jet axis) as a consequence of the deflection of outflows by infalling filaments. Also noticeable are the filaments of the X-ray emitting gas, which are spatially coincident with the cold gas filaments (see panels two and three). Panel four shows a spiral structure which gradually develops in the ICM over several episodes of AGN feedback, eventually turning into a cold front. We discuss these features in more detail in section~\ref{sec:observation}, where we compare them to observations.

\subsection{Physical Properties of the ICM} \label{sec:r_gas2}

Figure \ref{fig:phase} shows evolution of the ICM in the temperature-density phase space for run RT02. In this representation, the hot and dilute ICM lies in the bottom-right corner of the plot, and the vertical strip at $T\approx 10^4$\,K represents the temperature threshold below which hydrogen begins to recombine into atoms. In this phase space, the gas that is cooling passively (in absence of AGN feedback), travels along the diagonal to the top-left region of the plot. In presence of AGN feedback however, the distribution of temperatures and densities of the multi-phase ICM becomes noticeably wider. For example, during the feedback dominated episodes, at $t=0.65\,\text{Gyr}$ and $4.94\,\text{Gyr}$, the ICM is intensely heated by jetted feedback and radiation. These two distributions should be compared to the more quiescent episodes at $t=3.61\,\text{Gyr}$ and $8.35\,\text{Gyr}$. It is also of interest that the feedback dominated states are characterized by substantial amounts of high density gas, $\rho > 10^{-22}\,{\rm g\,cm^{-3}}$, with temperatures in the range $10^2 - 10^6$\,K. This ICM phase appears above and below the diagonal distribution and represents the extended cold gas filaments. Eventually the filaments settle into the gas disk, characterized by rotational velocities higher than $300\,{\rm km\,s^{-1}}$. When this happens, the gas occupies the top left corner of the phase space. We note that the phase plots for other simulations presented in this work are qualitatively similar and omit them for brevity.

Figure~\ref{fig:radial} illustrates the evolution of the gas density, electron number density, mass-weighted temperature, and entropy profiles in simulation RT02. We also show the evolution of properties of the ICM for other simulations from this suite in Appendix~\ref{sec:app_radial}, for comparison. The radial profiles are calculated as averages of the relevant properties in a sequence of nested spherical shells centered on the cluster core. For example, the average mass density at a given radius is calculated as volume weighted average over resolution elements enclosed within the shell, $\rho_{\rm shell} = \sum_{i} \rho_i\,V_i/\sum_{k} V_k$. This ensures proper weighting for resolution elements of different sizes, used with adaptive mesh refinement. Similarly, the mass-weighted temperature is calculated as $T_{\rm shell} = \sum_{i} T_i\,m_i/\sum_{k} m_k$, and is representative of the temperature of the bulk of the gas by mass.

The top left panel of Figure~\ref{fig:radial} shows the mass density profile calculated using the above procedure. The large enhancement in gas density at $r<10\,$kpc, that appears at $t=1.13\,$Gyr, indicates that once it forms out of infalling filaments, the rotationally supported cold gas disk dominates in this region at all times in the simulation. Beyond the extent of the disk, at $r>10\,\text{kpc}$, the gas density profile ``breathes" about the initial value following the heating or cooling dominated episodes in the cluster core. The top right panel shows the evolution of the electron number density. Because it is closely related to the ionization state of the gas, $n_e$ in the central 10\,kpc increases in AGN feedback dominated stages, and decreases in cooling dominated stages. At $r>10$\,kpc, $n_e$ decreases as a function of time due to the cooling of the ICM. An exception to this monotonic behavior is a powerful feedback episode at $t=9.19\,\text{Gyr}$ (red solid line), which leads to the ionization of the cold, atomic gas up to tens of kiloparsecs. In comparison, in simulations where feedback is dominated by emission of radiation (e.g., simulation TI02 shown in Appendix~\ref{sec:app_radial}, where $f_{\rm J}=0.1$), the cold gas disk is more compact, and $n_e$ varies by two orders of magnitude due to the ability of radiative feedback to quickly, albeit momentarily, ionize the gas.

The temperature and entropy profiles generally decrease during cooling dominated phases and increase during feedback dominated phases, as expected. The entropy is calculated using the electron number density and mass-weighted gas temperature, as $K\equiv k_\text{B}\,T\,n_e^{-2/3}$. Within the central 10\,kpc, $T$ and $K$ are dominated by the cold gas contributing to the massive rotationally supported gas disk. At $r>10\,\text{kpc}$, the entropy profile oscillates around $30\, \text{keV}\,\text{cm}^2$. This threshold is of relevance because clusters with central entropies below $30\,\text{keV}\,\text{cm}^2$ seem to show enhanced H$\alpha$ emission and presence of filaments in their cores \citep{Cavagnolo2008}. We will revisit this point in later sections to show that our simulations are in general agreement with this expectation. We also note that in simulations with low overall feedback efficiency, (e.g., simulation TI08 shown in Appendix~\ref{sec:app_radial}, where $\epsilon=10^{-4}$), the entropy profile beyond 10\,kpc does not noticeably deviate from the initial value. It follows that such low-level AGN feedback does not affect the properties of the ICM beyond the cluster core.

\subsection{Accretion \& Feedback Cycle} \label{sec:r_feedback}


\begin{figure}[t!]
\centering
\includegraphics[width=\linewidth]{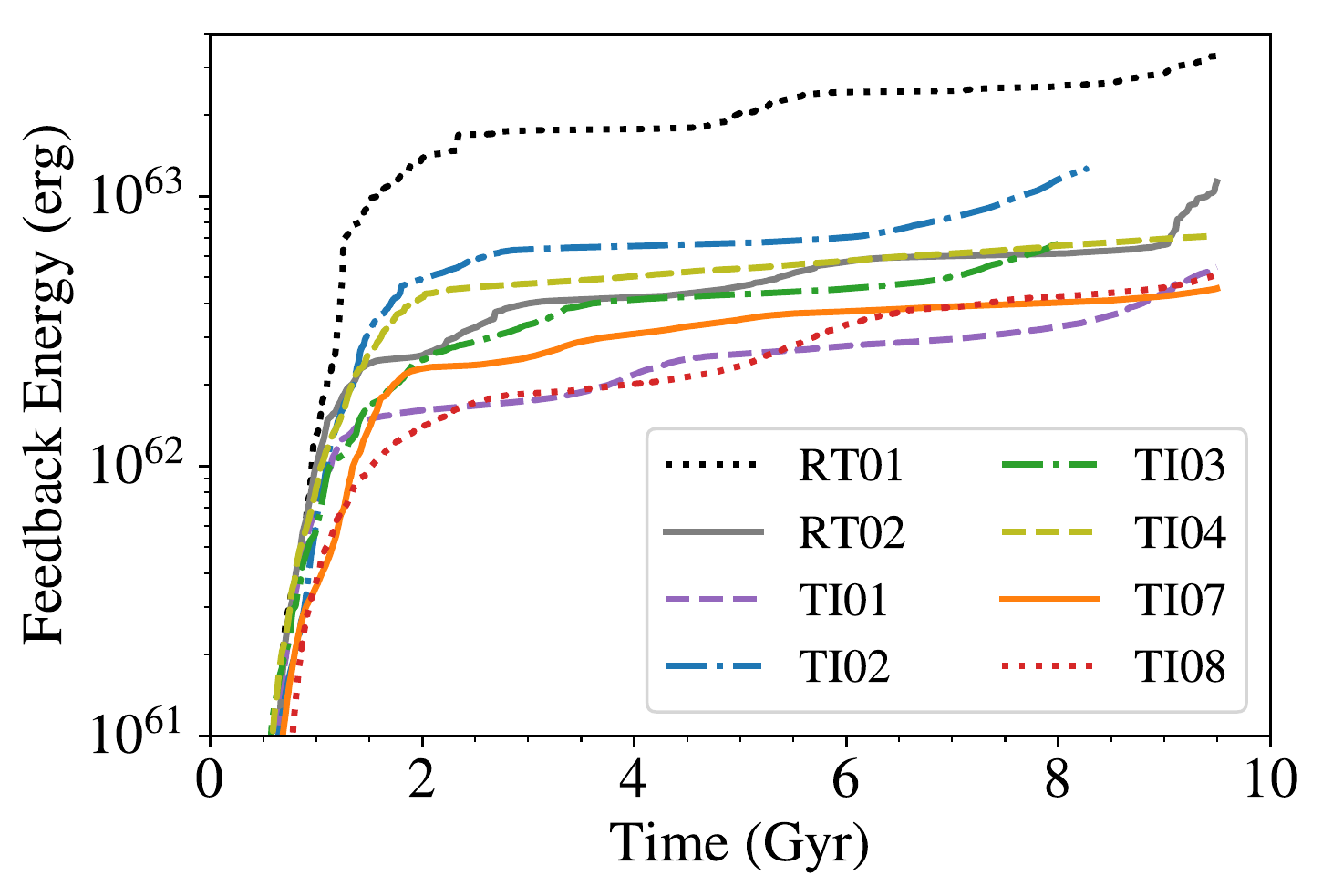}
\caption{Cumulative energy of AGN feedback as a function of time for the runs marked in the legend inset.}
\label{fig:energy}
\end{figure}

Figure~\ref{fig:energy} compares the cumulative energy of AGN feedback in different simulations. With the exception of RT01, all other simulations have energy output corresponding to $\sim 10^{62}\,{\rm erg\, s}^{-1}$, regardless of parameter choices and numerical resolution. Assuming a feedback efficiency of $\eta=10\%$, this implies a black hole mass growth of $10^{8-9}\,M_\sun$ over the course of $\sim10\,{\rm Gyr}$, which is consistent with masses of SMBHs in BCGs. It is worth emphasizing that while overall efficiency varies by two orders of magnitude ($\epsilon=10^{-4}{\rm ,\,}10^{-3}{\rm ,\,}10^{-2}$ in runs TI08, TI01, and TI04, respectively), the AGN energy output in most of our runs only varies within a factor of a few. This is because the accretion rate (determined by the amount of cold gas) adjusts to the accretion efficiency, resulting in a similar energy output. The high energy output in the outlier run RT01 on the other hand is a consequence of inefficient coupling of the radiative feedback in low resolution RT runs (see Appendix~\ref{sec:app_res} for discussion of this effect), leading to the departure from other runs.


\begin{figure*}[t!]
\includegraphics[width=.5\linewidth]{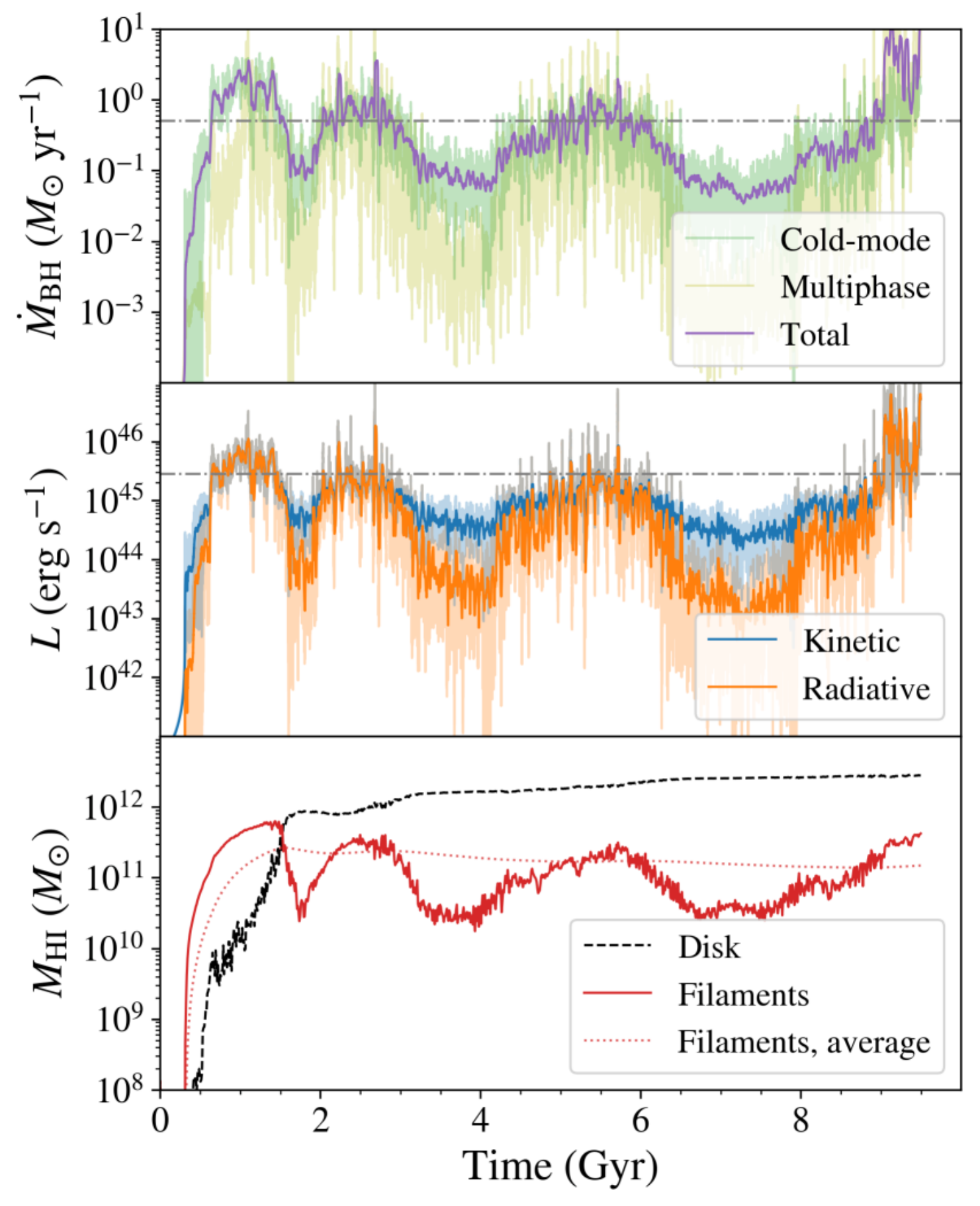}
\includegraphics[width=.5\linewidth]{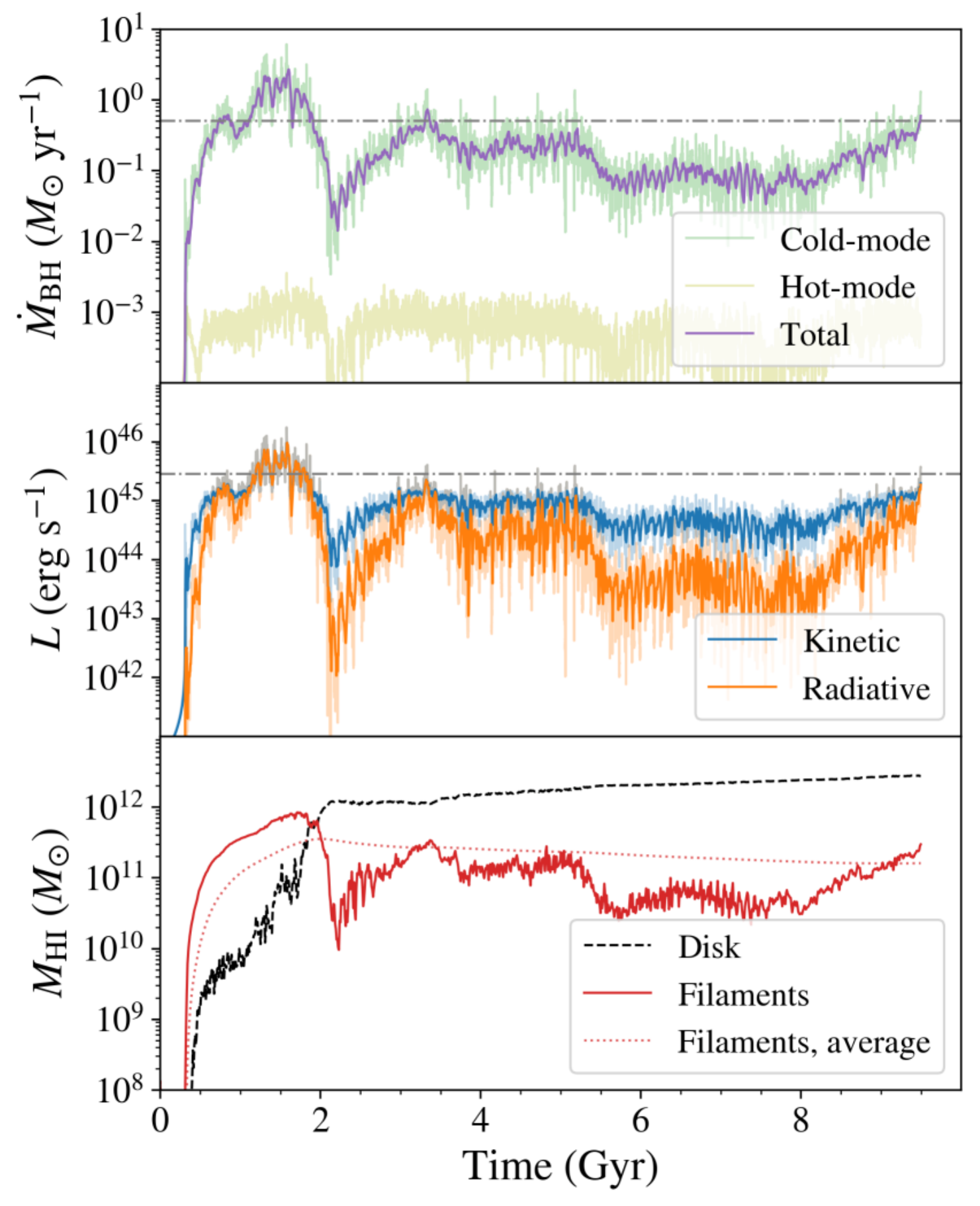}
\caption{Evolution of the accretion rate, AGN luminosity, and cold gas mass in simulations RT02 (left) and TI07 (right). {\it Top:} Different lines mark the cold-mode accretion rate (green), the multiphase or the hot-mode accretion rate (yellow), and the total SMBH accretion rate adopted in the simulation (purple). The total accretion rate is smoothed over $\sim40\,\text{Myr}$, while other shown accretion rates are instantaneous. The horizontal, grey dash-dot line marks the transition accretion rate, $\dot{m}_{\rm t}$, above which the SMBH is in radiatively efficient state. {\it Middle:} Different lines mark the power allocated to the kinetic (blue) and radiative feedback (orange). Lighter color lines illustrate instantaneous luminosities, while solid lines show the power averaged over $\sim20\,\text{Myr}$. The horizontal, grey dash-dot line marks the transition luminosity corresponding to $\dot{m}_{\rm t}$ above. {\it Bottom:} Mass of the rotationally supported, cold gas disk (black, dashed) and filaments (red, solid) traced by $\textsc{H\,i}$. Red dotted line marks the running average mass of the filaments.}
\label{fig:rt02}
\label{fig:TI}
\end{figure*}

In most simulations with AGN feedback, the accretion rate and feedback power exhibit a cyclic behavior. We examine it in this section using the high resolution runs, RT02 and TI07, carried out with our baseline choice of parameters (overall efficiency $\epsilon=10^{-3}$ and jet power fraction $f_\text{J}=0.5$ when $\dot{m}>\dot{m}_\text{t}$). The most important difference between the two runs is in the implementation of radiative feedback, where in TI07 it couples more efficiently with the cold gas in the cluster core, given that in this case 100\% of the energy released in radiation is deposited as thermal energy in the gas within the central $1\,$kpc.

Figure~\ref{fig:rt02} illustrates the SMBH mass accretion rate, the AGN (kinetic and radiative) luminosity, and the mass of the cold gas traced by neutral hydrogen in RT02 (left panels) and TI07 (right). The average values of $\dot{M}_{\rm BH}$ for these two and all other runs are reported in Table~\ref{tab:para}. RT02 is characterized by three cycles in $\dot{M}_{\rm BH}$, defined by the minima and maxima of the AGN feedback power. This cyclic behavior arises because each major feedback episode results in the heating of the ICM and suppression of the accretion rate on the SMBH. In RT02, $\dot{M}_{\rm BH}$ is 
determined as the larger of the cold mode and multiphase accretion rate. The resulting $\dot{M}_{\rm BH}$ is dominated by the cold-mode accretion at nearly all times, indicating that the reservoir of cold, atomic gas is never completely depleted in this run. The multiphase $\dot{M}_{\rm BH}$ on the other hand shows more variability and dips significantly below the cold-mode $\dot{M}_{\rm BH}$, because it reflects the drop in the average density and increase in the average temperature of the multiphase ICM caused by AGN feedback.

$\dot{M}_{\rm BH}$ in run TI07 (determined as the sum of the cold- and hot-mode accretion rate) is also dominated by the cold-mode accretion. In this case the hot-mode accretion rate, calculated for the gas with $T\geq 3\times 10^4\,\text{K}$, is negligible as it falls two orders of magnitude below the cold-mode $\dot{M}_{\rm BH}$. This difference can be understood by inspection of Figure~\ref{fig:phase}, which is based on simulation RT02 but also representative for TI07. It shows that the gas mass above this temperature threshold is dominated by the dilute, $\sim10^7\,$K ICM. We therefore find that the choice of the specific accretion model does not make a significant difference in our simulations, because the cold-mode accretion almost always dominates. Results from additional runs with different accretion models are presented in Appendix~\ref{sec:app_am} for completeness. 

One interesting property of the RT02 run is that in it the SMBH accretion rate oscillates about $\dot{m}_\text{t}=0.05$, implying that the AGN cycles between the radiatively efficient and inefficient states. Consequently, the fraction of power allocated to the jets (as opposed to radiation) ranges between 50\% and nearly 100\%, respectively. This transition occurs on timescales of a few billion years, with later cycles becoming longer. The peak kinetic and radiative luminosities exceed $\sim10^{45}$ erg s$^{-1}$, indicating that the SMBH in the radiatively efficient state corresponds to a radio-loud quasar. Alternatively, the SMBH in the radiatively inefficient state is characterized by kinetic luminosities of $\sim10^{44}$ erg s$^{-1}$ and radiative luminosities of $\sim10^{43}$ erg s$^{-1}$, more similar to a jet-dominated AGN.

In comparison, the accretion rate in the TI07 run remains below $\dot{m}_\text{t}$ after the first AGN outburst, and therefore, the radio-mode feedback dominates over radiation after the first 2\,Gyr. This difference in the evolution of the two runs is a consequence of more efficient heating of the cluster core in TI07, mentioned in the first paragraph of this section. The hotter ICM in TI07 results in lower and more uniform SMBH accretion rate, which in turn gives rise to a jet-dominated AGN. This implies that the photoionization heating calculated with the ray tracing algorithm in RT02 results in less thermal support to the core, since radiation emitted along some directions can escape to infinity without ever interacting with the cold gas in the cluster core. It also indicates that the radiation pressure, which is explicitly calculated in RT02 and neglected in TI07, does not play an important role in the suppression of $\dot{M}_{\rm BH}$.

\subsection{Correlation of Feedback Power with the Mass of Cold Filaments} \label{sec:r_impact}

The bottom panels of Figure~\ref{fig:rt02} show the mass of the cold gas disk and filaments as a function of time in runs RT02 and TI07. The disk, which is mostly composed of $\sim 10-100\,$K temperature gas, has mass of $\sim10^{12}\,M_\sun$ and extends up to $10\,\text{kpc}$ in radius. It is characterized by rotational velocities higher than $300\,\text{km\,s}^{-1}$, which corresponds to the circular velocity at $\sim1\,\text{kpc}$ from the cluster center. This property allows us to kinematically separate the cold filamentary gas in our simulations, which has lower rotational velocity compared to the disk and extends from the core up to $\sim100\,\text{kpc}$ (see Figure~\ref{fig:evo}).


\begin{figure}[t!]
\includegraphics[width=\linewidth]{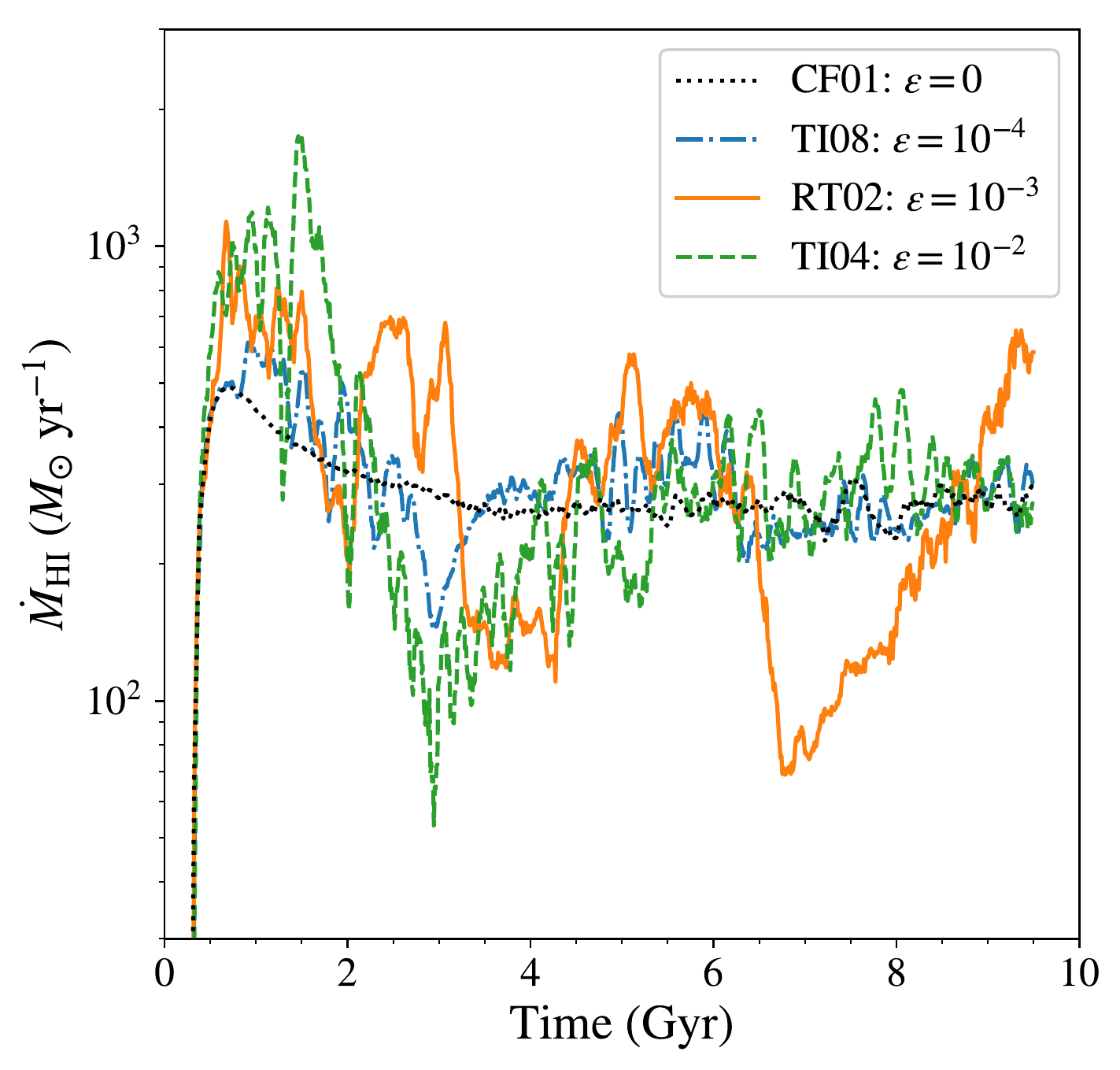}
\caption{Evolution of the mass cooling rate, measured as the rate of change in the mass of $\textsc{H\,i}$ over time in simulations with overall efficiency $\epsilon = 0$ (CF01; black dotted), $10^{-4}$ (TI08; blue dot dashed), $10^{-3}$ (RT02; orange solid) and $10^{-2}$ (TI04; green dashed). In runs with AGN feedback, $\dot{M}_\textsc{H\,i}$ is enhanced (reduced) during the high (low) AGN luminosity states.}
\label{fig:effi}
\end{figure}


\begin{figure*}[t!]
\includegraphics[width=.5\linewidth]{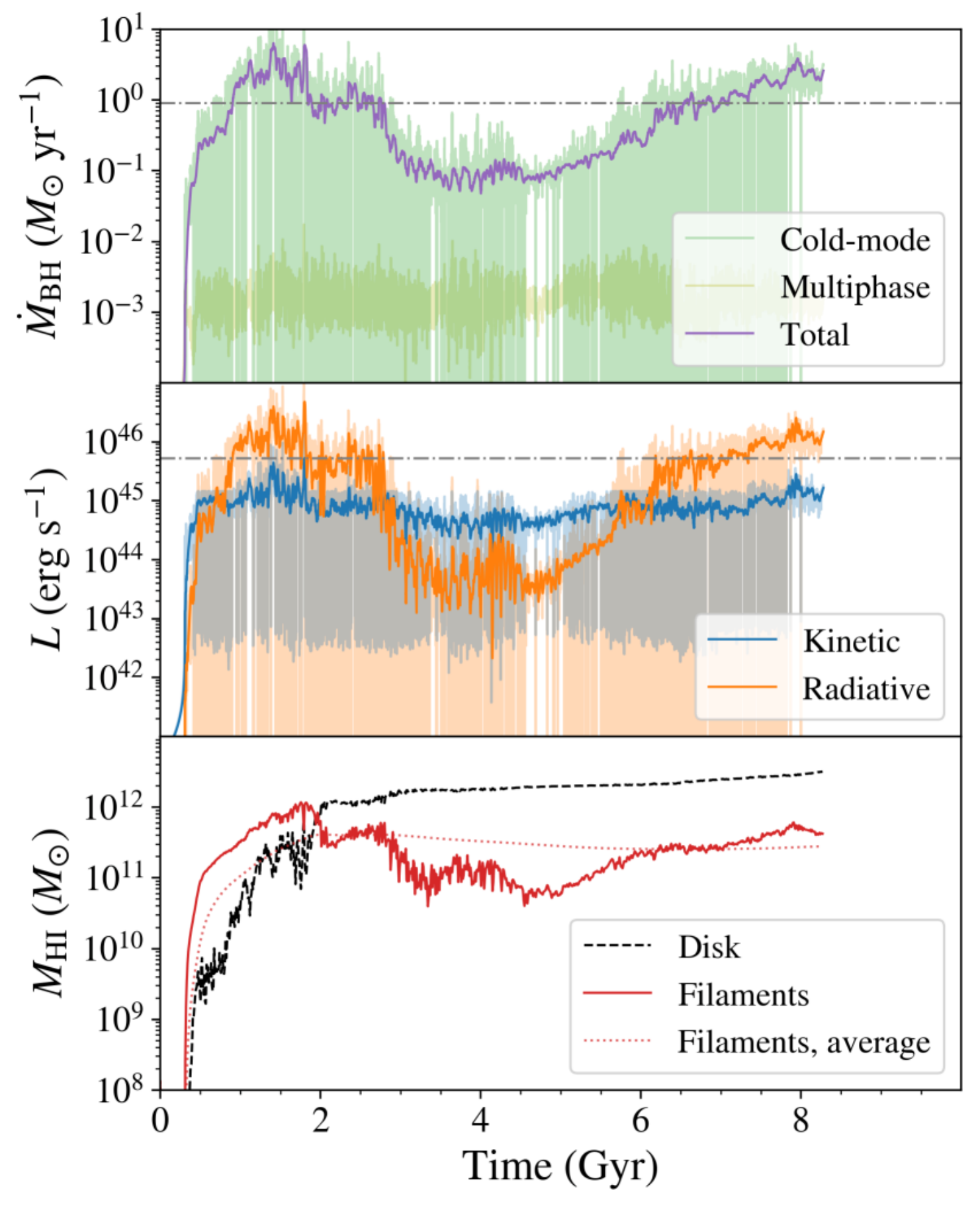}
\includegraphics[width=.5\linewidth]{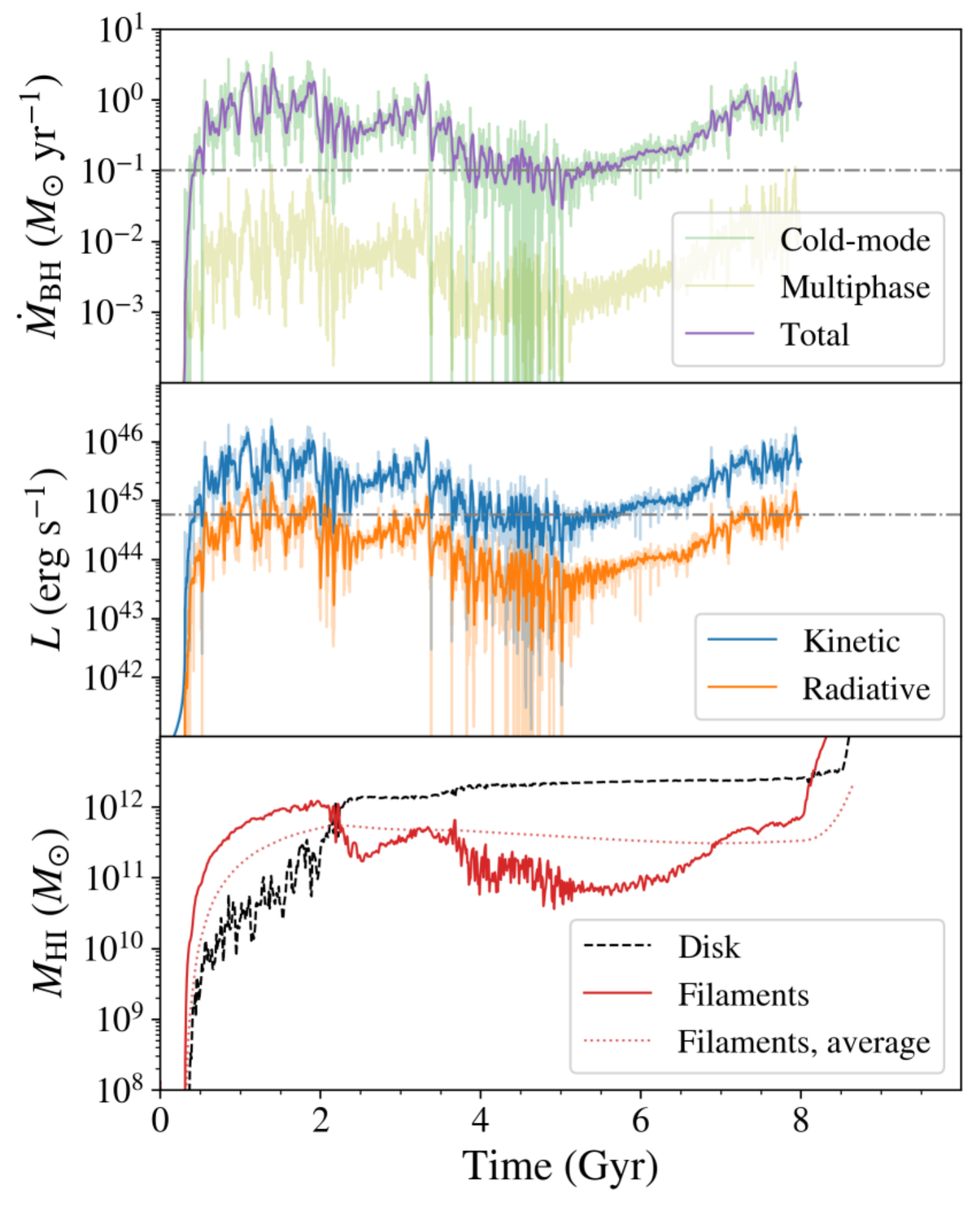}
\caption{Evolution of the accretion rate, AGN luminosity, and cold gas mass in simulations TI02 (left) and TI03 (right), in which $f_\text{J}=0.1\text{ and }0.9$, respectively. Note that the values of $\dot{m}_{\rm t}$ and corresponding transition luminosity are different for the two runs by design (horizontal, grey dash-dot lines). Different lines have the same meaning as in Figure \ref{fig:rt02}.}
\label{fig:fj}
\end{figure*}

Figure~\ref{fig:rt02} indicates that the mass in cold gas filaments, traced by $\textsc{H\,i}$\footnote{This component traces the gas that has temperature below a $\sim {\rm few}\times 10^4$\,K, and in reality also includes the molecular gas that is not modeled explicitly in our simulations.}, remains at the level of $\sim10^{11}\,M_\sun$, on average, with fluctuations of one order of magnitude in either direction. The reason why the total mass of the filaments does not increase with time, even though they are produced throughout the cluster evolution, is because they eventually fall into and become a part of the gradually growing cold gas disk. Comparable amounts of cold filamentary gas (within a factor of two) are encountered in all our simulations with AGN feedback. This indicates that the final, saturated state of the local thermal instability, that produces filaments in the ICM, is not particularly sensitive to the exact implementation of the AGN feedback, as long as the AGN is capable of triggering the instability by perturbing the ICM. The filament mass shows positive correlation with the SMBH mass accretion rate and the overall feedback luminosity, as shown in Figure~\ref{fig:rt02}. This is consistent with the picture in which AGN feedback promotes formation of the filaments rather than suppresses it, as also found by \citet{Revaz2008} and \citet{Li2014a}. These works find that the marginally thermally unstable gas is lifted and compressed by the AGN feedback, causing it to condense out of the ICM and fall back to the center, where it fuels the SMBH and the next AGN feedback episode. In this picture the cold gas that forms in the outflows and mixes with the hot ICM, further promotes cooling. This scenario is also supported by a recent study of 49 nearby elliptical galaxies by \citet{Lakhchaura2018}, who report a positive correlation between the $\text{H}\alpha+[\text{NII}]$ luminosity and AGN jet power. 

If the formation of cold filaments is stimulated by AGN feedback, then a strong correlation should also exist between the AGN feedback power and the instantaneous mass cooling rate of the filaments. Figure~\ref{fig:effi} shows the evolution of the mass cooling rate of atomic hydrogen, measured as the rate of change in the mass of $\textsc{H\,i}$ in the entire computation domain. The figure illustrates results for four different runs with overall feedback efficiencies of $\epsilon = 0.0$, $10^{-4}$, $10^{-3}$ and $10^{-2}$. The runs with AGN feedback show oscillation of $\dot{M}_\textsc{H\,i}$ over time about the value for the pure cooling flow, where the amplitude of the oscillation increases with the overall efficiency. For example, in run RT02 ($\epsilon=10^{-3}$) the mass cooling rate of peaks at about $10^3\,{M}_\sun\,\text{yr}^{-1}$ during the first feedback outburst and drops to $\sim100\,{M}_\sun\,\text{yr}^{-1}$ in low-luminosity stages of the feedback cycle. As the value of $\epsilon$ is decreased, $\dot{M}_\textsc{H\,i}$ asymptotes to the cooling flow value. This is evident in run TI08, in which the cooling rate is only mildly enhanced (or reduced) relative to the pure cooling flow, during the high (low) luminosity states. 

In summary, AGN feedback plays an important role in curbing the global cooling flow and in preventing the cooling catastrophe in CCCs, as established by many earlier works. Therefore, the impact of AGN feedback is negative in the context of the {\it global} thermal instability of the ICM. The fact that AGN feedback positively correlates with the mass cooling rate of the filaments means that at the same time it has a positive impact on the {\it local} thermal instability of the ICM. Namely, as shown in simulations by \citet{McCourt12} and \citet{Sharma12}, formation of filaments can only occur in an atmosphere that is globally marginally stable, and supported by a heating source. Otherwise, an unbridled global cooling flow (as in run CF01) is typically devoid of filamentary gas, as filaments become indistinguishable from the background flow. 

In addition to the results shown in this section we further quantify the correlation between the AGN feedback power and the mass or spatial extent of cold filaments, and present the analysis in a companion paper \citep{Qiu2019}. We point the reader to that paper for discussion of how this correlation can be used to probe the AGN activity in galaxy clusters.

\subsection{Relative Importance of Radiative \& Kinetic Feedback} \label{sec:r_inter}

In this section we investigate how the AGN feedback cycle changes as a function of the dominant feedback mode. As laid out in Section~\ref{sec:feedback}, our description of the relative prominence of the kinetic and radiative feedback is motivated by the \citet{Churazov2005} model, which is itself based on the phenomenology of the stellar X-ray binaries. The aspect of the model that we implement without changes is that kinetic feedback dominates in the radiatively inefficient state, when $\dot{m}\leq\dot{m}_\text{t}$. The modification to the model pertains to the radiatively efficient state of AGN, when $\dot{m}>\dot{m}_\text{t}$. In this regime we vary the fraction of the total feedback power allocated to jets ($f_\text{J}$) and radiation ($1-f_\text{J}$). This approach allows us to parametrize uncertainties related to the physics of jets and radiation in SMBHs accreting close to the Eddington rate.

Figure~\ref{fig:fj} shows the evolution of the accretion rate, AGN luminosity, and cold gas mass in simulations TI02 and TI03, in which $f_\text{J} = 0.1$ and 0.9, respectively. In both runs $\dot{M}_{\rm BH}$ is determined by accretion of the cold gas, as cold-mode dominates over the accretion rate of multiphase gas by $2-3$ orders of magnitude. The difference between the two runs is that the instantaneous cold-mode accretion rate shows significant variability around the average value in TI02 relative to TI03. This indicates that the cold gas reservoir in TI02 is ionized and heated to $T>3\times 10^4\,$K by the central AGN and then cools below this threshold on very short timescales. In TI03 on the other hand the cold gas reservoir remains at $T<3\times 10^4\,$K for most of the evolution (with the exception of the period around 4.5\,Gyr), which explains a relatively small spread in the instantaneous $\dot{M}_{\rm BH}$ for cold gas. Because in TI02 only 10\% of the feedback power is allocated to kinetic feedback, this flickering variability in $\dot{M}_{\rm BH}$ can be directly attributed to heating by radiative feedback. Therefore, radiative feedback is very efficient in rising the temperature of the gas, but it does not suppress the accretion rate for very long, as the dense gas readily cools through recombination. In TI03, jetted feedback dominates and results in $\dot{M}_{\rm BH} \lesssim 1\,M_\odot\,{\rm yr^{-1}}$, a factor of a few lower and more uniform than that in TI02, but the gas in the cluster core remains quite cold.

TI02 and TI03 runs can be compared to TI01 (shown in the right panel of Figure~\ref{fig:res} in Appendix~\ref{sec:app_am}), which is characterized by $f_\text{J} = 0.5$ and is the same in all other regards. The accretion rate in TI01 remains below $1\,M_\odot\,{\rm yr^{-1}}$ most of the time. It exhibits a shorter feedback cycle of $\sim 3\,\text{Gyr}$, relative to $\sim6\,\text{Gyr}$ in TI02 and TI03, estimated from the separation of the first two accretion rate and luminosity peaks. Run TI01 resembles TI02 in terms of a large spread in instantaneous $\dot{M}_{\rm BH}$ and feedback power, which as we noted above is a signature of intense radiative heating. On the other hand, AGN feedback in TI01 is dominated by jets over a large fraction of cluster evolution time, and more similar to TI03. We include the information about the average kinetic and radiative luminosity, as well as their standard deviations, for these and all other runs in Table~\ref{tab:para}.

In terms of the amount of cold gas, the massive disk in runs TI02 and TI03 reaches $10^{12}\,M_\odot$ already at $t=2\,$Gyr, whereas this happens somewhat later, at $t=3\,$Gyr in TI01.\footnote{The abrupt increase of the cold gas mass around 8\,Gyr in TI03 is a numerical artifact which arises when extended cold filaments reach the computational boundary.} Similarly, the average mass of cold filaments in TI02 and TI03 is $3-4\times10^{11}\,M_\odot$ and $2\times10^{11}\,M_\odot$ in TI01. Therefore, AGN feedback seems most efficient in suppressing the ICM cooling in the run TI01, although not by a large margin. 
 
In summary, we find evidence that evolution dominated by radiative feedback leads to higher values of $\dot{M}_{\rm BH}$ on average, and results in more dramatic ``boom and bust" feedback cycles, reflected in the variability of the AGN luminosity across a range of timescales. Conversely, kinetic feedback as a dominant mode appears more effective in suppressing the cooling catastrophe (as evidenced by the lower recorded $\dot{M}_{\rm BH}$) but is ineffective at uniformly heating the cold gas in simulations. Consequently, kinetic feedback results in a relatively uniform evolution of the SMBH accretion rate and AGN luminosity. This is consistent with results of \citet{Meece2017}, who find that the radiative feedback by itself is insufficient to prevent the cooling catastrophe, and must at best play a secondary role relative to the kinetic feedback. Finally, we find that AGN feedback appears to be most efficient in suppressing the cooling flow in runs in which both the kinetic and radiative feedback are present, and deliver comparable amounts of energy to the ICM.


\section{Comparison with Observations} \label{sec:observation}


\subsection{Atomic and Molecular Gas in CCCs} \label{sec:o_ha}

Observations of CCCs suggest that their cores contain large amounts of cold gas, typically dominated by the molecular component. This cold gas is thought to be associated with locally thermally unstable phase of the ICM, which condenses out of the hotter phase and falls toward the center of the cluster under the influence of gravity \citep{McCourt12, Sharma12, Voit2017}. For example, a study of 16 CCCs has revealed $10^9-10^{11.5}\,M_\odot$ of cold molecular gas within the radius of several tens of kiloparsecs of their BCGs \citep{Edge2001}. Similarly, H$_2$ and CO observations of the Perseus cluster have shown at least $5\times10^{10}\,M_\odot$ of warm \citep[$\sim10^3\,$K;][]{Hatch05, Lim08} and cold molecular gas \citep[$\sim10 -10^2\,$K;][]{Salome06, Salome2011}. Most of this gas forms a large scale system of filaments, of which at least some appear to be free falling into the center of their host cluster \citep{Lim08}. There is also some evidence for central, rotating molecular disk with mass $\sim 10^{10}\,M_\odot$ in NGC1275 \citep{Bridges98, Donahue00, Wilman2005}. 

While we do not explicitly model molecular gas, we note that a large fraction of the cold gas that occupies the inner 10\,kpc in our simulated clusters would in reality be in molecular state, given that its temperature can be as low as $10\,$K (see Figure~\ref{fig:phase}). As noted before, most of this gas is a part of the rotating disk with mass $\sim10^{12}\,M_\odot$. Furthermore, $\sim10^{11}\,M_\odot$ is in filaments that in some runs can extend as far as $100\,$kpc. Although we trace filaments in simulations as the $\textsc{H\,i}$ gas that recombines from the ionized ICM, they would in reality also be a mixture of atomic and molecular gas, as some fraction of cooling $\textsc{H\,i}$ would go on to form H$_2$.

While the total mass of the filaments measured from our simulations is comparable to that observed in other CCCs, the mass of the rotating disk is too large by about $1-2$ orders of magnitude. Based on this we infer that AGN feedback, as implemented in our simulations, is not as efficient in suppressing the formation of cold gas as it is in observed CCCs. We discuss in Section~\ref{sec:d_jet} why this may be the case and defer a more detailed investigation of the properties of molecular gas to a future study.

In addition to the molecular emission, one of the features commonly observed in cool-core clusters is the H$\alpha$ line emission associated with the filamentary gas with $T\sim10^4\,\text{K}$. In a study of 23 cool-core clusters \citet{McDonald2010} find that 65\% of the CCCs have detectable H$\alpha$ emission. Of those, 35\% of the CCCs exist in extended filamentary structures, while 30\% show compact, nuclear $\text{H}\alpha$ emission. A large scale system of H$\alpha$ filaments, surrounding the central galaxy of Perseus (NGC 1275), has been particularly well studied and found to have complex morphology and dynamics \citep[e.g.,][]{Conselice2001, GM18}. Furthermore, the molecular filaments in Perseus have been found to be spatially and kinematically associated with the H$\alpha$ filaments \citep{Hatch05, Salome06, Johnstone07}, and both are accompanied by the cooling X-ray filaments \citep[$k_{\rm B}T\sim 0.5$\,keV;][]{Fabian06, Lim08}. There is also evidence that some of the more massive filaments in the halo of NGC~1275 host compact star clusters with typical ages of a few Myr \citep{Canning14}. This complex landscape of multiphase gas and stars indicates that filaments are gravitationally unstable and that the most massive of them have recently collapsed and formed stars. 

The distribution of the H$\alpha$ filaments inferred from our simulations is similar to that observed in Perseus and other clusters (see first panel of Figure~\ref{fig:evo}). Specifically, we find that the filaments form for the first time during the first AGN feedback outburst: they expand radially out, stall, and then rain down toward the cluster center. Their kinematics is not necessarily that of a uniform outflow followed by an inflow, as some filaments are still rising while others are already falling, and some are being pushed sideways by the action of jets and bubbles. This picture is consistent with the predictions of the so-called {\it fountain} model, in which AGN feedback promotes the formation of filaments that in turn fuel the SMBH accretion \citep[e.g.,][]{Tremblay2018}. 

During the subsequent outbursts (e.g., at $\sim4.94\,\text{Gyr}$ in Figure~\ref{fig:evo}) the H$\alpha$ filaments are collimated along the jet axis by the cold gas disk. Hence, the filaments do not always trace the morphology of jets and jet-inflated bubbles but when they do, this may suggest the presence of a massive gas disk in the central galaxy.  As described in Section~\ref{sec:r_impact}, the mass of the filaments positively correlates with the AGN luminosity. We also find that for a given AGN luminosity, more collimated filaments tend to extend $3-4$ times further than those with nearly isotropic distribution, as illustrated in Figure~\ref{fig:evo}. Therefore, if the dynamics of simulated filaments is similar to that in real CCCs, then the filament mass, distribution and their spatial extent are additional observables that can be used to constrain the energetics of the AGN feedback cycle \citep{Qiu2019}.


\subsection{X-ray Emitting ICM} \label{sec:o_xray} 

Much of what we know about the properties of the ICM is enabled by the imaging telescope onboard the Chandra X-ray Observatory \citep{Weisskopf2000}. With its high angular resolution, the features of the X-ray emitting plasma, such as cavities, ripples, outflows, and cold fronts have been studied in great detail. Furthermore, recent results returned by the high spectral resolution telescope Hitomi \citep{Hitomi2016} provides a constraint on the motion and velocity dispersion of the ICM. In this section, we compare our simulation results with some aspects of the X-ray observations of galaxy clusters.

\begin{figure}[t!]
\centering
\includegraphics[width=\linewidth]{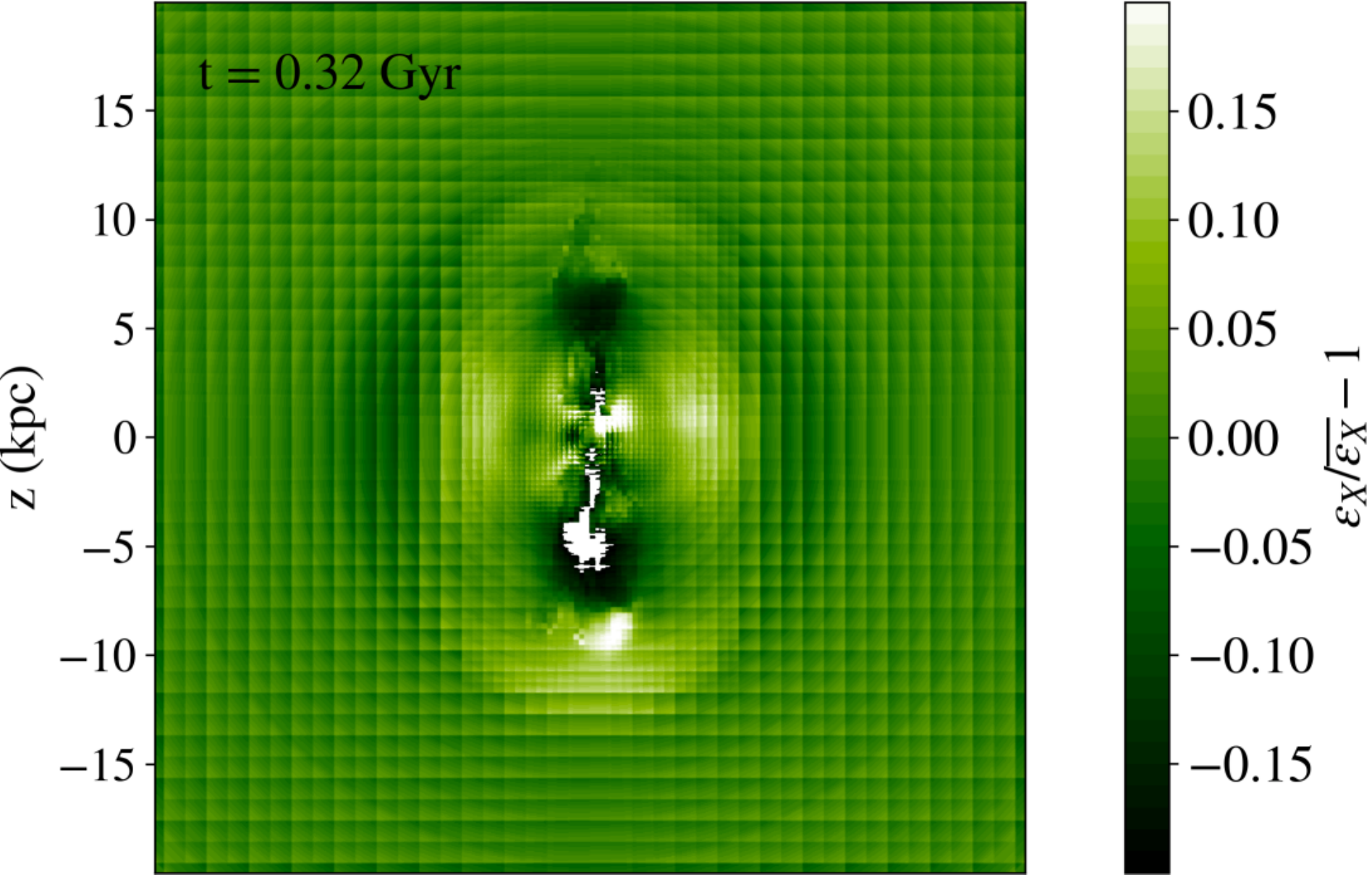}
\caption{A pair of cavities, shown as dark shadows, at $t=0.32$\,Gyr in run RT02. AGN jets are directed along the $z$ axis. The color marks the fractional variance from the azimuthally averaged X-ray surface brightness.}
\label{fig:cavity}
\end{figure}

\noindent {\it Cavities.} Figure~\ref{fig:cavity} shows the fractional variance of the X-ray surface brightness in run RT02 at $t=0.32$\,Gyr, shortly after the AGN was triggered. The image was created using the same procedure as in the bottom panels of Figure~\ref{fig:evo}. It shows two prominent cavities, visible as dark shadows inflated along the jet axis, which are easy to discern at early times because the ICM is still relatively undisturbed by the AGN feedback. At this point in time the diameter of each cavity is about 10\,kpc and the cumulative energy (kinetic+radiative) delivered by AGN feedback is $3.5\times10^{58}\,{\rm erg}$. This is comparable to the central AGN in the Perseus cluster, which has inflated cavities with radius $\sim7$\,kpc, delivering mechanical energy of about $1.2\times 10^{58}\,$erg per cavity \citep{Birzan2004}. The bottom panels of Figure~\ref{fig:evo} show the morphology of the X-ray emitting ICM at later times in the same simulation. At $\sim0.65\,\text{Gyr}$, the panel shows features resembling ripples and multiple cavities dotted around the central AGN, which are reminiscent of the Perseus cluster \citep{Fabian2011}.

Overall, the sizes of X-ray cavities in our simulations vary from a few to tens of kpc. Their shape is irregular compared to the cavities in Perseus, which appear to be rounder and have sharper edges. This may be a consequence of a simple image processing procedure that we use here, and the fact that we do not model the intracluster magnetic field, which can drape around the rising bubbles of the low density plasma to make them smoother and more resilient to instabilities \citep{Jones2005, Rusz2007, Dursi2008}. The scattered distribution of cavities arises naturally in our simulations because the cold filaments, when they fall towards the cluster core, tend to deflect the outflowing plasma in directions different from the jet axis \citep[this is also seen in simulations by][]{Li2014a}. Some tentative evidence for this conjecture is provided by \citet{Romney95}, who report deflection of jets on parsec scales in the central galaxy of Perseus, based on VLBI observations of the compact radio source.

An interesting implication of this phenomenon is that scattered distribution of cavities (as opposed to the series of cavities aligned with the jet axis) can be reproduced without invoking jet precession. The primary motivation for introducing jet precession in some simulations has been to heat the cluster core more uniformly, by having AGN jets sweep over a larger solid angle in the cluster core \citep[e.g.,][]{Meece2017}. While our simplified simulation setup does not capture the structure of accretion flow and jets on small scales (we keep the jet direction fixed along the $z$ axis), they indicate that AGN ``venting" in random directions may arise simply as a consequence of interaction of jets with the cold and dense gas in the BCG.

\noindent {\it Ripples.} As outflows and bubbles rise from the cluster core, they create ripples in the ICM. The ripples have been captured by X-ray observations, and are evidence for weak shocks and/or sound waves produced by the AGN feedback \citep[e.g.,][]{Sanders2007, Forman2007}. They are thought to carry large amounts of energy, and may be a significant heating mechanism that distributes the feedback energy throughout the cluster core. The bottom panels of Figure~\ref{fig:evo} illustrate several different generations of X-ray ripples in the simulated cluster core that extend up to 100s of kpc \citep[similar features are also seen in][]{Li2014b}. The ripples have a characteristic wavelength of $\sim10\,{\rm kpc}$ and the amplitude corresponding to $<20\%$ of the azimuthally-averaged X-ray surface brightness at a given radius, similar to the Perseus cluster \citep{Sanders2007}. It is worth pointing out that the ripples are present in our simulations at most times. They are most visible during the peak of the AGN feedback outbursts but are also present during the quiescent periods, when the X-ray cavities are not clearly defined. This suggests that the cluster core is continuously bathed in sound waves, as it responds to the variability in feedback power, even if no AGN bubbles are apparent.

\noindent {\it X-ray emitting filaments.} In addition to cavities and ripples, the second and third panels in the bottom of Figure~\ref{fig:evo} and Figure~\ref{fig:cavity} also show X-ray bright filaments extending along the jet direction. The filaments contain relatively cool plasma with $k_{\rm B}T\sim2\,{\rm keV}$ and are clearly associated with the filaments of the atomic hydrogen gas. This phenomenon has been observed in the Perseus cluster, where much of the cool X-ray gas ($\sim 10^9\,M_\odot$ at $k_{\rm B}T\sim 0.5$\,keV) is associated with the optical filamentary nebula \citep{Fabian06}. Similarly, observations of the jet in M87 reveal soft X-ray emission in the 0.5-2.5\,keV band but no apparent emission above 2\,keV, indicating that the outflows are mostly associated with the cooler X-ray gas \citep{Forman2007}. This picture supports the hypothesis that most of the cold filaments condense out of marginally unstable ICM plasma that is co-spatial with the AGN jets and cavities. AGN feedback provides both the initial perturbation necessary to seed the local thermal instability, as well as mixing of the cold gas with the ICM plasma. The mixing can promote adiabatic cooling of the soft X-ray gas (by lowering its average temperature) accompanied by little emission of thermal radiation, which may explain the lack of X-ray emission lines with characteristic energy $k_{\rm B}T<2\,{\rm keV}$ \citep{Peterson2003,Fabian2011b}. 

\noindent {\it Cold fronts}, characterized by a sharp discontinuity in the X-ray surface brightness and gas temperature, are commonly observed in CCCs \citep[see][for a review]{Markevitch2007}. In relaxed clusters, where there are no signs of recent major mergers, these features have been attributed to the sloshing of the ICM around the dark matter halo caused by encounters with small groups or subclusters \citep{Churazov2003,ZuHone2011}. In some of our simulations we nevertheless identify the presence of features that resemble cold fronts, even in the absence of mergers and encounters with subclusters. The bright spiral structure seen in the mock X-ray image in Figure~\ref{fig:evo} at $\sim8\,\text{Gyr}$ first appears at $\sim5\,\text{Gyr}$. This indicates that gas motion induced by the AGN feedback is also a viable way of stirring the ICM in the core and producing cold fronts that extend out to 100\,kpc. We defer more detailed analysis of this phenomenon to a future study. 

\begin{figure}[t!]
\centering
\includegraphics[width=0.9\linewidth]{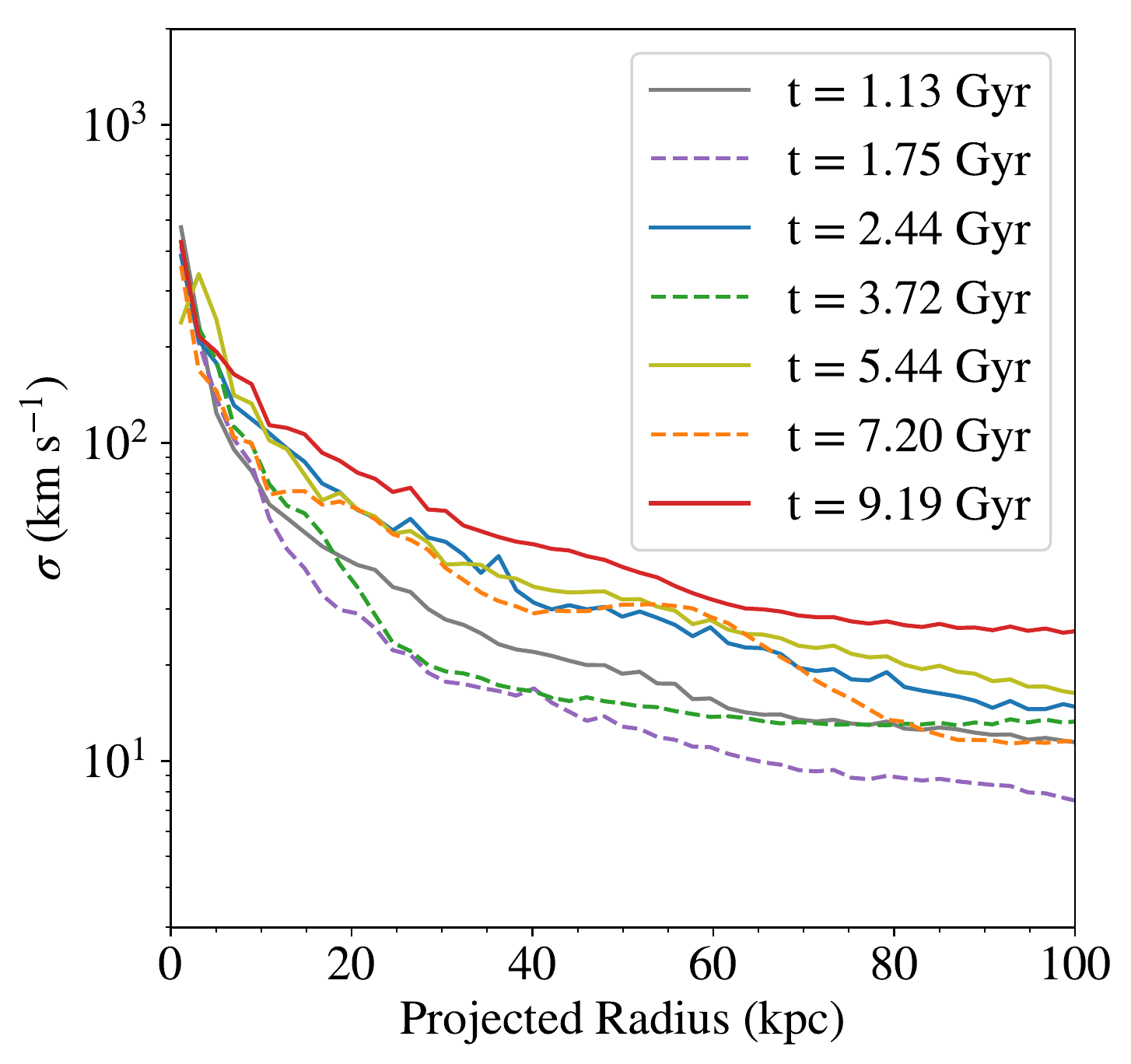}
\caption{X-ray emissivity-weighted velocity dispersion of the ICM in run RT02 at different times corresponding to maxima (solid) and minima (dashed) of AGN feedback luminosity shown in Figure \ref{fig:rt02}.}
\label{fig:sigma}
\end{figure}

\noindent {\it Velocity dispersion.} Recent results returned by the Hitomi X-ray Observatory provide another measure of the gas motion in the Perseus cluster. In this case, the line-of-sight velocity dispersion of the ICM between $30$ and $60\,\text{kpc}$ from the center has been inferred from the broadening of the X-ray emission lines to be $164\pm10\,\text{km\,s}^{-1}$ \citep{Hitomi2016}. In Figure~\ref{fig:sigma} we show the velocity dispersion of the ICM measured from run RT02 at several different epochs, as a comparison. In order to show a property that more closely corresponds to observations, we show the emissivity-weighted velocity dispersion as a function of the projected cluster radius, calculated as an average of components measured along the three axes, $\sigma=(\sigma_x+\sigma_y+\sigma_z)/3$. We find that $\sigma$ calculated in this way is above $100\,\text{km\,s}^{-1}$ in the inner $10\,\text{kpc}$ and varies between $10-100\,\text{km\,s}^{-1}$ at larger radii. Generally, $\sigma$ is higher in the high AGN luminosity states ($t=2.44$, 5.44, and 9.19\,Gyr), and lower during the low luminosity states ($t=1.75$, 3.72, and 7.20\,Gyr). Overall the values of $\sigma$ measured from our simulations are lower than that measured by Hitomi, similar to what \citet{Li2017} found. It is worth noting that \citet{Prasad2018} and \citet{Gaspari2018} report velocity dispersion values that match the Hitomi measurement, albeit using a different simulation setup. This suggests that stirring provided by the AGN jets in our simulations is too gentle or that there are other mechanisms which may result in increased velocity dispersion not captured by our simulations.


\subsection{Properties of Central AGNs in CCCs} \label{sec:d_var}

Another question of interest for both observations and simulations is what fraction of central AGNs in BCGs are quasars, or at least luminous enough that they are discernible as compact X-ray sources against the emission of their host clusters. This is of importance because it signals what fraction of central SMBHs is operating in the radiatively efficient mode, and has implications for the AGN feedback duty cycle. BCGs that host luminous AGNs are thought to be rare, but their precise fraction is challenging to determine from observations due to selection effects. Specifically, in shallow X-ray data both the central AGN and the host CCC have centrally peaked emission profiles, which are difficult to disentangle \citep{Pesce90}. As a result, the CCCs may be ignored or just classified as AGNs, especially at high redshift where this bias is more pronounced \citep{Green2017}.

For example, in a sample of $\sim1000$ clusters with $z<0.4$, \citet{Green2017} find only 7 AGN with X-ray luminosity comparable to its host cluster, implying $<1\%$ incidence of luminous AGN at low redshift. In another study based on Chandra observations, \citet{Hlavacek-Larrondo2013} show that many clusters with X-ray cavities at $z = 0.6$ have X-ray bright AGN. This is in contrast to the clusters of comparable luminosity at lower redshifts, without nuclear X-ray emission. They suggest that over the past $\sim5\,\text{Gyr}$, the central SMBHs in BCGs have evolved from radio-loud quasars (in which most of the power is emitted in radiation) to radio-loud AGN (in which kinetic luminosity of the jets dominates).

This observational evidence is consistent with a subset of our simulations in which the accreting SMBH powers a radio loud quasar in the first $1-2$\,Gyr of the cluster evolution and then switches to the radiatively inefficient regime, becoming a jet-dominated AGN (e.g., see run TI07 in the right panel of Figure~\ref{fig:rt02}). In order to quantify the prevalence of radio-loud quasars in our simulations, we measure the fraction of time that the accreting SMBH spends in the radiatively efficient regime as an AGN with radiative luminosity $>10^{45}\,\text{erg\,s}^{-1}$. We report this property for all our runs as $f_\text{QSO}$ in Table~\ref{tab:para}. 

With the exception of the passive cooling flow run, in which $f_\text{QSO} = 0$ by definition, $f_\text{QSO}$ varies between $4-47$\% in other simulations. Comparison of runs TI02, TI01, and TI03 shows that increasing the feedback power allocated to jets from $f_\text{J}=0.1$ to 0.5 to 0.9, leads to a decreasing $f_\text{QSO}$ from 0.46 to 0.24 to 0.04, respectively. We also find that varying the overall efficiency of feedback, from $\epsilon= 10^{-4}$ to $10^{-2}$ leads to a smaller degree of reduction, from $f_\text{QSO} = 0.23$ to $0.12$ in runs TI08 and TI04, respectively. Therefore, the most important factor that determines $f_\text{QSO}$ is the prevailing feedback mode (jets vs. radiation), and the total amount of energy delivered by the AGN feedback plays a lesser role, as long as it is sufficient to suppress the cooling flow. 

In terms of numerical effects, we find that increasing numerical resolution leads to a drop in $f_\text{QSO}$. For example, runs RT01 and RT02 correspond to the lower and higher numerical resolution simulations of the same scenario, and exhibit $f_\text{QSO} = 0.47$ and 0.30, respectively. This can be understood as in higher resolution runs the radiation has easier time penetrating and breaking up (smaller) clumps of cold gas, which increases the temperature of the gas and lowers the accretion rate onto the central SMBH. Consequently, the AGN in the higher resolution runs achieves a lower luminosity, on average. 

All other things being the same, $f_\text{QSO}$ is also smaller in thermal injection (TI) simulations compared with radiative transfer (RT) runs, due to the propensity of thermal feedback to efficiently heat the surrounding gas and reduce accretion. It is worth noting however that neither method provides an entirely correct description of interaction of the ionizing radiation and gas. Specifically, the TI method overestimates the heating of the gas by implicitly assuming that it absorbs 100\% of the radiation energy, while the RT method underestimates it because it does not account for photon trapping and diffusion in the optically thick gas. These two scenarios nevertheless bracket a range of physically plausible outcomes.

In summary, our idealized simulations suggest that {\it most} SMBHs in BCGs are likely to have been powerful radio-loud quasars at high redshift\footnote{One caveat to this statement is that our idealized simulations of isolated clusters may overproduce radio-loud quasars in the early stages of evolution, because they do not capture cosmological growth and mergers of clusters.}. If the scarcity of observed quasars in cluster BCGs at low redshift is determined by their duty cycle, then a transition from the radio-loud quasar to a jet-dominated AGN state must have occurred relatively early in the evolution of most CCCs (within the first 2\,Gyr). According to our simulations, this transition requires that the fraction of AGN feedback power allocated to jets is comparable to or larger than the fraction in radiative luminosity ($f_{\rm J} \geq 0.5$). It also requires very efficient thermalization of feedback energy, which can suppress the cold-mode accretion either through photo-heating of the ICM and/or through efficient thermalization of jet-driven shocks in the cluster core. If so, this implies that deeper X-ray surveys of higher redshift CCCs should discover an increasing fraction of radio-loud quasars in their BCGs. Determining that fraction would help test this hypothesis and understand how feedback operates.



\section{Discussion} \label{sec:discus}

In this section we discuss simplifying assumptions made in our simulations and compare our results with similar works in the literature. Following the example set by earlier works, we reiterate the most important aspects of the AGN feedback implementation which result in similarities and differences of our works. This is important given the complexity of contemporary simulations, as well as the ability of seemingly small variations in simulation setup to result in significant differences in the impact of AGN feedback \citep[e.g.,][]{Martizzi2018}.

\subsection{Simplifying Assumptions in Our Simulations} \label{sec:d_disk}

 Our simulations can be regarded as continued exploration of the Perseus cluster setup presented in \citet{Li2015} and \citet{Meece2017}, since all utilize the same numerical method and packages implemented in the code \texttt{Enzo}. The main differences in our work are that we explore the relative importance of radiative feedback, and introduce modifications to the implementation of kinetic feedback.

As described in Section~\ref{sec:feedback}, the overall feedback power in our simulations is allocated between the kinetic and radiative feedback as a function of the SMBH accretion rate, following the model proposed by \citet{Churazov2005}. An important modification made to this model however pertains to the behavior of SMBH in the radiatively efficient regime, which occurs when $\dot{m}>\dot{m}_\text{t}$. Instead of quenching jets in the radiatively efficient regime, as originally proposed by \citet{Churazov2005}, we allow jets to carry between $10-90$\% of the total feedback power. Based on this set of experiments we find that (a) jetted feedback must be present at high accretion rates, because radiative feedback alone cannot suppress runaway cooling, and (b) that jetted feedback likely accounts for $>10\,\%$ of the total AGN feedback power, since below this threshold BCGs in our simulations host radio-loud quasars about 50\% of the time (see Section~\ref{sec:d_var}). The latter number is a high fraction that is incompatible with a low incidence of luminous quasars observed in low redshift BCGs \citep{Green2017}.

This picture is supported by the recent radiation magneto-hydrodynamic simulations which measure the kinetic and radiative luminosity of SMBH nuclear accretion regions. These simulations show that the outflows powered by SMBHs in the radiatively efficient state have kinetic luminosities that are within a factor of a few of their radiative luminosities, for a wide range of SMBH accretion rates \citep{Sadowski2017, Gaspari2017a, Jiang2017}. The presence of outflows therefore seems to be ubiquitous, even at high accretion rates. 

Our suite of simulations does not capture self-gravity of the gas and does not follow gas through star formation. While the effect of self-gravity is negligible for the hot component of the ICM and less massive filaments \citep{Canning2014}, the cold disk in our simulations provides a large gas reservoir for star formation in the BCG. The energy injected by stellar feedback is not sufficient to prevent the radiative cooling of the ICM and alter its thermodynamics significantly, but it can deplete the cold gas disk by converting most of it into stars on a timescale of $1-2$\,Gyr \citep{Li2015}. Note however that the persistence of the cold gas disk in our simulations does not affect the accretion rate of the SMBH. This is because the bulk of the mass of the rotationally supported disk resides outside of the ``accretion region'' of 1\,kpc, where gas properties are used to derive $\dot{M}_{\rm BH}$. 

We also do not model magnetic fields or phenomena associated with them, such as anisotropic heat conduction. Conductive heating within the cluster cores has been found to compensate for up to $\sim 10$\% of the radiative losses for Perseus-like clusters \citep{Yang2016}, and is therefore expected to have a lesser impact in this class of CCCs. The same authors find that anisotropic conduction can nevertheless constitute an important heating source in more massive clusters, where it can compensate for $\sim 50$\% of radiative losses.

It is important to note that our simulations do not capture the interaction of cosmic rays (or relativistic electrons) with the magnetic fields or the ICM. Cosmic rays can provide additional pressure support to the ICM ($\sim 10\%$ of the thermal gas pressure), and can heat the ICM by exciting Alfv\'en waves and instabilities, through Coulomb interactions and hadronic collisions \citep{Guo2008}. In a recent work, \citet{Rusz2017} show that cosmic ray heating is indeed a viable channel for the thermalization of AGN kinetic feedback in clusters. Without the cosmic ray component of the jet plasma, the heating of the ICM in our and similar models is ``replaced" by the shock-heating of the outflows and photo-heating by the radiation. The exact physical mechanism for thermalization of AGN feedback is yet to be tested by these two groups of models, since at this time both appear to make predictions consistent with observations.

Finally, we do not model the cosmological evolution of clusters. Specifically, the spherically symmetric potential well of the cluster, BCG and the SMBH remains fixed in our simulations over the course of several to ten gigayears. While this is clearly an idealization, it is worthwhile considering its impact on the evolution of the ICM. Because our simulations feature a CCC with a fully developed potential well, the cooling rate of the ICM remains high, implying that AGN feedback must operate more vigorously in order to prevent the cooling catastrophe than in the scenario with an evolving potential well. Moreover, in reality, the assembly of galaxy clusters over cosmic time involves some number of minor and major mergers with other clusters and groups of galaxies. These perturb the underlying potential of the CCC and may lead to enhanced sloshing and periodic disruption of the cold gas reservoir \citep{Churazov2003,ZuHone2011}. We therefore expect that our choice not to model the cosmological context results in a cluster more prone to formation of the cooling flow. As a consequence, we may overproduce radio-loud quasars in the first few Gyr of evolution, before AGN feedback has had a chance to counter it.

\subsection{Impact of Numerical Scheme Used to Describe Kinetic Feedback} \label{sec:d_jet}

The most significant shortcoming of our simulations is that after a few Gyr the BCG accumulates large amounts of cold gas ($\sim 10^{12}\,M_\odot$), in the form of the rotationally supported disk that sometimes coexists with the extended cold gas filaments. While there is some observational evidence for the existence of molecular disks in central galaxies of CCCs, they tend to be $1-2$ orders of magnitude less massive than in our simulations. For example, \citet{Russell2017} report that the Phoenix cluster exhibits both a molecular torus and extended filaments with mass larger than $\sim 10^{10}\,M_\odot$. Observations also indicate that the BCG in Perseus hosts a rotating molecular disk of similar mass \citep{Bridges98, Donahue00, Wilman2005}. It is interesting that beyond these two well-known CCCs, central molecular disks and rings seem to be rare in other clusters and groups of galaxies \citep{Pulido2018}. This indicates that they either do not form in the first place, or that the depletion timescale of such disks is rather short (e.g., due to star formation).

Intriguingly, the massive gas disk has been a persistent feature of many numerical studies of the cooling flow problem \citep[e.g.,][]{Vernaleo2006,Gaspari2012,Li2014b,Prasad2015,Wang2018}. These studies have employed different codes and numerical methods, and have used a variety of sub-grid implementations of jetted feedback. The formation of such a disk appears to be a natural state in the evolution of CCCs and it supports the picture that the cooling flow in simulations can be reduced but never fully suppressed by AGN feedback. The cluster cores in simulations appear to be in the process of gentle circulation over billions of years \citep{Yang2016b}. This provides a more nuanced view of the cooling flow that goes beyond a simplified binary picture of the ``runaway cooling" vs. ``hot core" clusters.

It is worth noting that more recent hydrodynamic simulations of AGN kinetic feedback in CCCs have been successful in reproducing cold gas disks with mass consistent with that observed in the Perseus cluster \citep{Gaspari2012, Li2014b, Li2015}. Because the degree to which the cooling flow is suppressed in simulations with AGN feedback has been used as an important criterion for their success, it is worth comparing our assumptions to these works in some detail. For example, \citet{Gaspari2012} and \citet{Li2015} model outflows as plasma with sub-relativistic velocities ($\sim3-5\times 10^4\,\text{km\,s}^{-1}$), a component distinct from the relativistic, highly collimated jets \citep[see also][]{Omma2004}. In these simulations, the hot plasma outflows carry the gas at the rate $10-10^3\,M_\sun\,{\rm yr}^{-1}$, which is comparable to the rate of the inflow of cold gas, resulting in a low effective SMBH accretion rate. Once their kinetic energy is thermalized, such outflows are powerful enough to shock-heat the ICM to $T\sim10^{8-10}\,\text{K}$ \citep[see Figure~4 in][]{Gaspari2012} and prevent accumulation of more than $\sim 10^{11}\,M_\odot$ of cold gas in the cluster center.

In our simulations we too model the kinetic feedback as sub-relativistic outflows but adopt a different distribution of kinetic energy, where more massive gas clumps carry more energy (see Section~\ref{sec:m_r}). Because the cold and dense ICM is difficult to accelerate to high speeds, this jet-launching scheme results in outflows (up to $\sim 10^3\,M_\sun\,{\rm yr}^{-1}$) with initial velocity that does not exceed $3\times10^3\,\text{km\,s}^{-1}$. One consequence of the lower speed of the outflows in our simulations is that the temperature of the shocked ICM rarely exceeds $10^8\,\text{K}$ (see Figure \ref{fig:phase}). Consequently, outflows deliver less efficient shock-heating of the ICM. As a result, the total cold gas mass at the end of our simulations with feedback is not significantly reduced compared to the cooling flow run, both exceeding $10^{12}\,M_\sun$ after $\sim 5\,\text{Gyr}$. Therefore, the difference in the cold gas mass between our results and other similar works in the literature can largely be ascribed to different implementations of kinetic feedback.

It is interesting to note that with the exception of the cold gas mass, our simulations seem to reproduce many other features observed in CCCs (see Section~\ref{sec:observation}), which are completely absent from the fiducial, pure cooling flow model. We therefore surmise that there is a continuum of possible outcomes for simulations of CCCs in terms of the cold gas mass and that our simulations are, for reasons given above, at the lower end in terms of the efficiency of coupling of the AGN feedback to the ICM. 

We also draw several conclusions relevant for the numerical schemes of kinetic feedback used in simulations of CCCs. Firstly, from a numerical point of view, simulations require the plasma launched in the outflows to be warmer than the filaments of cold gas that fall in the cluster center. This is because the filaments are too massive and heavy to be lifted and relaunched by the outflows and they instead lead to ``clogging" and failed jets in simulations. In our simulations, the thermal content of the filaments in the accretion region is increased by the radiative heating from the central AGN. Secondly, the energy carried by the warm outflows must be efficiently thermalized (via shock-heating, cosmic ray streaming, etc.) as the outflows mix with the ICM. It is worth noting that observations of CCCs seem to find little evidence for the existence of large amounts of ICM plasma above $10^8\,\text{K}$ \citep[e.g.,][]{McNamara2012,Hitomi2018}, so whichever mechanism leads to the thermalization of the jet energy should be gentle, yet effective, leading to a high degree of coupling of jet kinetic energy to the ICM.


\section{Conclusions} \label{sec:conclusion}

We perform a suite of 3D radiation-hydrodynamic simulations of a CCC, modeled on the Perseus cluster, with an aim to explore the joint role of kinetic and radiative feedback powered by accretion of cold gas onto the SMBH in the central cluster galaxy. We model radiative feedback as a central source of ionizing radiation and kinetic feedback as jet-driven outflows. Our main findings are as follows:

1. One of the key features of our model is the presence of radiative feedback, which is prominent at high SMBH accretion rates. We find that radiative feedback alone is incapable of staving off the cooling catastrophe, and must be accompanied by kinetic feedback at both the low and high accretion rates. This numerical setup produces radio-loud (jet-dominated) AGN at low accretion rates, and radio-loud quasars at high accretion rates.

2. In this work we model AGN radiative feedback using either the ray-tracing radiative transfer or thermal energy injection. While both methods lead to qualitatively similar cluster evolution, thermal injection results in more efficient heating of the cluster core and leads to the lower average SMBH accretion rate. Consequently, a fraction of time that the AGN spends as a high luminosity quasar is smaller in thermal injection runs. It is worth noting that both, the ray tracing and thermal injection, provide an approximate description of radiative feedback and that together they bracket a range of physically relevant scenarios.

3. The AGN feedback in our simulations transitions between radiatively efficient and inefficient states on timescales corresponding to a few Gyr. When CCC evolution is dominated by radiative feedback, it leads to higher values of SMBH accretion rate on average, and results in more dramatic “boom and bust” feedback cycles, reflected in the variability of the AGN luminosity across a range of timescales. Conversely, kinetic feedback as the dominant mode results in a relatively uniform evolution of the SMBH accretion rate and AGN luminosity. 

4. The fraction of time during which the central AGN reaches and maintains quasar-like radiative luminosity ($\gtrsim10^{45}\,{\rm erg\,s^{-1}}$) varies from $f_{\rm QSO} = 4-47$\% in our simulations. The most important factor that determines this fraction is the prevailing feedback mode (jets vs. radiation), whereas the total AGN luminosity plays a lesser role, as long as it is sufficient to partially suppress the cooling flow. Specifically, we find that jetted feedback likely accounts for $>10\,\%$ of the total AGN feedback power. Below this threshold BCGs in our simulations host radio-loud quasars about 50\% of the time, a fraction that is incompatible with a low incidence of luminous quasars observed in low redshift BCGs. 

5. We find a positive correlation between the AGN feedback power and the mass of the cold gas filaments. Based on this we confirm that AGN feedback promotes the formation of cold gas filaments in CCCs. If so, this indicates that CCCs that are hosts to massive and spatially distributed ${\rm H}\alpha$ filament networks are likely to have undergone a powerful feedback episode within the past ${\rm few}\times10\,\text{Myr}$. Conversely, the filament mass and their spatial extent may be used to place an additional observational constraint on the energetics of the AGN feedback cycle. 

6. Our simulations indicate that intermittent feedback from the central AGN is capable of producing the X-ray cavities and ripples (similar to those reported in the Perseus cluster) that are scattered around the cluster core, even in absence of jet precession. Furthermore, we find that AGN feedback can induce gas sloshing in the central $\sim100\,\text{kpc}$ strong enough to produce cold fronts similar to those observed in some CCCs.

Simulations presented here can be regarded as continued exploration of the Perseus cluster setup presented in \citet{Li2015} and \citet{Meece2017}, albeit with a different implementation of feedback. With the exception of the mass of the cold gas disk, our simulations seem to reproduce many features observed in CCCs, which are completely absent from the fiducial, pure cooling flow model. We conclude that there is a continuum of possible outcomes for simulations of CCCs in terms of the resulting cold gas mass, and that our simulations are at the lower end in terms of the efficiency of coupling of the AGN feedback to the ICM. In the future we plan to examine how changing this efficiency affects the observable properties of simulated CCCs, and how features like ${\rm H}\alpha$ filaments and X-ray cavities can be used as a joint measure of AGN feedback.


\acknowledgments

We thank the anonymous referee for a careful reading of the manuscript and thoughtful comments that helped improve this work. We also thank David Buote, Ena Choi, Megan Donahue, Elena Gallo, Massimo Gaspari, Liyi Gu, Brian McNamara, Mark Voit, Karen Yang, and John ZuHone for useful comments and discussions. Support for this work was provided by the National Aeronautics and Space Administration through Chandra Award Number TM7-18008X issued by the Chandra X-ray Center, which is operated by the Smithsonian Astrophysical Observatory for and on behalf of the National Aeronautics Space Administration under contract NAS8-03060. TB and JHW acknowledge support from the National Science Foundation under grant No. NSF AST-1333360 during the early stages of this work. JHW was supported by National Science Foundation grants AST-1614333 and OAC-1835213, NASA grant NNX17AG23G, and Hubble theory grant HST-AR-14326. Numerical simulations were performed on the high-performance computing cluster PACE, administered by the Georgia Tech Office of Information and Technology. Computations described in this work were executed and analyzed using the publicly-available \texttt{Enzo} code and \texttt{yt} toolkit \citep{Turk2011}, which are the products of collaborative efforts of many independent scientists from numerous institutions around the world. Their commitment to open science has helped make this work possible.


\appendix


\section{Radial acceleration by components}\label{sec:app_acc}


\begin{figure}[ht!]
\centering
\includegraphics[width=0.5\linewidth]{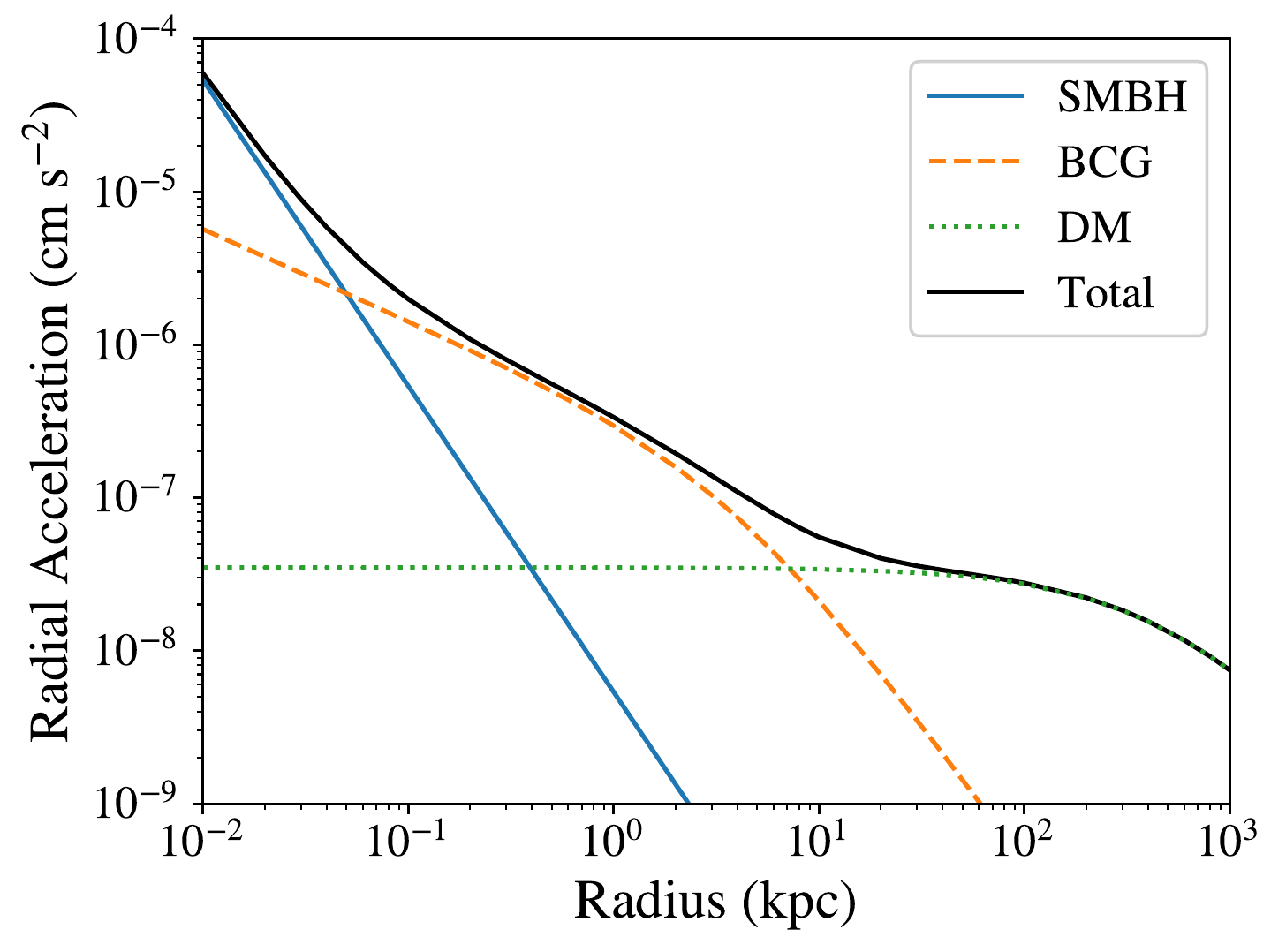}
\caption{Radial acceleration of the ICM due to the gravitational influence of the SMBH (blue solid), BCG (orange dashed), and dark matter halo (DM; green dotted), as adopted in our simulations. The black solid line marks the total of all three components of acceleration.}
\label{fig:acc}
\end{figure}

As noted in Section~\ref{sec:m_c}, the background gravitational potential of the cluster is assumed to be static (i.e., it does not evolve over time) and it includes three components: the dark matter halo, the stellar bulge of the BCG, and the central SMBH. We show contributions to the potential of different components in Figure~\ref{fig:acc}
and note that it is similar (albeit not identical) to that used by \citet{Li2012}. Specifically, the SMBH dominates at $r\lesssim0.1\,\text{kpc}$, the BCG dominates in the range $0.1\lesssim r \lesssim 10\,\text{kpc}$, and the influence of dark matter halo is important beyond $\sim10\,\text{kpc}$. The dark matter density distribution is modeled as the NFW profile \citep{Navarro1996}
\begin{equation}
 	\rho^{\text{NFW}}(r)=\frac{\rho_0^{\text{NFW}}}{\left(\frac{r}{r_s}\right)\left(1+\frac{r}{r_s}\right)^{2}} \,\,.
\end{equation}
Here, $\rho_0^{\text{NFW}}=8.475\times 10^{14}\,M_\sun\,\text{Mpc}^{-3}$, $r$ is the radius from the center of the cluster, and $r_s=0.494\,\text{Mpc}$ is the scaling radius. Note that this $\rho_0^{\text{NFW}}$ is a factor $\delta=1.13$ higher than in \citet{Li2012}. We apply this scaling factor to all components of acceleration, resulting in a slightly deeper potential well. In this setup the ICM (defined by the density and temperature profiles given in Section~\ref{sec:m_c}) is close to being in hydrostatic equilibrium at the beginning of the simulations. We have verified this by carrying out simulations in which radiative cooling of the ICM, SMBH accretion and AGN feedback were disabled, thus allowing the cluster to settle into a permanent hydrostatic equilibrium.

The spherically-averaged radial acceleration due to the BCG at the center of the Perseus cluster is described as \citep{Mathews2006}
\begin{equation}
	\frac{GM_*(r)}{r^2}=\delta\left[\left(\frac{r_{\rm kpc}^{0.5975}}{3.206\times 10^{-7}}\right)^s+\left(\frac{r_{\rm kpc}^{1.849}}{1.861\times 10^{-6}}\right)^s\right]^{-1/s}\,\text{cm\,s}^{-2},
\end{equation}
where $r_{\rm kpc}$ is in units of kpc, $s=0.9$, and $M_*(r)$ is the enclosed stellar mass at radius $r$. We also account for the contribution to the gravitational potential from the SMBH with mass $3.8\times 10^8\,M_{\sun}$. This is a factor of $\delta$ higher than the mass of the SMBH in the center of the Perseus cluster, as reported by \citet{Wilman2005}. It is worth noting that there may still be a considerable uncertainty about the mass of the SMBH in the central galaxy of Perseus, NGC~1275 \citep{Sani2018}. We do not expect this to affect our results since the SMBH potential dominates on scales $\lesssim100\,$pc, which are unresolved in this work.


\begin{figure}[t!]
\centering
\includegraphics[width=0.49\linewidth]{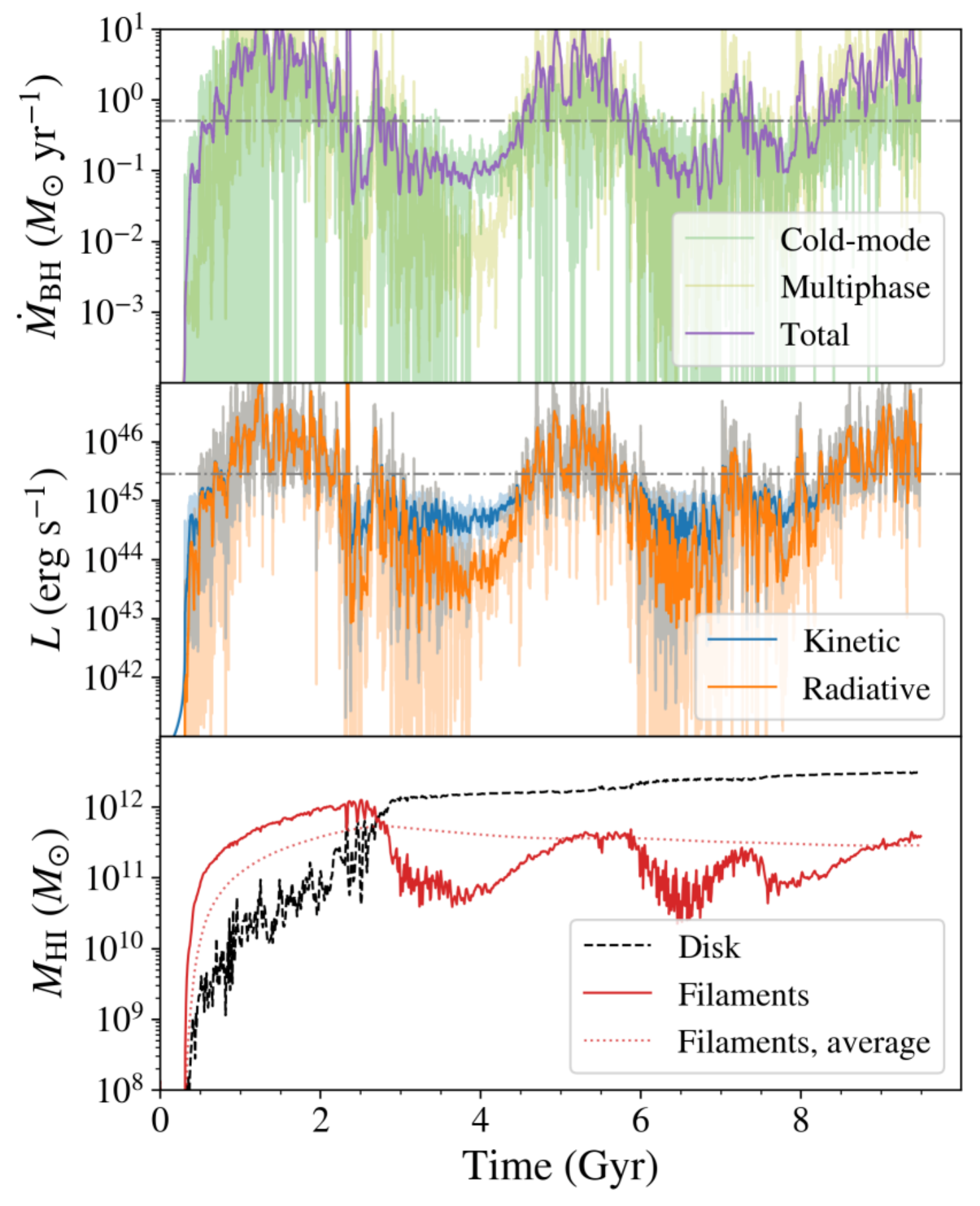}
\includegraphics[width=0.49\linewidth]{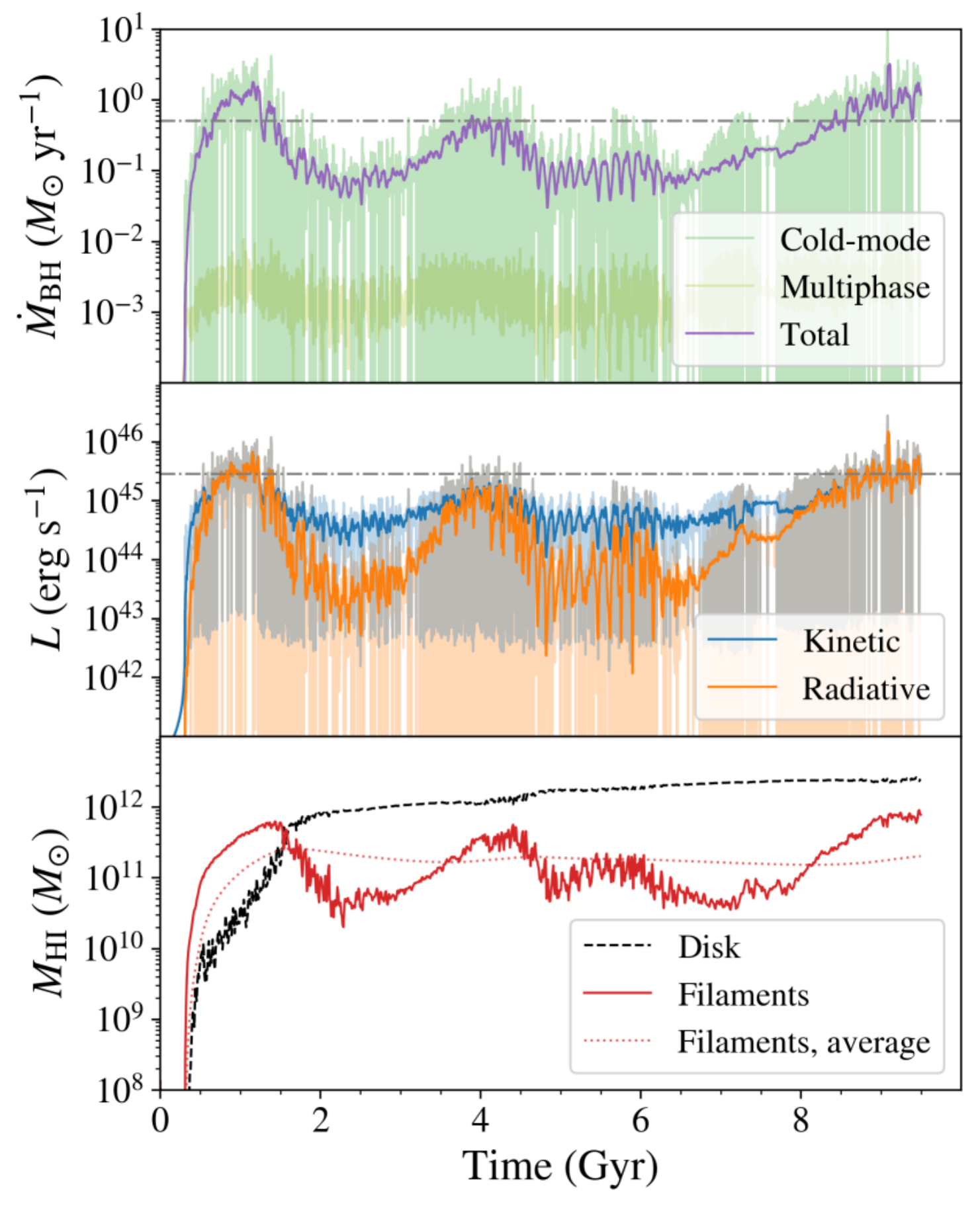}
\caption{Evolution of the accretion rate, AGN luminosity, and cold gas mass in simulations RT01 (left), TI01/AM01 (right). Different lines have the same meaning as in Figure~\ref{fig:rt02}.}
\label{fig:res}
\end{figure}


\section{Resolution Study}\label{sec:app_res}


\begin{figure}[t!]
\centering
\includegraphics[width=0.49\linewidth]{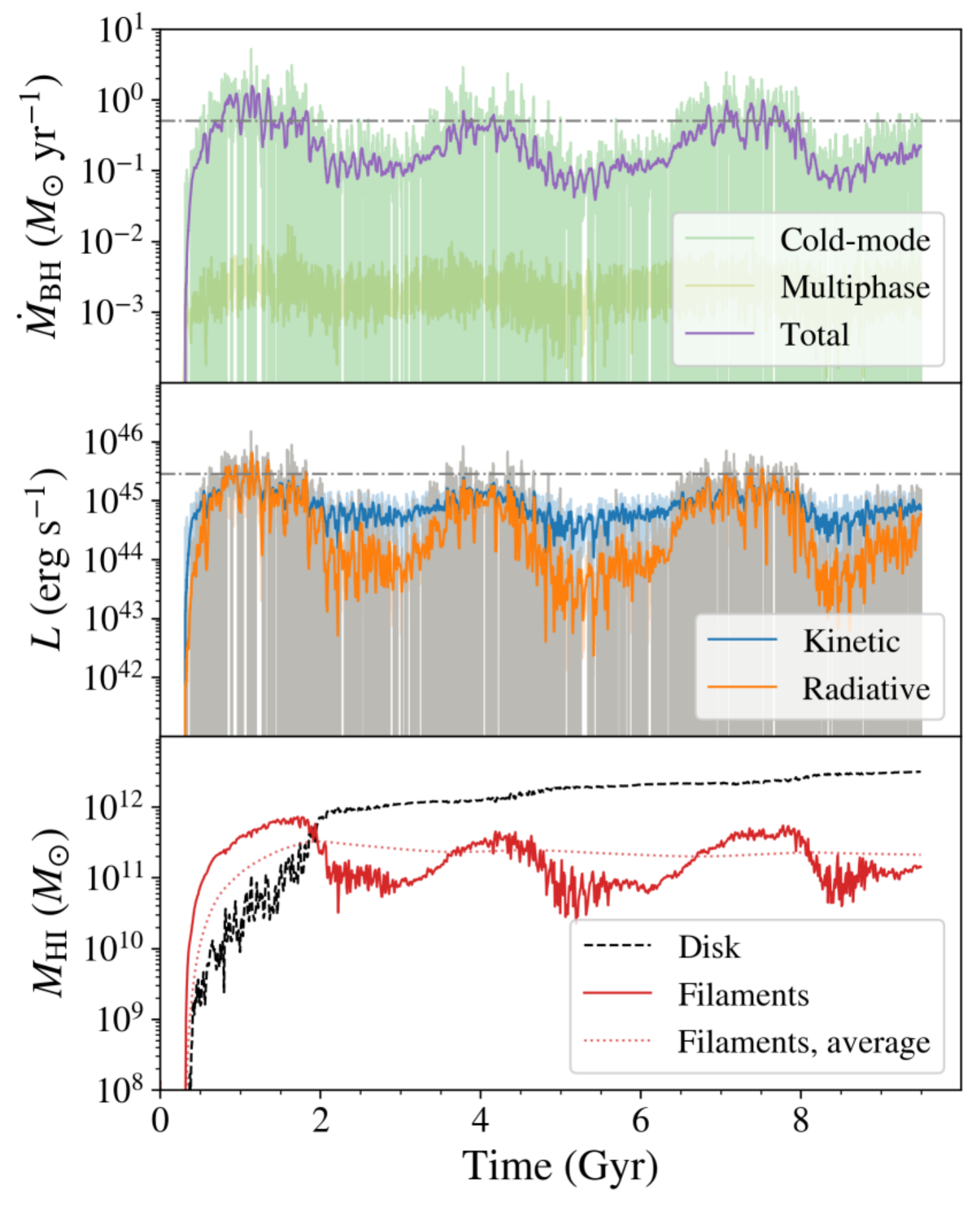}
\includegraphics[width=0.49\linewidth]{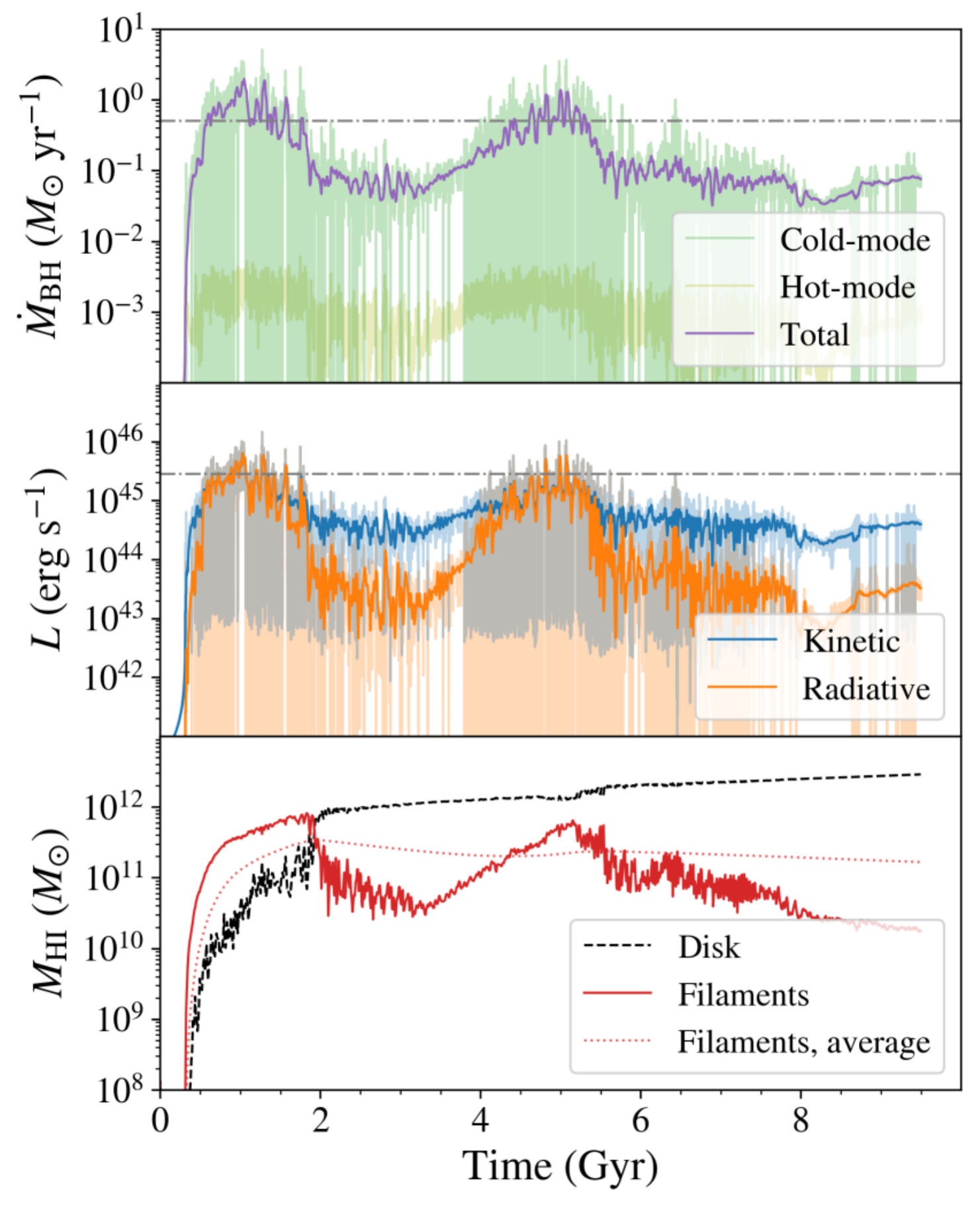}
\caption{Evolution of the accretion rate, AGN luminosity, and cold gas mass in simulations AM02 (left) and AM03 (right). Different lines have the same meaning as in Figure~\ref{fig:rt02}.}
\label{fig:am}
\end{figure}

In this section we summarize the results of a resolution study, carried out in order to understand the impact of numerical resolution on our simulations. As shown in Table~\ref{tab:para}, our simulations are carried out with resolutions of 0.24\,kpc or 0.49\,kpc (corresponding to the size of the smallest resolution element), which are comparable to other recent works in the literature \citep{Gaspari2012, Li2015, Prasad2015}. In this context, we compare two sets of runs at different resolutions: those in which radiative feedback is calculated with explicit radiative transfer (RT) and those in which it is implemented as thermal feedback (TI). The intention is to test how these two different implementations of radiative feedback depend on numerical resolution, in addition to all other processes which are present in both sets of runs. 

Figure~\ref{fig:rt02} shows the higher resolution runs RT02 and TI07, and Figure~\ref{fig:res} shows the lower resolution counterparts RT01 and TI01. The lower resolution runs qualitatively reproduce the accretion, feedback, and cold gas evolution of the higher resolution runs. There are however several differences worth pointing out: in RT01 the SMBH accretion rate and AGN feedback power are higher than those in RT02. Both are a consequence of less efficient radiative heating of the ICM in lower resolution runs, an effect which arises for the following reasons. In the radiative transfer module of the code \texttt{Enzo}, the absorbers in a given computational cell (for e.g., hydrogen atoms) can only be ionized by photons once in each photon time step, $dt_\text{P}$. As mentioned in Section~\ref{sec:m_code} $dt_\text{P}$ is set by limiting the change of $\textsc{H\,i}$ density in each cell to $<10\%$, {\it or} by the light crossing time of the smallest cell, whichever is greater. Because the former can in principle be many orders of magnitude smaller than the latter in the central 1\,kpc of our computational domain, it can lead to a dramatic slowdown of the simulation. To mediate this effect, we implement a floor to the photon time step set by the light crossing time of the smallest computational cell, $dx_\text{min}/c$. Because in this case a coarser numerical resolution results in larger $dt_\text{P}$, the amount of photon energy deposited in some volume for the same length of time is reduced relative to the higher resolution simulations. This resolution dependence can be removed if $dt_\text{P}$ is set everywhere by the requirement that the change in the amount of \textsc{H\,i} from one time step to another is $<10\%$, as shown in Figure~40 of \citet{Wise2011}.

In TI runs all radiative energy is deposited in the central accretion region, which is larger than the resolution limit of our simulations and therefore, independent of it. Consequently, the time averaged properties in TI01 show no significant difference from the higher resolution runs TI07 and RT02. The lower resolution run TI01, however, has a larger variance of the instantaneous amplitude of the accretion rate and feedback power. Note that the choice of the specific accretion model (``Max" or ``C+H") does not make a difference in the context of numerical resolution. This is because the cold-mode accretion dominates in all cases, and so all runs show the same behavior with resolution. This point is illustrated in the next section.


\section{Test of Different Accretion Models}\label{sec:app_am}

Because our simulations do not resolve the nuclear accretion region of the SMBH, we use properties of the gas around the SMBH to estimate its accretion rate. In this section, we describe the impact of different accretion prescriptions, introduced in Section~\ref{sec:m_mdot}, on results of our simulations. For this purpose we select runs AM01, AM02, and AM03, which have different accretion prescriptions and are identical in all other regards (see Table~\ref{tab:para} for a description of their parameters). In AM01, the accretion rate is calculated as the larger of the cold-mode accretion rate and the accretion rate for the warm, multiphase gas, as shown in equation~\ref{eqn:mdot}. In AM02, only the cold-mode accretion rate is used in the simulation. In AM03, the sum of the cold-mode ($T<3\times 10^4\,$K) and hot-mode ($T\geq 3\times 10^4\,$K) accretion rates is employed. 

The resulting evolution of the accretion rate, AGN luminosity, and cold gas mass for these three runs is shown in the right panel of Figure~\ref{fig:res} (for AM01) and in Figure~\ref{fig:am} (AM02 and AM03). In all three simulations, the AGN cycles through 2-3 outbursts over the course of 10\,Gyr. The overall evolution is very similar, because accretion of cold gas determines the accretion rate in all runs. In simulations where the accretion of the warm and hot gas are involved (AM01 and AM03, respectively), the variability in the amplitude of jet power has a lower limit of $\sim10^{43}$ erg s$^{-1}$, set by the Bondi accretion rate. This lower limit is not present in AM02, where only cold-mode accretion rate is considered. In this case, when there is no cold gas around the SMBH, the feedback switches off completely.

\section{Physical Properties of the ICM in Additional Runs}\label{sec:app_radial}


\begin{figure}[t!]
\flushright

\includegraphics[trim=0.2cm 0.3cm 0.3cm 0.3cm, clip=true, height=0.231\linewidth]{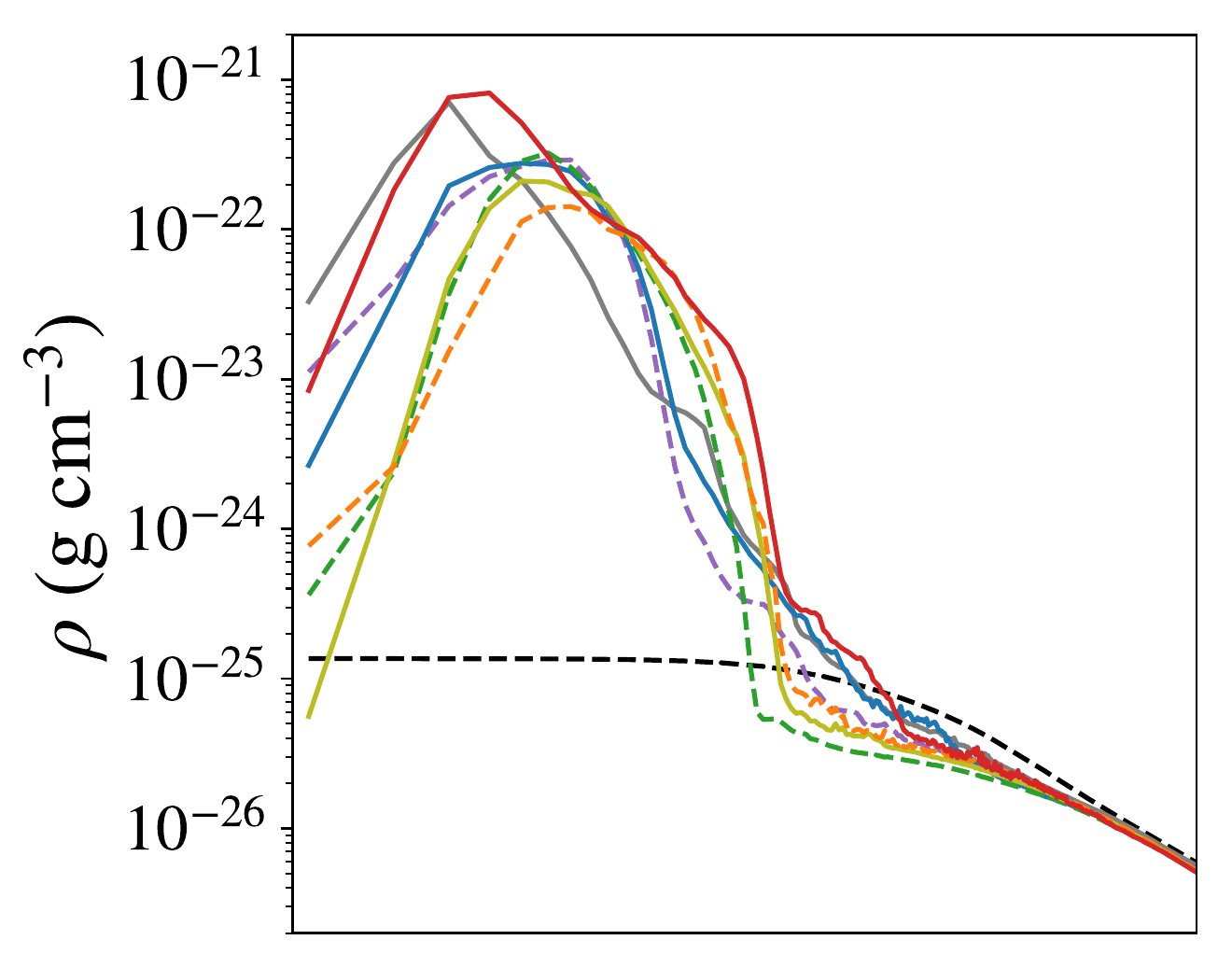}\includegraphics[trim=0.3cm 0.3cm 0.3cm 0.3cm, clip=true, height=0.231\linewidth]{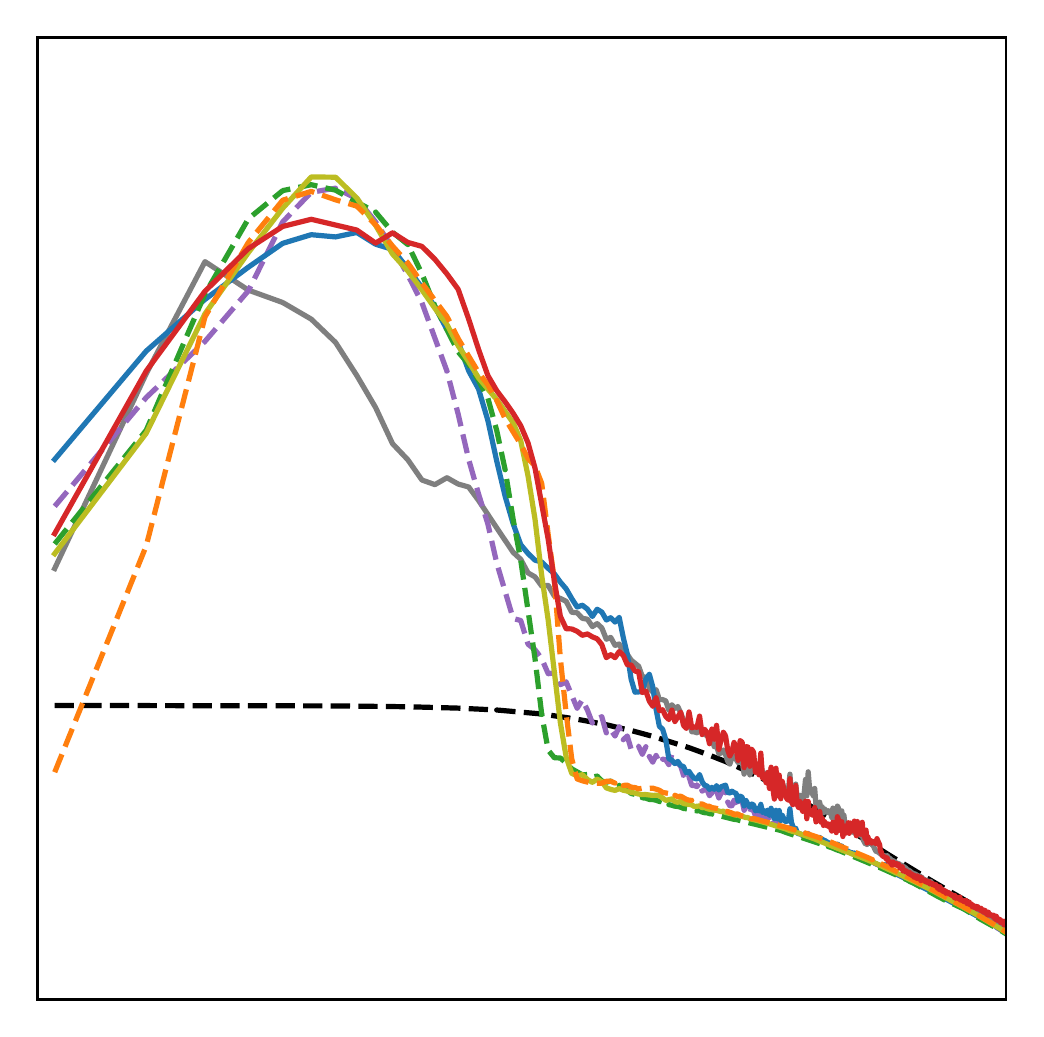}\includegraphics[trim=0.3cm 0.3cm 0.3cm 0.3cm, clip=true, height=0.231\linewidth]{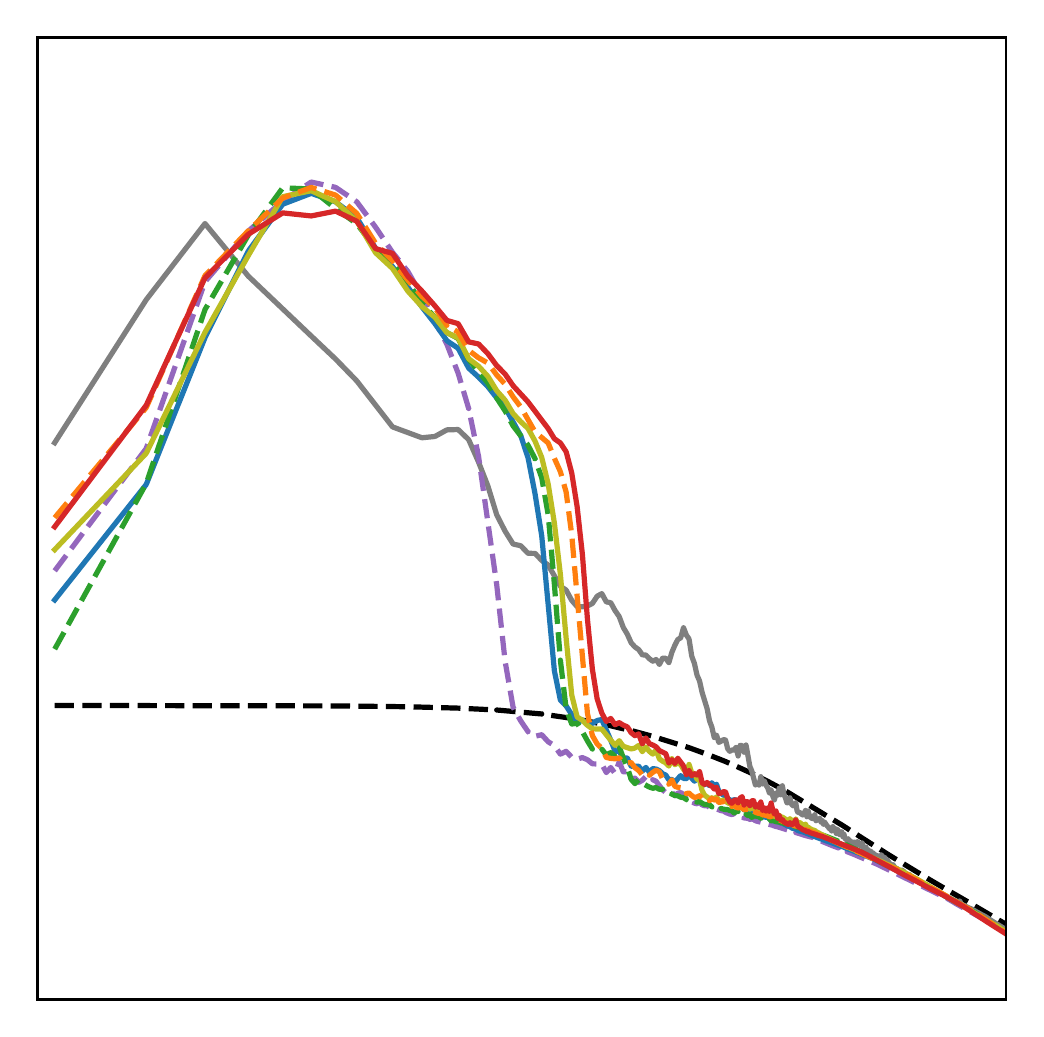}\includegraphics[trim=0.3cm 0.3cm 0.3cm 0.3cm, clip=true, height=0.231\linewidth]{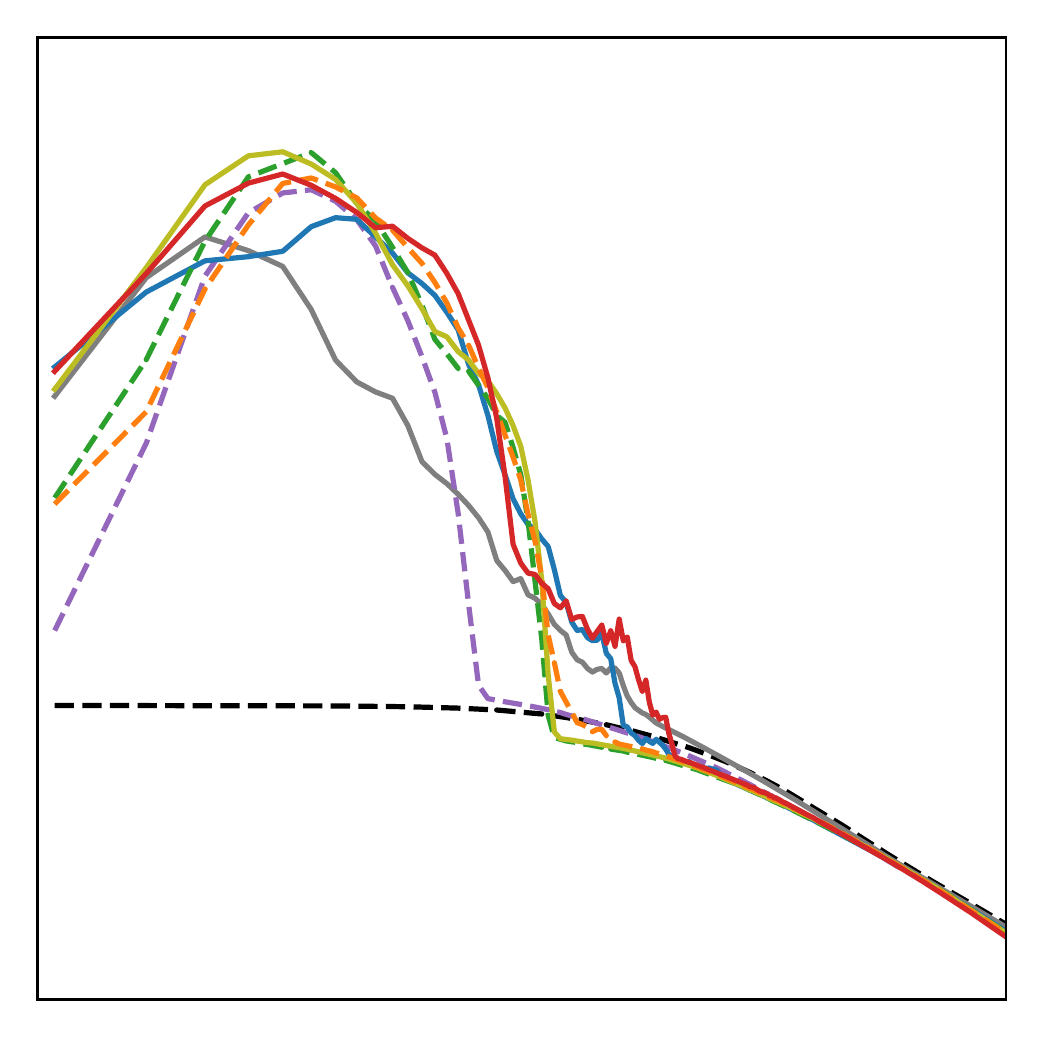}

\includegraphics[trim=0.2cm 0.3cm 0.3cm 0.3cm, clip=true, height=0.231\linewidth]{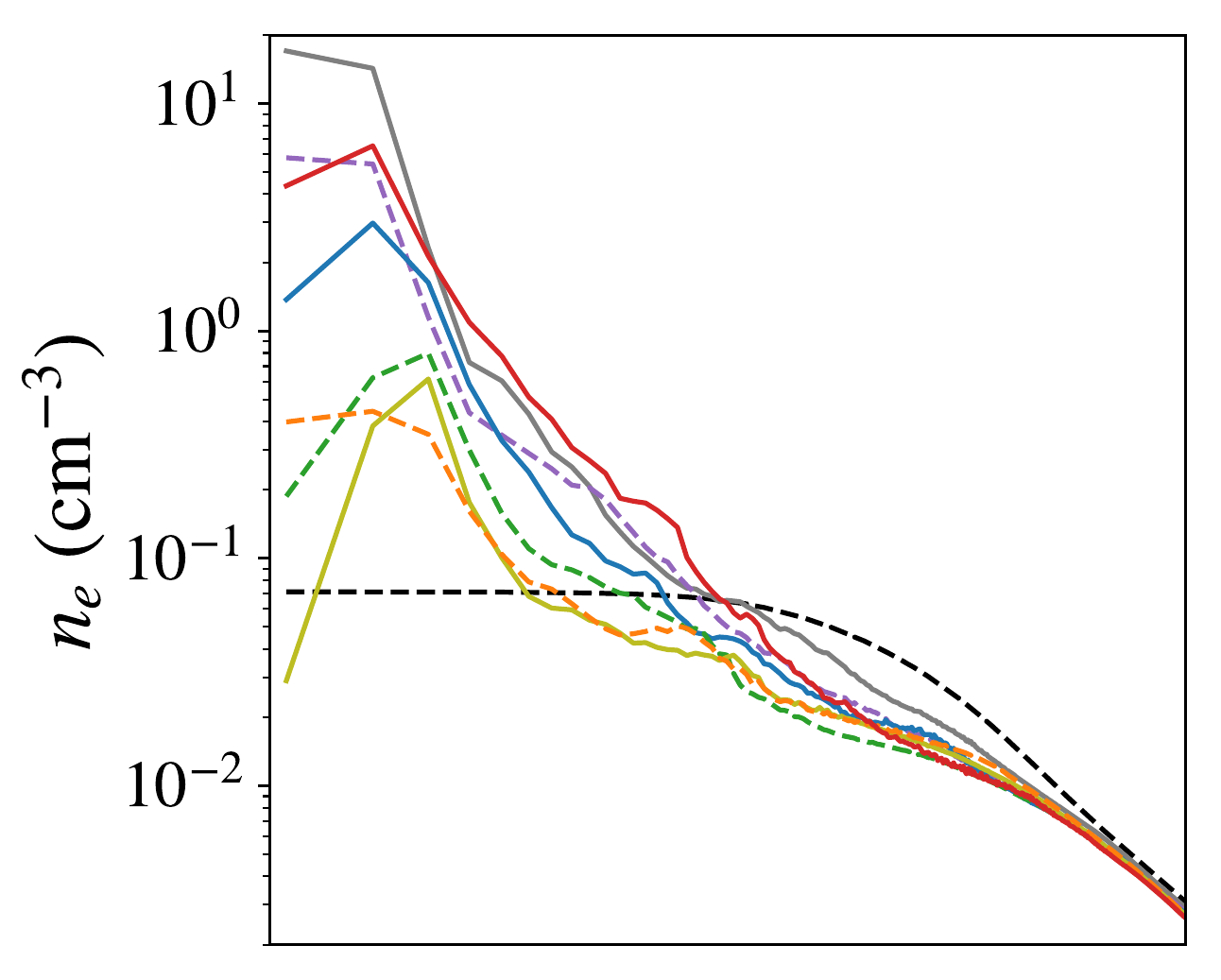}\includegraphics[trim=0.3cm 0.3cm 0.3cm 0.3cm, clip=true, height=0.231\linewidth]{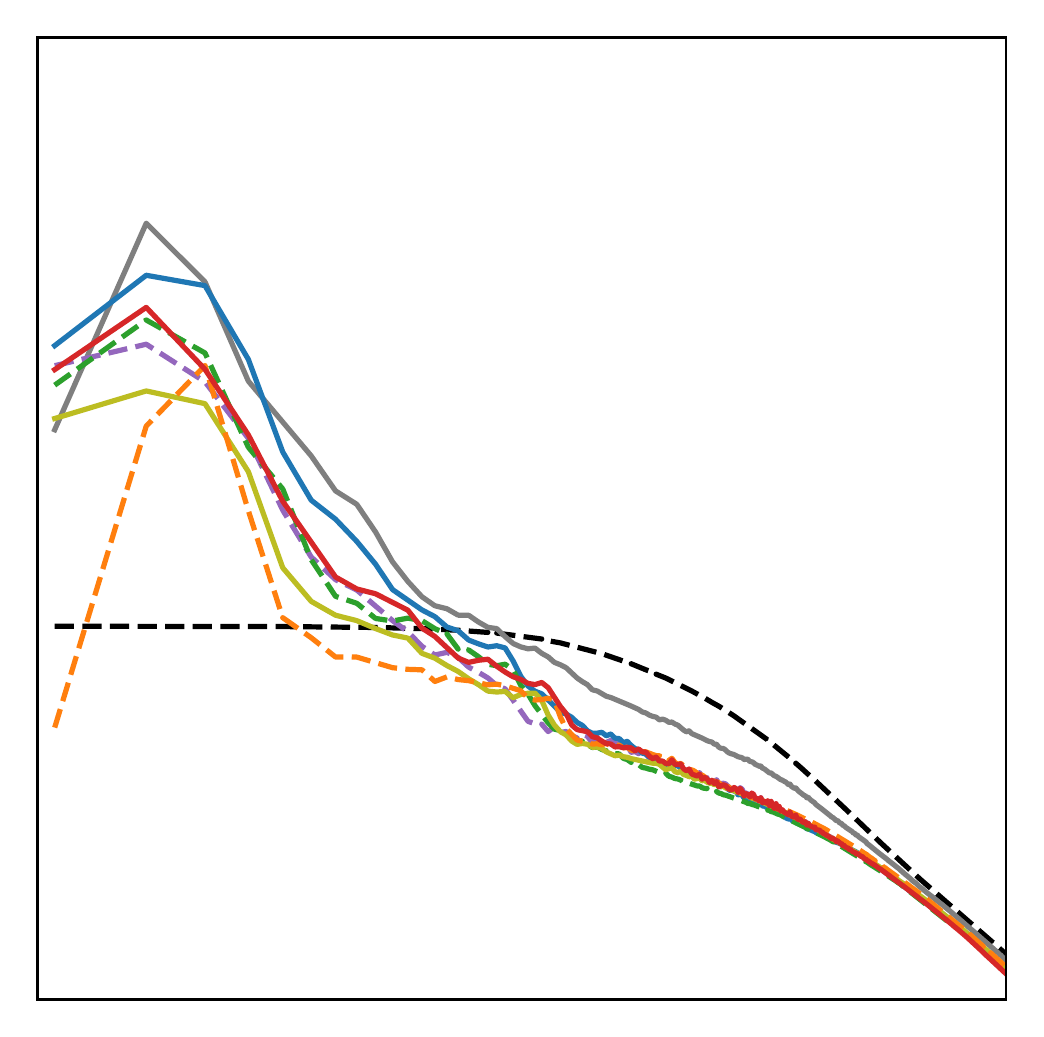}\includegraphics[trim=0.3cm 0.3cm 0.3cm 0.3cm, clip=true, height=0.231\linewidth]{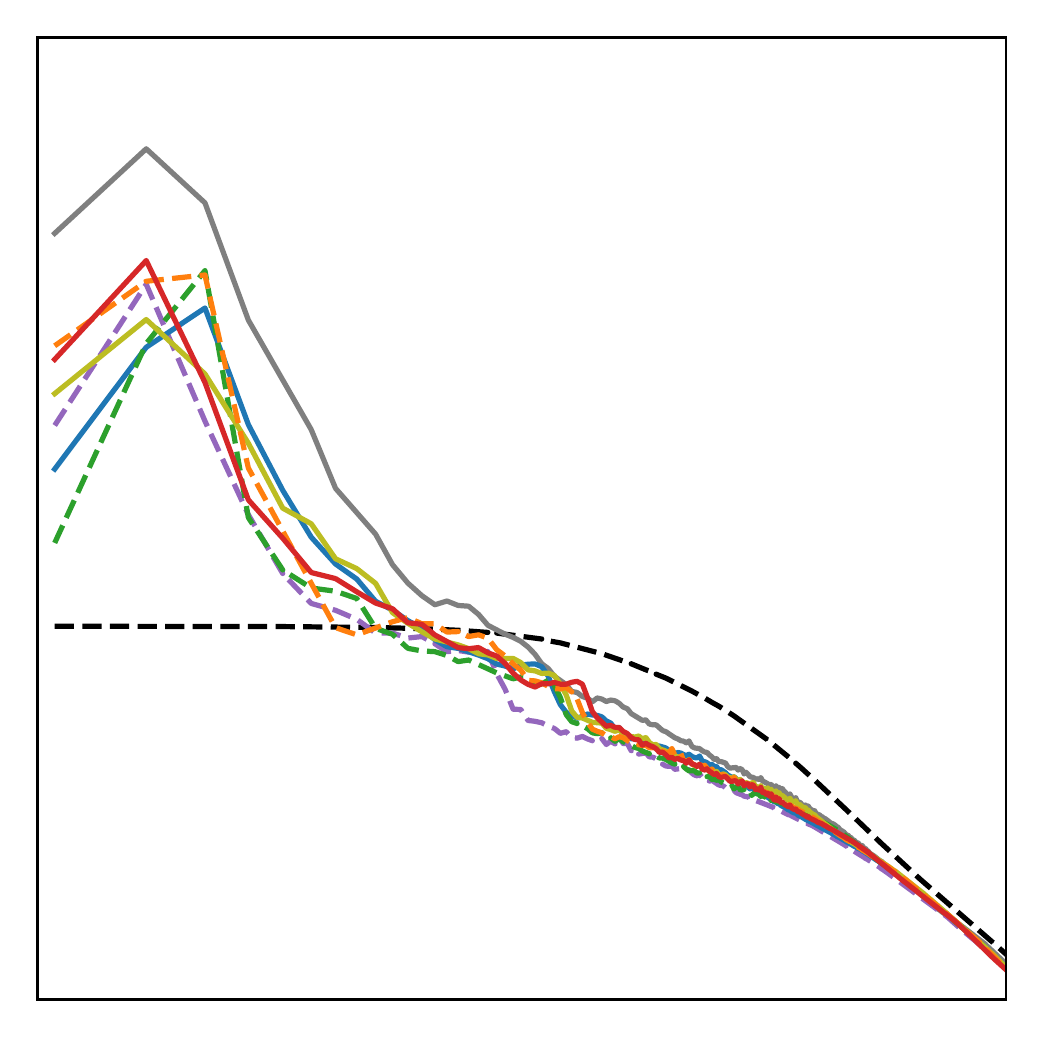}\includegraphics[trim=0.3cm 0.3cm 0.3cm 0.3cm, clip=true, height=0.231\linewidth]{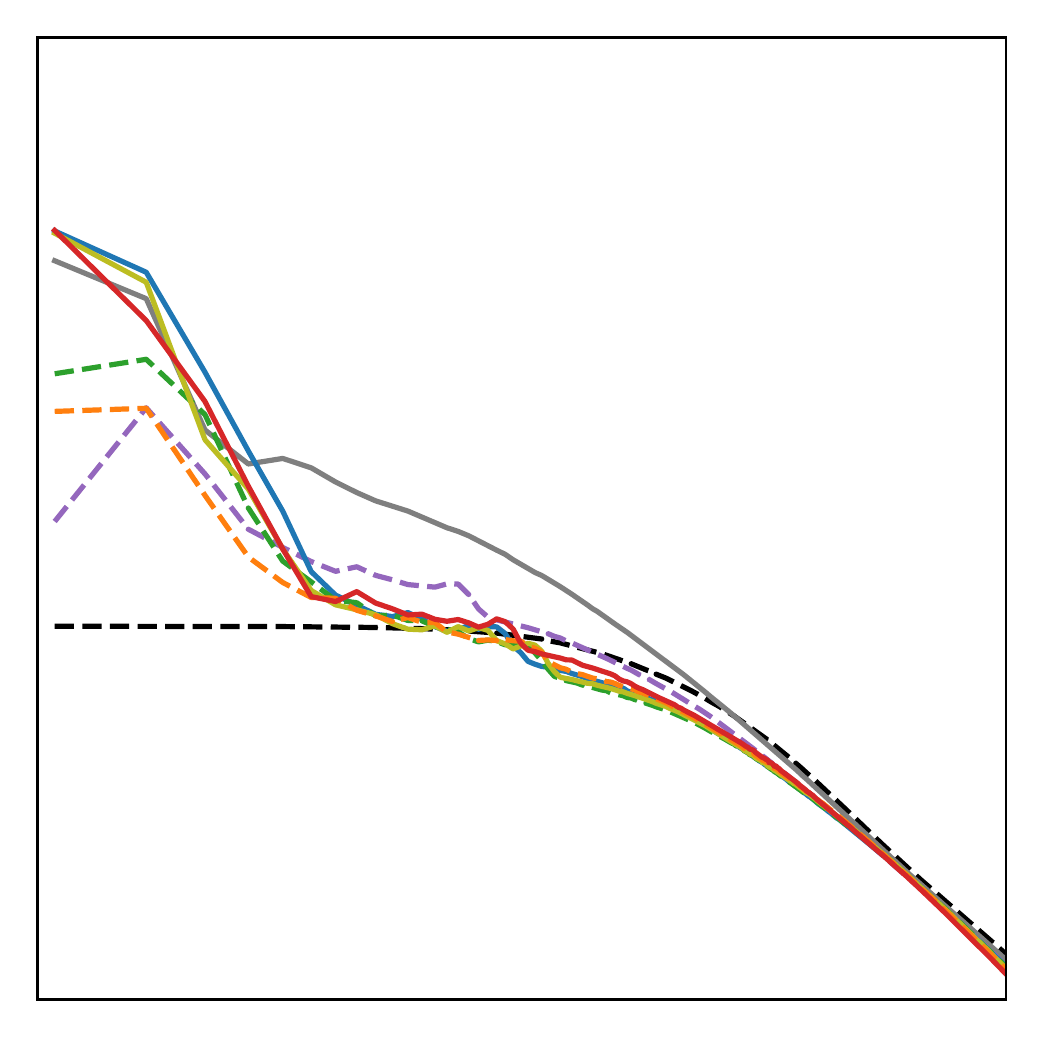}

\includegraphics[trim=0.2cm 0.3cm 0.3cm 0.3cm, clip=true, height=0.2705\linewidth]{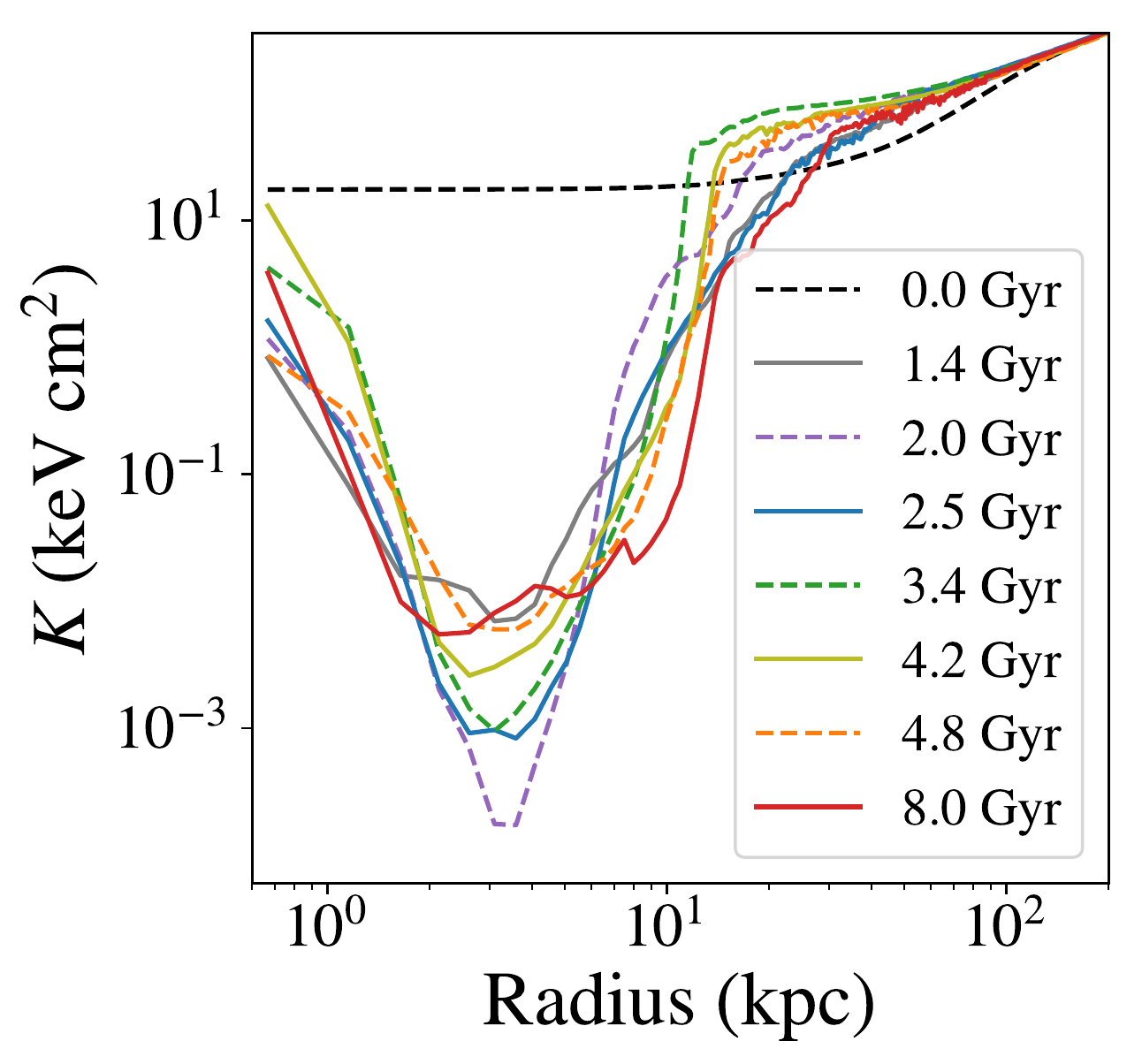}\includegraphics[trim=0.3cm 0.3cm 0.3cm 0.3cm, clip=true, height=0.2705\linewidth]{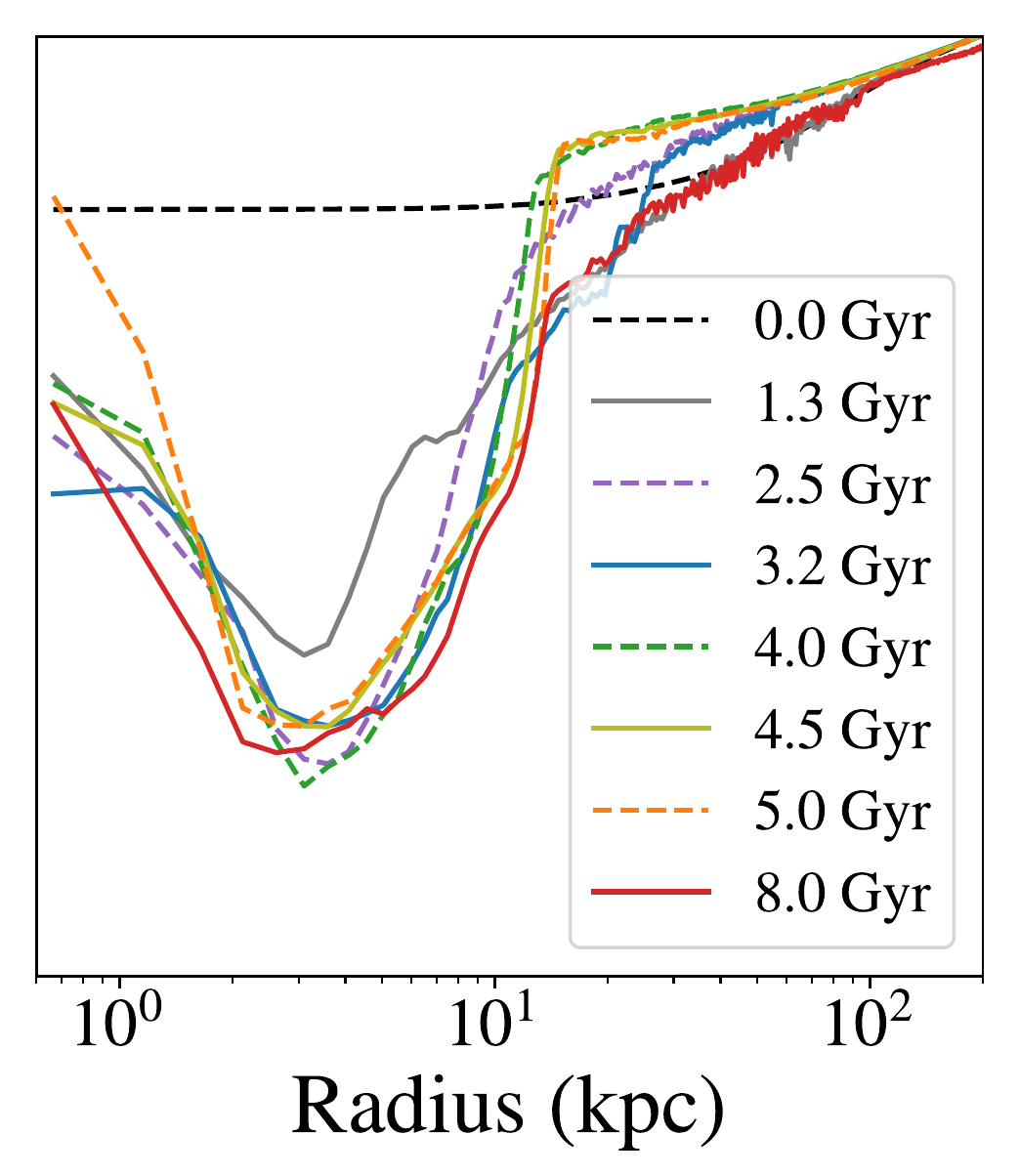}\includegraphics[trim=0.3cm 0.3cm 0.3cm 0.3cm, clip=true, height=0.2705\linewidth]{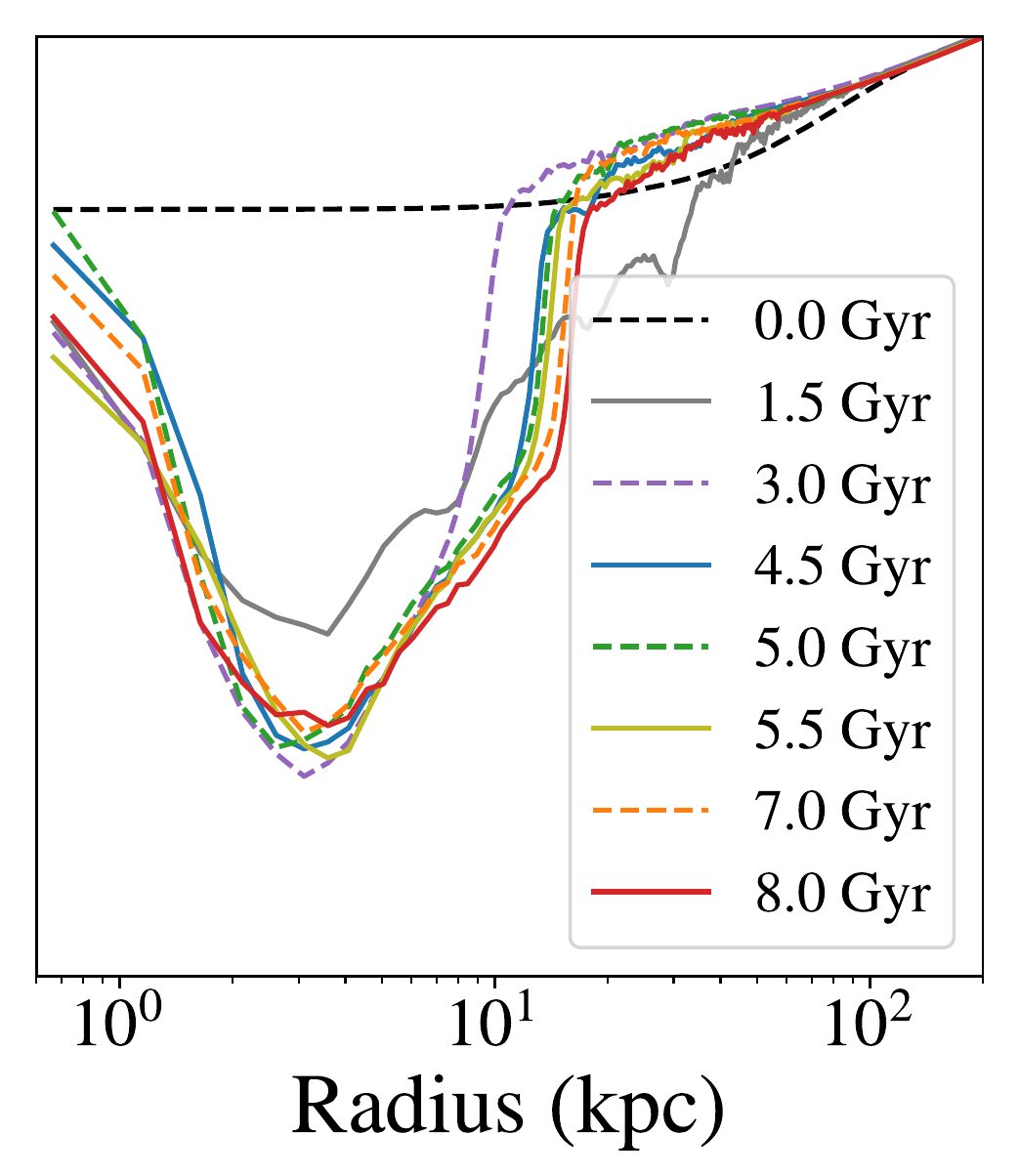}\includegraphics[trim=0.3cm 0.3cm 0.3cm 0.3cm, clip=true, height=0.2705\linewidth]{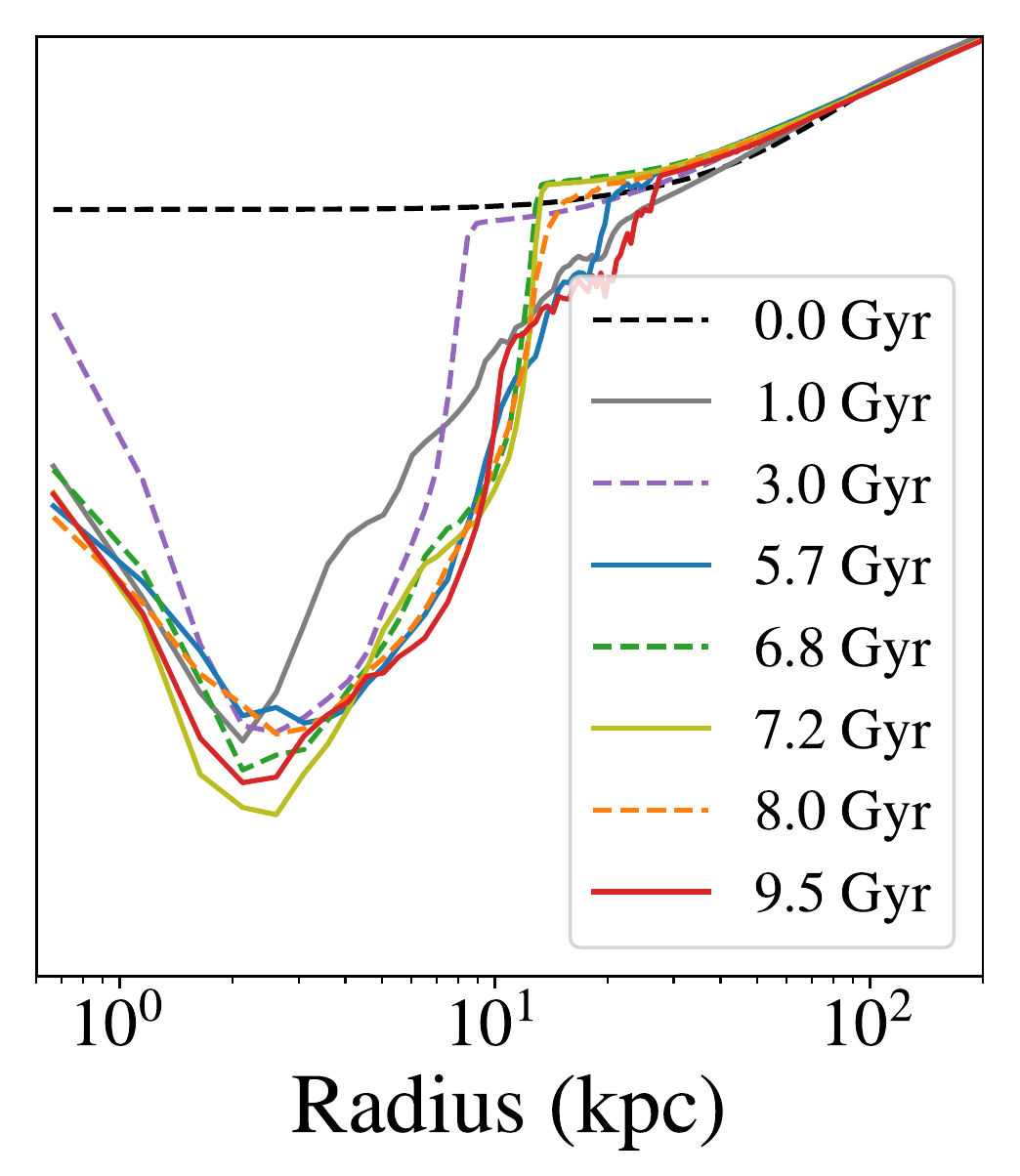}

\caption{Radial profiles of density, electron number density, and entropy in different simulations. From left to right, the columns show evolution of radial properties of the cluster in TI02 ($f_{\rm J}=0.1$), TI03 ($f_{\rm J}=0.9$), TI04 ($\epsilon=10^{-2}$), and TI08 ($\epsilon=10^{-4}$). Colors represent different times associated with the local maxima (solid) and minima (dashed) of the AGN feedback power.}
\label{fig:add_profile}
\end{figure}

In this section we describe the properties of the ICM measured from additional simulations, which can be compared to the RT02 run discussed in Section~\ref{sec:r_gas2} and shown in Figure~\ref{fig:radial}. Figure~\ref{fig:add_profile} shows the radial profiles of mass density, electron number density, and entropy in simulations TI02 ($f_{\rm J}=0.1$), TI03 ($f_{\rm J}=0.9$), TI04 ($\epsilon=10^{-2}$), and TI08 ($\epsilon=10^{-4}$), with key parameters listed in the parentheses. In all runs the central region is dominated by cold gas and beyond 10\,kpc the ICM properties oscillate around the initial values. One noticeable difference with the run RT02 is that due to the different implementation of radiative feedback in these runs (modeled as thermal injection), all exhibit higher temperature and electron number density in the inner few kpc region than RT02. The differences diminish beyond a few kpc as the outflows mix with the ICM. Therefore, the implementation of radiative feedback used in this work (ray tracing vs. thermal injection) does not have a very strong impact on the evolution of the ICM beyond the central few kpc.

For simulations with lower $f_{\rm J}$, illustrated by TI02, most of the feedback power is allocated to radiative feedback, which strongly heats and ionizes the inner few kpc but is not very effective at lowering the density of cold gas beyond this radius. Specifically, the density of the cold gas reaches maximum in this simulation and peaks at smaller radii relative to TI03 and other runs, indicating a more compact cold gas disk. The entropy of the central 10\,kpc also exhibits larger variations over time in TI02, due to the ``boom and bust'' feedback cycle dominated by radiative feedback.

In simulations with different values of $\epsilon$, represented by TI04 and TI08, the variation in ICM properties relative to the initial value illustrates the spatial "reach" of the AGN feedback. On the one hand, in the high-efficiency TI04 run, the properties of the ICM are affected by the AGN feedback out to 100\,kpc. On the other hand, in TI08, the ICM is only weakly affected by the AGN feedback beyond 10\,kpc. As noted in Section~\ref{sec:r_feedback}  and shown in Figure~\ref{fig:energy}, despite very different efficiencies, the cumulative energy of AGN feedback for these runs is similar, pointing to the self-regulation of accretion rate in response to AGN feedback.


\bibliographystyle{aasjournal}
\bibliography{apj-jour,myrefs}

\end{CJK}
\end{document}